\documentclass[aps,prb,twocolumn,superscriptaddress,floatfix,english,twocolumn,amsmath,amssymmb,longbibliography]{revtex4-2}

\usepackage{graphicx} 
\usepackage{dcolumn}
\usepackage{bm}
\usepackage{hyperref}
\usepackage{color}
\usepackage[T1]{fontenc}
\usepackage{amsmath}
\usepackage{amssymb}
\usepackage{esint}
\usepackage{comment}

\usepackage{amsbsy}
\usepackage{amsfonts}
\usepackage{babel}
\usepackage{enumerate}
\usepackage{epsf}
\usepackage{esint}
\usepackage{epsfig,psfrag}
\usepackage{float}
\usepackage{graphicx}
\usepackage{latexsym}
\usepackage{mathbbol}
\usepackage[usenames,dvipsnames]{xcolor}
\usepackage[mathscr]{eucal}
\usepackage{wrapfig} 
\usepackage[normalem]{ulem}

\definecolor{amaranth}{rgb}{0.9, 0.17, 0.31}

\begin{abstract}
    We consider a smooth interface between a topological nodal-line semimetal 
    and a topologically trivial insulator (e.g., the vacuum) or another semimetal with a nodal ring of different radius.
    Using a low-energy effective Hamiltonian including only the two crossing bands, 
    we show that these junctions accommodate a two-dimensional zero-energy level and
    a set of two-dimensional dispersive bands, 
    corresponding to states localized at the interface. 
    We characterize the spectrum, identifying the parameter ranges in which these states are present, 
    and highlight the role of the nodal radius and the smoothness of the interface. 
    We also suggest material-independent ways to detect and identify these states, 
    using optical conductivity and infrared absorption spectroscopy in magnetic field.
\end{abstract}
    
\begin{document}

\title{Dispersive Drumhead States in Nodal-Line Semimetal Junctions}

\author{Francesco Buccheri} 
\email{francesco.buccheri@polito.it}
\affiliation{Politecnico di Torino, Corso Duca degli Abruzzi, 24, I-10129 Torino, Italy}
\affiliation{Institut f\"ur Theoretische Physik, Heinrich-Heine-Universit\"at, Universit\"atsstr. 1, D-40225  D\"usseldorf, Germany}

\author{Reinhold Egger} 
\email{egger@hhu.de}
\affiliation{Institut f\"ur Theoretische Physik, Heinrich-Heine-Universit\"at, Universit\"atsstr. 1, D-40225  D\"usseldorf, Germany}

\author{Alessandro De Martino}
\email{Alessandro.De-Martino.1@city.ac.uk}
\affiliation{Department of Mathematics, City, University of London, Northampton Square, EC1V 0HB London, United Kingdom}

\maketitle


\section{Introduction}
\label{sec:Introduction}

Nodal-line semimetals (NLS) are a class of materials characterized by a gapped spectrum 
everywhere in the Brillouin zone (BZ), with the exception of one or more non-degenerate (Weyl) 
or doubly-degenerate (Dirac) band crossings, which occur on a one-dimensional manifold (an ellipse or a ring) 
in proximity of the Fermi energy. The toroidal Fermi surface and the high carrier mobility 
are associated with strong and characteristic thermoelectric and magnetotransport response 
\cite{Syzranov2017,Laha2020,Barati2020,Yang2022,Lau2021,Wang2022b}. A prominent example is 
$\mbox{Ca}_3\mbox{P}_2$,
whose synthesis was reported in \cite{Xie2015}. Intense experimental 
and theoretical activity has allowed to isolate several other compounds 
\cite{Abedi2022,Yu2017,Weng2016,Mele2019,Wu2022,Gao2023,Chang2019,Hu2017,Laha2019,Emmanouilidou2017}, 
among which \mbox{ZrSiS} \cite{Pezzini2017}, $\mbox{Zr}\mbox{Te}_5$ 
\cite{Wang2022a,Wang2022b}, \mbox{ZrGeSe} \cite{Guo2019}, 
\mbox{ZrSiTe} \cite{Hu2016,Stuart2022}, and to detect characteristic transport signatures, 
as well as the topological surface states via quasiparticle interferometry 
or angle-resolved photoemission spectroscopy. 
Synthetic materials, such as phononic \cite{Deng2019} 
and photonic crystals \cite{Park2022,Bercioux2023}, 
have also recently provided valuable experimental realizations of the spectrum.

The presence of a nodal line is associated with topological "drumhead" states (TDS) 
when the NLS has an ideally sharp interface with a trivial insulator, such as the vacuum. 
It is known that these states are exponentially localized on the surface, 
and that their support in momentum space
is enclosed by the projection of the nodal line on the surface Brillouin zone \cite{Burkov2011}.
\begin{figure}
    \centering
    \includegraphics[width=0.9\columnwidth]{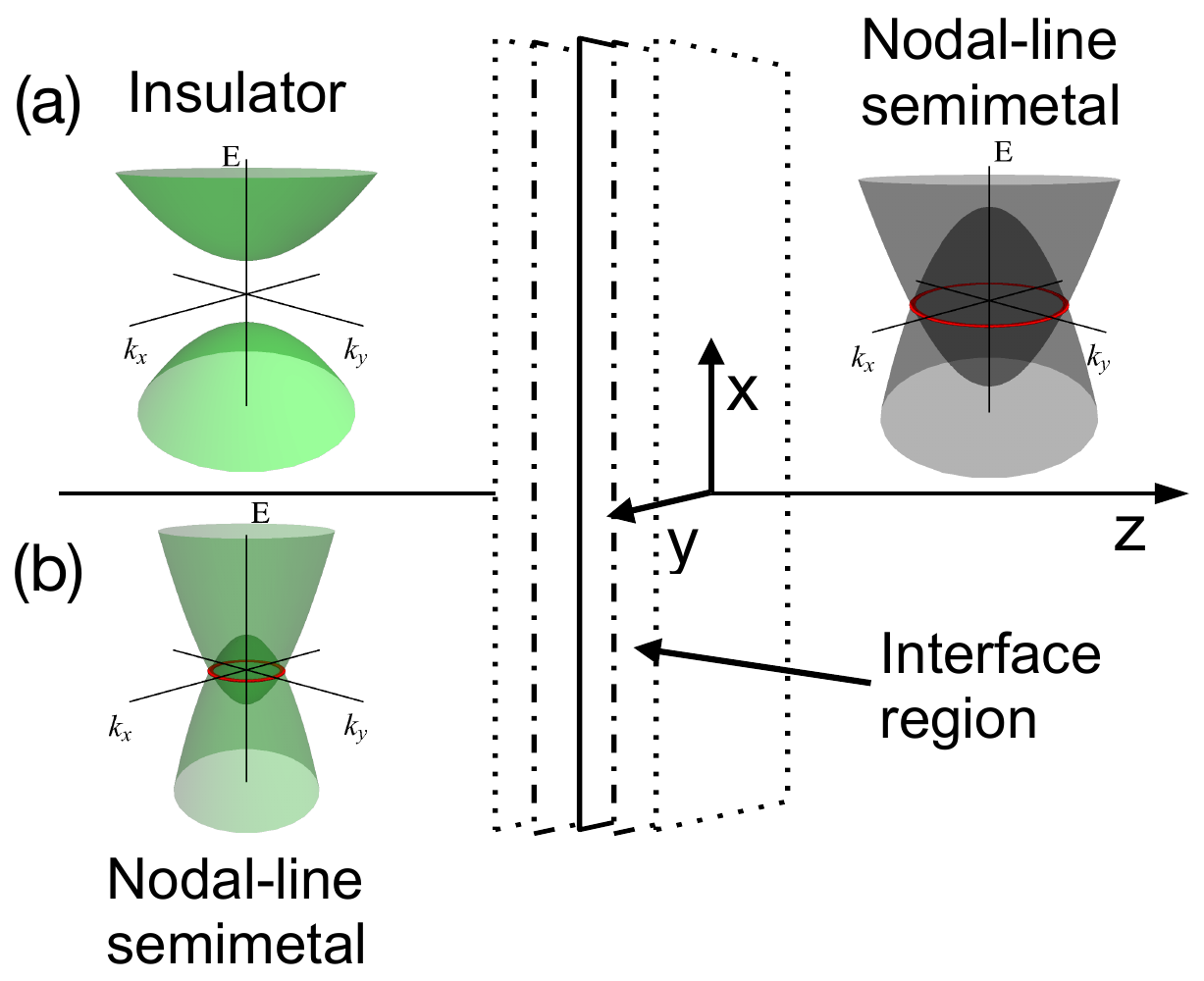}
    \caption{Schematic representation of the system under study. 
    The nodal-line semimetal on the right is separated from the insulator (a) or from 
    another nodal-line semimetal with smaller nodal radius (b) on the left 
    (represented in the plot by their respective band structures) 
    by an extended intermediate region, in which the nodal line gradually
    shrinks and, in case (a), a band gap opens.
    }
    \label{fig:interface}
\end{figure}
Real interfaces can, however, differ from the ideal interfaces in many aspects. 
Apart from lattice mismatch effects between the materials, 
which we will not consider in this work, electrostatic effects and non-uniformity 
of the surface on the atomic scale can have remarkable consequences. 
In topological insulators, 
the electrostatic bending of the bands can determine the presence of additional 
surface states of non-topological origin \cite{Bianchi2010,Chen2012,Alspaugh2020}. 
Moreover, it is well-known that a discrete spectrum of finite-energy dispersive surface states appears if a band inversion occurs smoothly in space over a length $\ell>2\hbar v/\Delta$, 
with $v$ being the Fermi velocity of the bulk excitations
and $\Delta$ the band gap
\cite{Volkov1985,Volkov1986,Lu2019}.
Such smooth interfaces can be realized in heterostructures
as a consequence of the interplay of an electrostatic potential 
and the strain effects from a substrate \cite{Inhofer2017}.
Moreover, the situation in which the two materials are both present in
an intermediate region, as in specially synthesized samples via chemical substitution \cite{Bermejo2022},
provides another example of a smooth junction. 
These so-called Volkov-Pankratov states have been detected 
via their signatures in absorption spectroscopy and AC transport measurements.
These ideas can promptly be translated to topological semimetals. 
The presence of relatively smooth interfaces, in fact,
may be a result of the geometry of the heterostructure, 
or a way of modeling a surface inhomogeneous on the order of several unit cells, 
or present from the synthesis stage (e.g., from blending with a minority component \cite{Xie2015}), 
or engineered in the sample preparation stage via chemical substitution \cite{Nilforoushan2021}.

In technological applications, interfaces are not only unavoidable, but even welcome, 
because topologically protected surface states are at the core of mechanisms that limit 
phonon- or impurity-induced dissipation in a wide spectrum of topological materials 
\cite{zhu2013,Gorbar2016,Breitkreiz2019,DeMartino2021,Buccheri2022a}. 
In NLS, the TDSs are associated with specific transport phenomena, 
such as spin-flipped reflection of electrons within the region of the BZ enclosed 
by the nodal line, and can be used to probe the topological nature of the sample \cite{Chen2018}. 
What is more, interface states have recently been proposed as a building block 
of highly tunable Landau-level laser sources \cite{Goerbig2023}. 
Understanding the precise nature of the surface states and their interactions is 
therefore of great importance for the advancement of the field.

In this work, using an effective Hamiltonian that describes the physics of a NLS
in the vicinity of the nodal line, we show that additional states besides the TDSs
can be present at the interface with a topologically trivial insulator (including the vacuum), in the configuration of Fig. \ref{fig:interface}.  
We call these additional interface states ``dispersive drumhead states'' (DDSs). They
are similar to the TDSs, but they instead exist at finite energy and are 
not protected by a topological index. 
 One may consider the DDSs as 
 analogues of the so-called Volkov-Pankratov states, which 
 have been investigated in various related contexts, e.g., in graphene \cite{vandenBerg2020,Kawakami2023}, 
topological insulators \cite{Tchoumakov2017,Lu2019,Bercioux2020} and Weyl semimetals \cite{Mukherjee2019}.
Indeed, for a fixed value of the momentum in the plane of the nodal line,
one can see the DDS as originating from the gap inversion described in \cite{Volkov1985,Volkov1986}. 
However, an important difference is that the DDSs have a finite support in momentum space, 
bounded by the nodal line.
We provide bounds on the parameters describing the sharpness of the surfaces 
for such states to appear and relate the number of states in the spectrum 
to three dimensionless parameters encoding the relevant characteristics of the sample and the external field.
We provide the exact energies and eigenfunctions of the localized states and point out 
that magneto-optical absorption spectroscopy in the infrared region and the related 
frequency-dependent optical conductivity offer a most effective way to observe the presence 
of the localized states in experiments \cite{Inhofer2017,Bermejo2022}.

This work is structured as follows: in Sec.~\ref{sec:Model} we introduce
our effective model and describe the general strategy to solve the
associated Schr\"odinger problem. Subsequently, in Sec.~\ref{sec:SpectrumNoField},
we describe the spectrum in the absence of an external magnetic field.
We then study the problem in a magnetic field in Sec.~\ref{sec:SpectrumB}
and describe the optical absorption lineshapes in Sec.~\ref{sec:Absorption}.
We summarize the main characteristics of the spectrum and discuss the
detection possibilities in Sec.~\ref{sec:Conclusions}. Technical details can be found 
in various Appendices. We often use units with $\hbar=1$.

\section{Model}
\label{sec:Model}

We consider the low-energy Hamiltonian describing the electronic degrees of freedom in the vicinity of a band crossing in a generic nodal-line semimetal \cite{Chan2016a}
\begin{equation}
H (\boldsymbol{k} )  =  
v_{z}k_{z}\tau_{y} +  M(k_{p})\tau_{z} + f(k_{p})\tau_{0},
\label{eq:H}
\end{equation}
where the $\tau_{j}$  ($j=x,y,z$) are the Pauli matrices,
\mbox{$\boldsymbol{k}=\left(\boldsymbol{k}_{p},k_z\right)$},
\mbox{$\boldsymbol{k}_{p}=\left(k_{x},k_{y}\right)$} is the component of
the momentum in the plane of the nodal line and $k_{p}$ its modulus. The
``mass'' function is 
\begin{equation}\label{eq:MassFunction}
    M(k_p) = D_p\left( k_p^{2} - a \right),
\end{equation}
and the particle-hole symmetry breaking term is
$$
f(k_p) = D_{0} \left( k_p^{2}-a \right) + V_{0},
$$
where $v_{p}=D_p/\sqrt{\left|a\right|}$ and $v_{z}$ are (positive) velocities 
and $V_{0}$ is a chemical potential. 
With the two bands having opposite eigenvalues under inversion,
the latter is a symmetry of the Bloch Hamiltonian~\eqref{eq:H}, as
$$
H(\boldsymbol{k}) = \tau_{z} H(-\boldsymbol{k})\tau_{z}.
$$
We will often consider, as a reference, 
$\mathrm{Ca}_{3}\mathrm{P}_{2}$ 
\cite{Xie2015}, which exhibits a band crossing in a window of $\pm10$ meV around the Fermi energy. 
In this case, the Pauli matrices act on an orbital degree of freedom, 
and the model possesses full $SU(2)$ spin rotation invariance, as it is diagonal in the
spin degree of freedom. Then the time-reversal operator $\mathcal{T}$
squares to the identity and is represented as the complex conjugation $\mathcal{T}=K$  \cite{Chan2016a}. 
It is a symmetry of~\eqref{eq:H}, as 
$$
H(\boldsymbol{k})=H^{*}(-\boldsymbol{k}).
$$
The simplified model with $f(k_{p})=0$ has an additional
particle-hole symmetry that exchanges valence and conduction bands,
implemented by the transformation 
$$
\tau_{x}H^{*}\left(-\boldsymbol{k}\right)\tau_{x}=-H\left(\boldsymbol{k}\right).
$$
Parameter estimates for the example mentioned above can be found
in \cite{Chan2016a}. Here we only quote the values 
$\hbar v_z\approx 2.5$\,eV\,\AA\, and $D_p\approx 4.34$\,eV\AA$^2$\,for later use. 
The effective two-band model~\eqref{eq:H}
is possibly the simplest model  
describing a ring-shaped band crossing in a generic NLS. 
Indeed, the spectrum consists of the two bands 
\begin{equation}
E_{\pm}(\boldsymbol{k}_p,k_z)=f(k_{p})\pm\sqrt{v_{z}^{2}k_{z}^{2} + M^{2}(k_{p})}.
\label{bulkeigenvalues}
\end{equation}
Its defining feature is that, for $a>0$, the bands touch on the nodal line
defined by $k_{z}=0$, $k_{p}=\sqrt{a}$, which is generically allowed
if a mirror symmetry is present \cite{Fang2015}. Conversely, whenever
$a<0$, a gap of magnitude $2D_p\left|a\right|$ appears. 
The eigenstates of the Bloch Hamiltonian are denoted by 
$\left|u_{j}\left(\boldsymbol{k}\right)\right\rangle $, with $j=\pm$ 
denoting the conduction/valence band, and can be used to define Zak's phase
\begin{eqnarray}
\mathcal{P}_-\left(\boldsymbol{k}_{p}\right) & = & i\intop_{-\pi}^{\pi}dk_{z}
\left\langle u_{-}\left(\boldsymbol{k}\right)\right|\partial_{k_{z}}\left|u_{-}\left(\boldsymbol{k}\right)\right\rangle. 
\label{eq:Berryphase}
\end{eqnarray}
This integral is quantized in integer multiples of $\pi$
as a consequence of inversion or mirror symmetry along the $z$ axis  \cite{Bernevig2013,Chiu2014}, 
and can therefore be seen as the topological invariant of a one-dimensional system with two parameters, 
the momentum components $\boldsymbol{k}_{p}$. While the value of this integral is not directly computable 
within the framework of the low-energy Hamiltonian \eqref{eq:H}, 
one can easily compute the difference in this value when the in-plane momentum $\boldsymbol{k}_p$ 
lies inside and outside the nodal line by deforming the contour to a loop enclosing the nodal line.
The Berry phase along this loop is the topological invariant of the NLS defined in \cite{Fang2015,Fang2016}. 
While the Berry curvature itself is vanishing in the BZ, 
the Berry phase acquired on a loop enclosing the nodal line is not, 
due to the singularity introduced by the band crossing.
In fact, the Berry phase is quantized by the presence of a discrete symmetry, 
namely, the mirror symmetry with respect to the plane of the nodal line
\cite{Chiu2014}. In this work,
we also restrict ourselves to the class of models which possess
time-reversal symmetry in the bulk, implemented as described above.

\begin{figure}
\centering{}
\includegraphics[width=0.9\columnwidth]{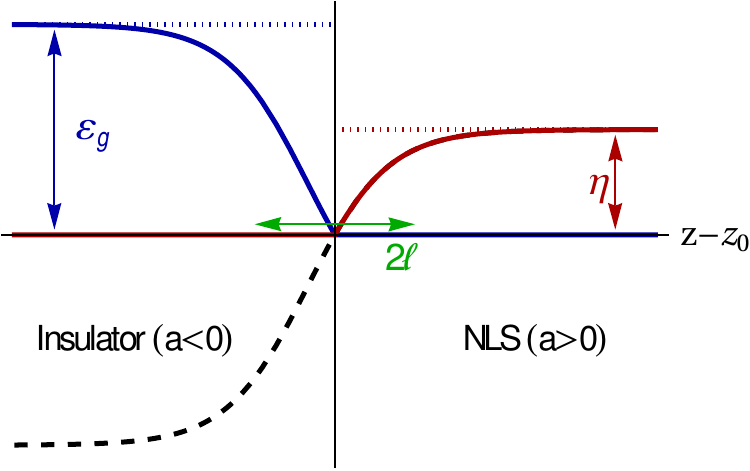}
\caption{Schematic plot of the position-dependent energy gap (solid blue line) 
and of the nodal-line radius (solid red line), as defined from the function \eqref{eq:a(z)}, 
or mass function \eqref{eq:MassFunction} computed at $k_p=0$ (dashed line).  
The parameters $\eta$ and $\varepsilon_\text{g}$ are defined in Eq.~\eqref{eta}.
The dotted lines depict instead the limit in which the interface is sharp, 
with a sudden gap opening at $z=z_0$.
}
\label{fig:profile}
\end{figure}

In order to model an insulator, we need to open a band gap, but the symmetries prevent the insertion 
of terms proportional to $\tau_x$ or $\sim k_j\tau_x$. We can, however, select a negative 
value of the parameter $a$, thus creating a gap of size $2D_p|a|$. We can then model an interface between 
a trivial insulator and a nodal-line semimetal by defining a function $a=a\left(z\right)$, 
which breaks translation invariance in the $z$ direction. 
We here study the case where the interface plane and the nodal-line plane are parallel.
We also assume that this symmetry 
breaking takes place locally, in a region of size $\sim\ell$, while far from the interface 
the bulk translation invariance is restored. 
The shape of the function $a$ must be selected in such a way that
1) the sharpness of the interface can be tuned by a characteristic
length parameter $\ell$, and 2) it asymptotically reproduces an insulator
with a defined band gap or a semimetal with a defined nodal line.
We conveniently select a function that satisfies these requirements
and allows full analytical progress in the form:
\begin{equation}
a (z) = a_0 + a_1 \left[ \tanh\left( \frac{z}{\ell} \right) -1 \right],
\label{eq:a(z)}
\end{equation}
with positive parameters $a_0$ and $a_1$. 
The interface profile is illustrated in Fig.~\ref{fig:profile}.
We see that on the far right ($z\gg\ell$), 
we have a NLS with nodal radius $k_0=\sqrt{a_0}$. 
The nodal line shrinks to a point when $z\to z_0$,
with \mbox{$z_0 = \ell\, \text{arctanh}\, (1-a_0/a_1)$}, and
a gap opens when $z<z_0$, as long as $2a_1>a_0$. On the far left ($z\ll-\ell$), 
we then have an insulator with gap \mbox{$2E_\text{g} = 2D_p(2a_1 - a_0)$},
see also Fig.~\ref{fig:interface}.
This model allows us to consider also the case of an interface between two NLSs
with different nodal radii, illustrated in Fig.~\ref{fig:interface} too. 
This is realized when $E_\text{g}<0$, such that 
for $z\ll-\ell$ we have a NLS with nodal radius $k_0'=\sqrt{a_0-2a_1}$.

Ignoring for the moment the particle-hole symmetry-breaking term, 
the model contains three energy scales: the bulk band edges at $\boldsymbol{k}=0$ 
on the right and on the left, respectively $E_0=D_pa_0$ 
and $E_\text{g}=D_p(2a_1-a_0)$, 
and the characteristic energy scale associated with the inhomogeneity 
in the $z$ direction, $\frac{\hbar v_z}{\ell}$. 
The physics of the interface problem is then fully determined by
the dimensionless ratios
\begin{equation}
    \eta =\frac{\ell D_pa_0}{\hbar v_z}, 
    \quad \varepsilon_\text{g} = \frac{\ell D_p (2a_1-a_0)}{\hbar v_z}.
    \label{eta}
\end{equation}
It is convenient to introduce also their combination 
\begin{equation}
    \lambda =\frac{\ell D_pa_1}{\hbar v_z} = \frac{\varepsilon_\text{g}+\eta}{2}, 
       \label{eq:smoothness}
\end{equation}
which we refer to  as the "smoothness" parameter.
We will see below that 
the characteristics of the interface states depend on $\lambda$  and $\eta$.

We use the energy scale $\hbar v_z/\ell$ to define 
the reduced Hamiltonian $\mathcal{H} =\frac{\ell H}{\hbar v_{z}}$ 
and the corresponding energy relative to the 
chemical potential $\varepsilon=\frac{\ell\left(E-V_{0}\right)}{\hbar v_{z}}$, 
and rescale the longitudinal coordinate $z$ by the junction characteristic length $z/\ell \rightarrow z$. 
The momenta in the nodal line plane are instead more conveniently rescaled as $q_p=k_p/\sqrt{a_0}$.
With these conventions, the reduced Hamiltonian assumes the form
\begin{equation}
\mathcal{H} = -i\tau_{y}\frac{\partial}{\partial z} +
\frac{\tau_{z}+
\gamma\tau_{0}}{\sqrt{1-\gamma^{2}}} \left[  d(q_p ) - \lambda \tanh z\right],
\label{eq:hgamma}
\end{equation}
where we have defined the auxiliary function of the transverse momentum
\begin{equation}
d(q_p)=  \eta q_{p}^{2} - \eta +\lambda.
\label{eq:deltabar0}
\end{equation}
Translation invariance in the direction perpendicular
to the interface ($z$ direction) is broken, hence, the Bloch Hamiltonian \eqref{eq:h} is parametrized
by the two surviving momenta $q_x$ and $q_y$ in the plane defined by the interface.
The parameter $\gamma=\frac{D_0}{D_p}$ in Eq.~\eqref{eq:hgamma} quantifies the 
particle-hole symmetry breaking, with $\gamma=0$ being the particle-hole symmetric case. 
For the sake of simplicity, we consider throughout this paper the model with $\gamma=0$. 
The general case is considered in App.~\ref{sec:gamma}, where we 
show that the results obtained for $\gamma=0$ can be adapted to the case $\gamma\neq 0$. 
Since for $\gamma=0$ the spectrum is particle-hole symmetric, 
we restrict our discussion to the non-negative part, 
being understood that each positive energy state has a negative energy partner. 

Before closing this section, we note that,
while the nodal line in real materials does not necessarily form a circle in the BZ,
more general shapes can be accommodated in our model by using an appropriate coordinate 
system in the nodal plane.
These material-specific features are not expected to produce qualitative changes to the 
picture of DDSs presented here.

\section{Spectrum of the interface problem} 
\label{sec:SpectrumNoField}

\subsection{The Dirac-Schr\"odinger equation and its general solution}
\label{gensol}

Here, we study the Dirac equation $\mathcal{H} \psi = \varepsilon  \psi$ 
with the Hamiltonian \eqref{eq:hgamma} and $\gamma=0$:
\begin{equation}
\mathcal{H} =  -i\tau_{y} \partial_z + \left[ d(q_p)- \lambda \tanh z \right] \tau_{z}.
\label{eq:h}
\end{equation}

The asymptotic behavior of the solutions for $z\to\pm \infty$ is determined just 
by the bulk gaps and is independent of the specific details of the interface profile.
We can identify three spectral regions, characterized by the parameters 
\begin{equation}
\kappa_{\pm}\left(q_p\right) =\sqrt{\left[d(q_p)\mp\lambda\right]^{2}-\varepsilon^{2}},
\label{kappapm}
\end{equation}
which control the behavior of the solutions for $z\to\pm\infty$. 
As they play an important role in the following, it is worth writing 
these two quantities explicitly as
\begin{align}
\kappa_{+} & = \sqrt{( \eta q_p^2-\eta)^2 -\varepsilon^2},
\label{kappap}
\end{align}
in which the radius of the nodal ring appears through $\eta$, 
and
\begin{align}
\kappa_- &= \sqrt{(\eta q_p^2 + \varepsilon_\text{g})^2 -\varepsilon^2} ,
\label{kappapmexplicit}
\end{align}
in which the gap parameter is explicit. 

In the energy--transverse momentum region where $\kappa_\pm$ are both real,  
the solutions describe states localized at the interface, exponentially decaying 
on both sides of the junction with different decay lengths, given by $\kappa_+^{-1}$ 
on the right  and $\kappa_-^{-1}$ on the left. 
These solutions represent  interface states, similar to the interface arcs 
of Weyl semimetals \cite{Mukherjee2019,Buccheri2022b}, 
with the important difference that in the present situation they form two-dimensional bands,
as opposite to the one-dimensional interface arcs. 
We discuss these solutions in Sec.~\ref{subsec:drumheadstates}.
In the region where $\kappa_+$ is imaginary
and $\kappa_-$ real, the solutions describe propagating states 
in the NLS which are perfectly reflected by the interface, with evanescent waves in the 
left side of the junction. Finally, if $\kappa_\pm$ are both imaginary, we have propagating states 
in both sides, with finite transmission across the interface. These last two types of 
solutions are discussed in Sec.~\ref{scatteringstates}.

The choice of the interface profile~\eqref{eq:a(z)} allows for the complete analytical 
solution of \eqref{eq:h} in all spectral regions, which we present in App.~\ref{sec:Equations}. 
We summarize here the basic steps. First, we make the ansatz 
\begin{equation}
\psi = \begin{pmatrix}
\psi_{1} \\ \psi_{2}
\end{pmatrix} 
= \phi_+ | + \rangle + \phi_- | - \rangle,
\label{phi}
\end{equation}
where $\left|\pm\right\rangle $ denote the normalized eigenstates of the Pauli matrix $\tau_{x}$. 
Then we introduce the variable
\begin{equation}
u  = \frac{1-\tanh z}{2},
\label{eq:u}
\end{equation}
which ranges between $0$ and $1$. In terms of this variable, 
the right asymptotic region $z\to + \infty$ is mapped into the limit $u \to 0$,  
while the limit $u \to 1$ corresponds to the left asymptotic region $z \to -\infty$.
Based on the asymptotic considerations above, we factorize the wave function as
\begin{align}
\begin{pmatrix}
    \phi_+\\
    \phi_-
\end{pmatrix} =  u^{\kappa_{+}/2}\left(1-u\right)^{\kappa_{-}/2}
\begin{pmatrix}
    \chi_+ \\
    \chi_-
\end{pmatrix}.
\label{eq:xidef}
\end{align}
In App.~\ref{sec:Equations} we show that the functions $\chi_{\pm}(u)$ satisfy 
an hypergeometric equation, see App. \ref{sec:Hypergeometric}, and provide the details of the solution.  
In the following, we discuss in turn the spectrum of bound states and of propagating states.

\subsection{Interface states}
\label{subsec:drumheadstates}

First, we consider the energy-momentum domain in which $\kappa_\pm$ are both real:
\begin{equation}
    \varepsilon^2< \min \left\{(\eta q_p^2-\eta)^2, (\eta q_p^2+\varepsilon_\text{g})^2 \right\}.
 \label{domainBS}
\end{equation}
In this region, the spectrum consists of states exponentially localized around the junction. 

From general considerations based on the bulk-boundary correspondence, 
we expect a topologically-protected localized state at the interface between 
samples with different values of a bulk topological index \cite{Hasan2010,Burkov2011,Chiu2014,Fang2016}. 
Going back to our Dirac equation \eqref{eq:h}, it follows indeed that  a unique TDS state solution with zero energy can be found in the form
\begin{eqnarray}
\psi_{0}\left(z\right) & = & 
\mathcal{N}_{0}\frac{e^{d(q_p)z}}{\cosh^\lambda (z)}\left|+\right\rangle,
\label{eq:Psi0}
\end{eqnarray}
where the normalization coefficient $\mathcal{N}_{0}$
is provided in Eq.~\eqref{eq:N0}. The state~\eqref{eq:Psi0} exists 
in the range  
\begin{equation}
    \max \left\{0, -\varepsilon_g/\eta  \right\} < q^2_p < 1,
    \label{range0}
\end{equation}
and has inverse decay lengths $\kappa_{+,0} = \eta(1-q_p^2)$ for $z>0$ and  
$\kappa_{-,0} = (\eta q_p^2+\varepsilon_\text{g})$ for $z<0$.
If $\varepsilon_\text{g}>0$, it has the form of a drumhead band~\cite{Chan2016a},
while for $\varepsilon_\text{g}<0$, i.e., if the interface separates two NLSs with different nodal radii, 
this energy level has support in an annular region of transverse momentum.

In the spectral region~\eqref{domainBS}, in addition to this zero-energy interface band, 
finite-energy  localized states emerge when the junction is smooth enough (see below).
As shown in App.~\ref{sec:Equations}, these states are described by the wave function 
\begin{equation}
    \begin{pmatrix}
    \chi_+ \\
    \chi_-
    \end{pmatrix} =
\begin{pmatrix} 
  F(\kappa-\lambda, \kappa+\lambda+1 ; \kappa_{+}+1 ; u) \\
 \frac{\kappa_+ + d - \lambda}{ \varepsilon} F( \kappa+\lambda,\kappa-\lambda+1;\kappa_{+}+1;u )
\end{pmatrix}
\label{eigenstates}
\end{equation}
with the hypergeometric function $F(a,b;c;u)$ \cite{NIST:DLMF}, 
$d$ given in Eq.~\eqref{eq:deltabar0}, and $\kappa= \frac{\kappa_+ +\kappa_-}{2}$. 
The condition of normalizability requires that either of the first two arguments of the hypergeometric
functions equals a non-positive integer, and we arrive at the quantization equation 
\begin{equation}
    \kappa -\lambda = -m, \quad m \in \mathbb{N}_0,
    \label{quantcond}
\end{equation}
where $m=0$ corresponds to the zero-energy state ~\eqref{eq:Psi0}.
We thus find a family of interface levels labeled by the integer $m=1,2,\ldots,m_\text{max}$ 
(where the maximal value $m_\text{max}$ is discussed below), with dispersion
\begin{align}
\varepsilon_{m} (q_p) &= 
\sqrt{ m\left( 2\lambda - m \right) \left[ 1-\left(\frac{\eta q_p^2-\eta +\lambda}{\lambda-m}\right)^2\right]}.
\label{eq:em}
\end{align}
These states are localized at the interface, with different inverse 
localization lengths on the two sides of the interface, given by
\begin{align}
\kappa_{\pm,m} & 
 = \left\{ \begin{array}{c}
  \frac{\lambda}{\lambda-m} \left( \eta -\eta q_p^2 - \varepsilon_m^{(0)} \right)   \\
  \frac{\lambda}{\lambda-m} \left( \eta q_p^2   + \varepsilon_\text{g}- \varepsilon_m^{(0)} \right)
\end{array} 
\right. ,
\label{eq:kappapm}
\end{align}
where
\begin{equation}
 \varepsilon^{(0)}_m = m \left( 2 - \frac{m}{\lambda} \right)
 \label{mergingENLS}
\end{equation}
is the energy at which the  $m^\text{th}$ interface band merges into the bulk band of the NLS on the right.

In order to describe the characteristics of the interface bands~\eqref{eq:em}, it is convenient to
distinguish three scenarios, depending on the relative sizes of the gaps on the right 
and the left of the junction. 

\begin{figure}
\centering{}
\includegraphics[width=0.9\columnwidth]{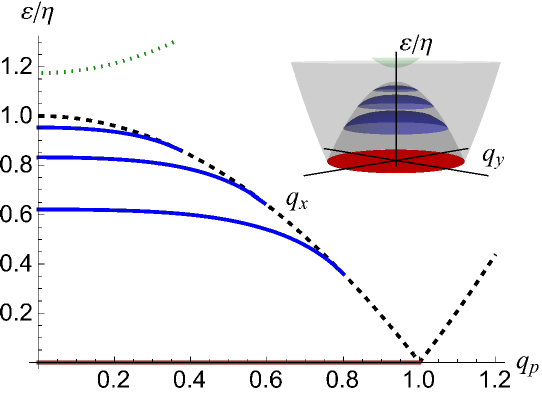}
\caption{
Dispersion relations of the interface states~\eqref{eq:em} 
as a function of the radial momentum $q_p$ in the plane of the nodal line 
for $\eta=5.06$ and $\lambda=5.5$. These parameter values correspond to the large gap regime, 
see Eqs.~\eqref{vanHovesing}-\eqref{mmax1}.
The black dashed line and the green dotted line describe the edge of the bulk bands
on the right and on the left of the interface respectively.
The DDSs (blue lines) always lie at finite energy below the continuum threshold. 
The red line at zero energy describes the TDS, 
with support in the interval \mbox{$0 < q_p < 1$}. 
The TDS is the only interface state that survives in the limit of a sharp junction.
Inset: dispersion relation in the $q_xq_y$ plane.
}
\label{fig:BSdispersionI}
\end{figure}

In the {\em large gap regime} $\varepsilon_\text{g}> \eta$, 
the interface dispersions are decreasing functions of the transverse momentum $q_p$, 
and feature a maximum at $q_p=0$ with energy
\begin{equation}
\varepsilon^{(1)}_m = \sqrt{  m\left(2\lambda - m \right) 
\left[ 1-\left(\frac{\lambda - \eta}{\lambda-m}\right)^2\right]}\;.
\label{vanHovesing}
\end{equation}
The condition of positivity  of $\kappa_{\pm,m}$ in \eqref{eq:kappapm} 
implies that the $m^\text{th}$ interface band has support 
in the range of $q_p$ given by 
\begin{equation}
 0 <  q_p^{2} < 1 - \varepsilon_m^{(0)}/\eta ,
\label{eq:support}
\end{equation}
so it is always a full disc, a "drumhead". These surface levels are illustrated in 
Fig.~\ref{fig:BSdispersionI}.
Because of their shape, we refer to them as dispersive drumhead states.
We stress that these states are found \emph{in addition} to the flat TDSs. 
For a sharp interface, only the latter are present.  
A curvature of the TDS band can actually be originated by an energy dispersion 
of the nodal line in the bulk \cite{Burkov2011}. This is however not the case of 
our Hamiltonian~\eqref{eq:h}, and the TDS band is indeed flat, as opposed to the DDSs.
The additional interface bands can only appear in a junction with a wide 
enough intermediate region. In fact,
Eq.~\eqref{quantcond} implies that the interface hosts the topological zero-energy state 
for every value of $\lambda$, but,  in addition to it, for a smooth junction 
there are $m_\text{max}$ DDSs, where
\begin{equation}
m_\text{max}= \left \lfloor 
\lambda \left( 1-\sqrt{1-\frac{\eta}{\lambda}} \right) \right \rfloor ,
\label{mmax1}
\end{equation}
and $\lfloor x \rfloor$ is the floor function. Furthermore, 
from Eqs.~\eqref{eq:em} and~\eqref{eq:support} we see that higher values of the integer $m$ 
correspond to DDSs with higher energies and smaller radii of the support.
In App.~\ref{sec:Vacuum} we give explicit results for the special case of an infinite gap 
$\varepsilon_\text{g} \to +\infty$ (i.e., $\lambda \to + \infty$ at fixed $\eta$), describing
the case of an interface with the vacuum, where some simplifications emerge. 

\begin{figure}
\centering{}
\includegraphics[width=0.9\columnwidth]{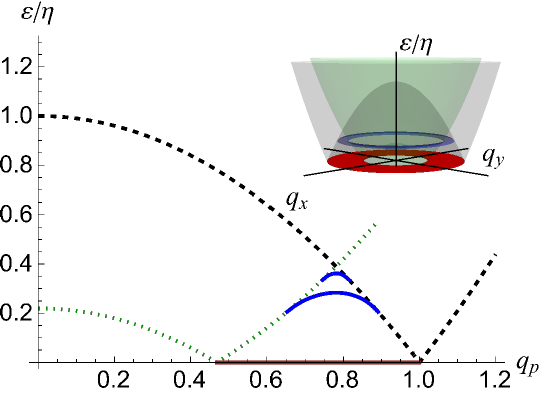}
\caption{
Dispersion relations of the interface states~\eqref{eq:em} in a junction 
between two NLSs of different nodal ring radii, for $\eta=8.2$ and $\lambda=3.2$, 
corresponding to the negative gap regime, 
see Eqs.~\eqref{eq:support2} and~\eqref{evHring}.
The energy of the DDSs  lies below the continuum threshold of \emph{both} materials, 
represented as black dashed and green dotted lines. 
The zero-energy TDS state (red line)
has an annular support and is present also for a sharp junction.
Inset: dispersion relation in the $q_xq_y$ plane. 
Only one DDS is shown.
}
\label{fig:BSdispersionNLS}
\end{figure}

In the {\em negative gap regime} $-\eta<\varepsilon_g<0$ (i.e., $\lambda<\eta/2$), 
the interface separates two NLSs. In a practical realization, they can be 
different materials, each with a nodal line determined by its chemical composition.
Note that in this regime there is no gap on either side of the junction. 
Equations~\eqref{eq:kappapm} imply 
that the support of all interface states is the ring 
\begin{equation}
(\varepsilon_m^{(0)}  - \varepsilon_\text{g})/\eta  <  q_p^{2} < 1 - \varepsilon_m^{(0)}/\eta.
\label{eq:support2}
\end{equation}
In this scenario, the maximum of the dispersion  is shifted to 
$q_p^2= 1-\frac{\lambda}{\eta}$, at energy
\begin{align}
\varepsilon^{(1)}_m=\sqrt{ m\left( 2\lambda - m \right)}, 
\label{evHring}
\end{align}
and the number of interface states which the interface can accommodate is given by 
$m_\text{max}=\left\lfloor\lambda\right\rfloor$.
This regime is exemplified in Fig.~\ref{fig:BSdispersionNLS}.

\begin{figure}
\centering{}
\includegraphics[width=0.9\columnwidth]{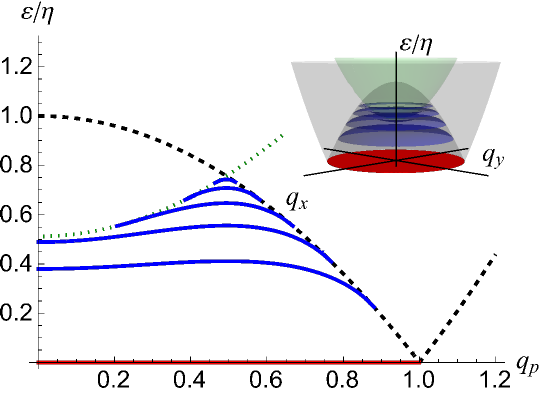}
\caption{
Energies of the interface states \eqref{eq:em} in a NLS-semiconductor junction 
for $\eta=8.2$ and $\lambda=6.2$, representing the small gap regime.  
The dispersion of the higher DDSs merges with the states 
above the semiconductor gap (green dotted line) on the left 
side of the junction, and with the bulk states of the NLS (black dashed line) on the right. 
Inset: dispersion relation in the $q_xq_y$ plane. The highest energy DDS is not shown.
}
\label{fig:BSdispersionS}
\end{figure}

Finally, there is a third scenario, {\em the small gap regime} $0<\varepsilon_\text{g} \leq \eta$,
in which the material occupying the half-space $z<0$  has a band gap equal or 
smaller than the characteristic bulk energy of the NLS. In practice, 
the numerical value of the gap is determined by the nature of the material, and its
chemical (composition, doping, \ldots) and physical (pressure, strain, \ldots) properties.
Typical parameters \cite{Chan2016a} yield an estimate of $0.2\mbox{ eV}$ for this energy, i.e., 
we are considering the interface between a NLS and a semiconductor. 
As illustrated in Fig. \ref{fig:BSdispersionS}, this situation is intermediate between 
the previous two: on the one hand, the states with the lower values of $m$ 
have drumhead-shaped support; on the other hand, as soon as the inequality 
$\varepsilon_m^{(0)}>\varepsilon_\text{g}$ is satisfied, 
we have states with annular support as in Eq.~\eqref{eq:support2}.

We conclude this section by emphasizing the decisive role played 
by the smoothness parameter $\lambda$ in determining the interface spectrum. 
In the \emph{sharp interface} limit $\lambda < 1$,
the change of sign in the hyperbolic tangent
term takes place on a lengthscale $\ell$ which is short compared
to $\frac{\hbar v_{z}}{D_p a_{1}}$. In this limit, no interface states
other than the TDS at $\varepsilon=0$ are allowed, irrespective of the 
nodal radius in the NLS. 
In the opposite \emph{smooth interface} limit $\lambda \gg 1$, the
transition between the insulator and the topological semimetal 
is very smooth, with a characteristic scale $\ell\gg\frac{\hbar v_{z}}{D_p a_1}$.
In this case, a large number ($\sim \lambda$) of interface states exists, 
as long as $ \lambda < \eta$. If, however, $\lambda$ exceeds $\eta$, the number of 
states decreases with increasing $\lambda$ and it is ultimately
limited by the nodal radius, see Eqs.~\eqref{mmax1} and~\eqref{infinitegapmmax}.

As we saw that the number of DDSs is bound by the parameter $\lambda$, 
it is worth considering the consequences in our prototype $\mbox{Ca}_3\mbox{P}_2$, 
which has a unit cell size $a = 5.31\mbox{\AA}$ \cite{Ca3p2materials}. 
Even with a relatively sharp junction, characterized by a junction length $\ell=3a$ 
and $\varepsilon_\text{g}=\eta$,
one obtains \mbox{$\lambda \approx 1.12$}, i.e., the material is in an intermediate regime 
with respect to its junction length and an additional surface state should appear.

\subsection{Scattering states}
\label{scatteringstates}

Let us now consider the $\varepsilon-q_p$ domain in which \mbox{$\kappa_+=ik_+$} is imaginary:
\begin{equation}
    \varepsilon^2 > (\eta q_p^2-\eta )^2.
    \label{domainST}
\end{equation}
In this region, the spectrum consists of propagating states in the NLS 
side which for $z\to+\infty$ are made of a superposition 
of a left-moving wave incident on the junction from the right 
and a right-moving wave reflected back. 
The exact expression of these states is given in Eq.~\eqref{propatatingphi}.
Their asymptotic form is 
\begin{equation}
    \begin{pmatrix}
        \phi_+ \\
        \phi_-
    \end{pmatrix}    
     \sim  \varphi_+(k_+) e^{-ik_+z}  + \mathcal{R} \, \varphi_+(-k_+) 
     e^{ik_+z}, 
    \label{eq:reflasy}
\end{equation}
with the spinors $\varphi_\pm$ given 
in Eq.~\eqref{asymptotic+spinors}, in agreement with Eq.~\eqref{asymptotics+},
and reflection amplitude 
\begin{equation}
\mathcal{R} =  \frac{ \Gamma(ik_+) \, \Gamma( \kappa' - \lambda) 
\, \Gamma( \kappa' + \lambda + 1 ) }
{ \Gamma(-ik_+) \,\Gamma( \kappa  - \lambda) \,
\Gamma( \kappa + \lambda +1 )},
\label{reflectionampl}
\end{equation}
where $\Gamma(x)$ is the gamma function~\cite{NIST:DLMF},
$\kappa = \frac{\kappa_-+ik_+}{2}$ and $\kappa' = \frac{\kappa_--ik_+}{2}$. 
If $\kappa_-$ is real, which occurs if $\varepsilon^2<(\eta q_p^2+\varepsilon_\text{g})$,  
the wave function decays exponentially in the left side of the junction, 
and indeed we find $|\mathcal{R}|^2=1$. 
If instead $\kappa_-$ is also imaginary, with $\kappa_-=-ik_-$, 
the state is partially transmitted towards $z\to-\infty$
with the transmission coefficient
\begin{align}
|\mathcal{T}|^2 &=\frac{\cosh(2\pi k')-\cosh(2\pi k)}{\cos(2\pi\lambda)-\cosh(2\pi k)},
\label{transmission}
\end{align}
where we use the notation 
\begin{equation}
    k=\frac{k_++k_-}{2},\quad k'=\frac{k_--k_+}{2}.
\end{equation}

\subsection{Density of states}
\label{subsec:DOS}

\begin{figure}
\centering{}
\includegraphics[width=0.9\columnwidth]{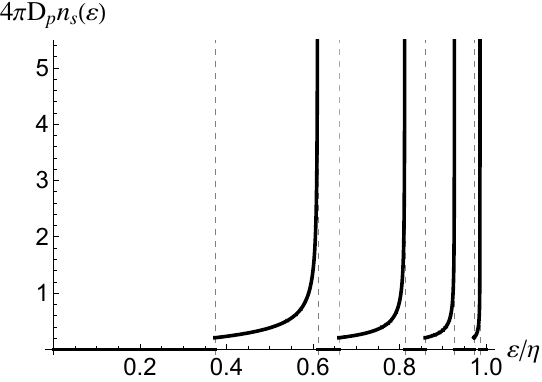}
\caption{Interface density of states~\eqref{surfaceDOS}
in units of $\frac{1}{4\pi D_p}$ for $\lambda=\eta=4.8$, as a function of the energy. 
(The zero-energy delta peak is not shown.) It shows finite jumps and van Hove 
singularities, respectively at the thresholds $\varepsilon_{m}^{(0)}$ and $\varepsilon_{m}^{(1)}$ 
of the localized DDS bands \eqref{eq:em}, identified by the dashed vertical lines in the plot.}
\label{fig:looney}
\end{figure}

Within the energy window $|\varepsilon|<\eta$, where topological states can exist,
the bulk density of states (DOS) of a NLS is a linear function of the energy:
\begin{equation}
n^{(\text{b})}  = \frac{|\varepsilon|}{4\pi  D_p \ell}. 
\label{BulkDOS}
\end{equation}

The interface states are instead characterized by the two-dimensional DOS given by
(for $\varepsilon\geq 0$)
\begin{align}
&n^{(\text{i})}  =  
 \frac{\eta}{4\pi D_p} \delta(\varepsilon) + \nonumber \\
& + 
\frac{1}{4 \pi D_p}  \sum_{m=1}^{m_\text{max}}
 \frac{ \left( \lambda-m\right)\varepsilon \, \Theta( \varepsilon_m^{(1)} - \varepsilon) \,
 \Theta( \varepsilon - \varepsilon_m^{(0)}) } {  \sqrt{\lambda \varepsilon_m^{(0)} 
 (\lambda \varepsilon_m^{(0)}  - \varepsilon^2)  } }   ,
\label{surfaceDOS}
\end{align}
where $\Theta(x)$ is the Heaviside function. (For simplicity, we give here only 
the expression valid for $\lambda \geq \eta$.)
In the large gap case $\lambda>\eta$, $n^{(\text{i})}(\varepsilon)$
has a singularity only at zero energy, because
$\varepsilon_m^{(1)}< \sqrt{ \lambda \varepsilon_m^{(0)}}$. In the other regimes ($\lambda\leq \eta$), 
instead, $\varepsilon_m^{(1)}=\sqrt{ \lambda \varepsilon_m^{(0)}}$
and $n^{(\text{i})}$ features square-root singularities.
This is illustrated in Fig.~\ref{fig:looney}, where we show the case $\lambda=\eta$.
The DOS exhibits steps at $\varepsilon_m^{(0)}$ and van Hove singularities at $\varepsilon_m^{(\text{1})}$,
where the energy matches the lower or upper edge of one of the
localized bands \eqref{eq:em} respectively. This behavior suggests that the contribution 
from interface states to the observables that depend on the DOS, e.g., the AC conductance, 
can be comparable to the one of the bulk when inter-band transitions are allowed.
This is analogous to what is observed for type-I Weyl semimetals in \cite{Mukherjee2019}, 
with the interesting difference, arising from the shape of the dispersion relation, 
that the support of the drumhead states is always compact (see Fig.~\ref{fig:BSdispersionI}). 

In the presence of an interface with the vacuum, such features of the Fermi surface could
be directly detected by tunneling microscopy, as done in single- and multi-layer graphene \cite{Li2009,Brihuega2009}. 
This type of interface can be described within our formalism 
by considering the infinite-gap limit in the insulating side, 
while keeping the radius of the nodal ring finite. Details are presented 
in App.~\ref{sec:Vacuum}. When the interface is with another material, instead, 
we expect that one is still able to detect them using absorption spectroscopy. 
In particular,  we will identify the dependence of the energy levels 
on the magnetic field in Sec.~\ref{sec:SpectrumB}.

\section{Spectrum in magnetic field}
\label{sec:SpectrumB}

In this section, we introduce a magnetic field and identify the characteristic signatures of the spectrum. While an external radiation induces transitions between localized states also in the absence of a magnetic field, we show below that the dependence of the spectrum on the magnetic field strength offers a precise way of identifying and classifying the localized states.
Apart from the orbital effects, the magnetic field $\mathbf{B}$ also couples with the spin of the electrons 
via a Zeeman term $H_z=\mu_B\intop d\mathbf{r}\mathbf{B}\cdot\mathbf{S}$ 
which breaks time reversal symmetry, if present. 
Focussing for definiteness on the case of $\mbox{Ca}_3\mbox{P}_2$, 
the $SU(2)$ spin degeneracy is lifted and one obtains two spin-polarized copies of a NLS, 
separated in energy by an amount 
$\sim 1.2$\,meV for a magnetic field of $1$\,T \cite{Chan2016a,Liu2018}. 
For temperatures well below $\sim15$\,K, only the lowest copy is relevant 
for magnetic fields of this order. Conversely, the occupation of the highest 
copy becomes non-negligible for temperatures of this order.
However, as the electron-photon matrix elements do not couple opposite spins 
and the spacing between the energy levels is the same for both copies, 
one does not find any new absorption peak. For this reason, we focus in the following 
on a single spin polarization, for definiteness, along the magnetic field, 
and leave implicit the associated spin selection rule. 

Specifically, we introduce a magnetic field $\mathbf{B}=B\hat{e}_z$
perpendicular to the plane of the nodal line and the interface. The magnetic 
field introduces a new length scale, the magnetic length $\ell_{B}=\sqrt{\hbar/eB}$, which 
we use to rescale the coordinates in the $xy$ plane as $X=x/\ell_B$ and  $Y=y/\ell_B$.
The momentum $\boldsymbol{k}$ is replaced by
the gauge-invariant momentum $\boldsymbol{\Pi}= -i \hbar \nabla +e \mathbf{A}$.
The dimensionless Hamiltonian~\eqref{eq:h}
takes the form
\begin{equation}
\mathcal{H}  =   -i\tau_ y \partial_{z} +
\tau_z \left[ \mathcal{H}_\text{LL} - \eta - \lambda(\tanh z-1)\right],
\label{eq:dimensionlessHB}
\end{equation}
where 
\begin{equation}
\mathcal{H}_\text{LL} = \frac{\alpha}{2} \left( \Pi_X^2 + \Pi_Y^2 \right)
\label{hLL}
\end{equation}
is the Hamiltonian of two-dimensional non-relativistic fermions in a perpendicular magnetic field, 
and $\alpha $ is the rescaled cyclotron energy
\begin{equation}
\alpha =  \frac{2\ell D_p}{\hbar v_{z}\ell_{B}^{2}}. 
\label{eq:alphalambda0}
\end{equation}

The Hamiltonian~\eqref{eq:dimensionlessHB} is separable in a transverse 
and a longitudinal part, and its eigenstates assume the factorized form
\begin{equation}
\Psi_{n,s,m} (X,Y,z)  =  \Phi_{n,s}(X,Y)\, \psi_{n,m}(z),
\label{eq:Factorizedstate}
\end{equation}
where $n=0,1,2,\dots$ labels the Landau levels and the quantum number $s$ 
is the degenerate index of the Landau band.
The form of the wave functions $\Phi_{n,s}$ depends on the gauge but it is not needed here. 
The index $m$ labels the interface states occurring in each Landau band $n$.
The wave functions $\psi$ are the eigenstates of the Hamiltonian
\begin{equation}
\mathcal{H}  =   -i\tau_ y \partial_{z} +
\tau_z \left[  \omega_n - \eta - \lambda(\tanh z-1)\right],
\label{eq:dimensionlessHB2}
\end{equation}
which has the same form as~\eqref{eq:h}, except for the replacement 
$\eta q_p^2\to \omega_n$, where $\omega_n$ are the non-relativistic Landau level energies 
\begin{equation}
\omega_{n}  =  \alpha \left( n+\frac{1}{2} \right).
\label{eq:omegan}
\end{equation}
Therefore, all the results concerning the spectrum of the interface problem discussed
in Sec.~\ref{sec:SpectrumNoField} can be transferred at once to the present case, 
by simply replacing  $\eta q_p^2$ with $\omega_n$. The discrete nature of $\omega_n$
has some interesting implications, which we discuss below.

Once again, the spectrum can be divided into two sectors, which can be interpreted
as propagating waves reflected at the interface and interface bound states. For the first
category, we observe the formation of Landau bands with one-dimensional dispersion 
along the magnetic field direction, similarly to what happens in Weyl semimetals \cite{Mukherjee2019}.  
As long as the energy is below the gap of the insulator $\varepsilon_\text{g}$, 
we obtain solutions decaying exponentially for $z<0$, but behaving asymptotically for $z\gg 1$ 
as a superposition of incoming and outgoing waves in the form \eqref{eq:reflasy}, 
with the reflection amplitude $\mathcal{R}$ given 
in~\eqref{reflectionampl} with the substitution $\eta q_p^2\to\omega_n$. 

\begin{figure}
\centering{}
\includegraphics[width=0.9\columnwidth]{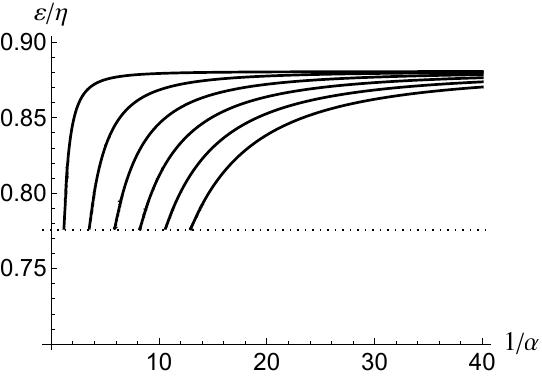}
\caption{Energies of the interface bound states in a magnetic field with $m=1$ and $n=0,\dots,5$ 
(from left to right)
as a function of $\alpha^{-1}$ for $\lambda=\eta=1.9$
(where $\alpha$ is the rescaled cyclotron energy in Eq.~\eqref{eq:alphalambda0}). 
The horizontal dotted line denotes the energy $\varepsilon_1^{(0)}$ at which the state merges into the bulk. 
The threshold values of $\alpha$ at which a new Landau band accommodates the interface state can be found 
from Eq.~\eqref{mmaxMF} with $m^{(n)}_\text{max}=1$.
}
\label{fig:ene}
\end{figure}

In addition to these states in the continuum, each transverse Landau level
can accommodate, depending on the values of the parameters $\eta$, $\lambda$, and $\alpha$,
a longitudinal zero-energy state in the form \eqref{eq:Psi0}, and a
discrete set of longitudinal finite-energy bound states, once again labeled 
by the integer $m>0$. The dispersion of the DDS is given by \eqref{eq:em} with $\eta q_p^2$  
replaced by $\omega_n$. The quantum number $s$ does not appear in the expression~\eqref{eq:em}, 
and the spectrum inherits the Landau level degeneracy.
The energies are shown in Fig.~\ref{fig:ene} as a function 
of (the inverse of) the rescaled cyclotron energy $\alpha\propto B$.
One sees that the number and the form of the dispersion of the DDS vary with these parameters, as
derived in Sec.~\ref{subsec:drumheadstates}. In the ultra-quantum
limit $B\gg\frac{\hbar^{2}v_{z}}{eD_p\ell}$, on the left of the
figure, only one state is present, while in the low-field limit $B\ll\frac{\hbar^{2}v_{z}}{eD_p\ell}$,
on the right of the figure, the states group in dense bands, 
which are still separated one from the other. 
The spectrum can accommodate a large number of states, growing as $1/\alpha$.

Interestingly, there is a critical value of the magnetic field above 
which no zero-energy normalizable solution exists, corresponding to the 
magnetic breakdown of the topological semimetal \cite{Ramshaw2018}. 
Indeed, the condition~\eqref{range0} translates into  
\begin{equation}
 \max \left\{0, -\varepsilon_g \right\} < \alpha 
 \left( n+\frac{1}{2} \right)  < \eta.
\label{alpha<2q0^2}
\end{equation}
Therefore, if $\alpha>2\eta$, no Landau level can support a zero-energy interface state.
If $\alpha<2\eta$, instead, all Landau bands $n=0,1,\dots,n_\text{max}^{(0)}$ accommodate
one such state, with $n_\text{max}^{(0)}$ the largest integer smaller than $\frac{\eta}{\alpha}-\frac{1}{2}$.
This condition can be written as $\ell_Bk_0>1$, hence it
is independent of the interface smoothness (the scale $\ell$)
and holds even in the limit of an interface with the vacuum ($\varepsilon_\text{g}\to +\infty$).

For a very smooth interface with $\lambda \gg 1$ (focussing for simplicity on the case $\lambda\geq \eta$),
the number of localized states supported by Landau band $n$ is given by
\begin{equation}
m^{(n)}_\text{max} = \left\lfloor \lambda \left( 1 -\sqrt{1-\frac{\eta-\alpha(n+1/2)}{\lambda}} \right) \right\rfloor.
\label{mmaxMF}
\end{equation}
This number decreases for increasing $n$ or increasing magnetic field $B\sim\alpha$. Then, 
for each Landau band $n$, there is a critical field 
above which there are no localized interface states given by
\begin{equation}
   \alpha^{(n)}_\text{th} = \frac{(\eta -2 +1/\lambda)}{n+1/2}.
\end{equation}

\begin{figure}
\centering{}
\includegraphics[width=0.9\columnwidth]{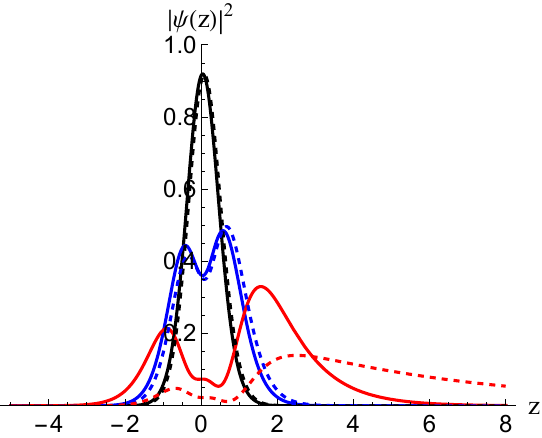}
\caption{Probability density for the interface states in the lowest Landau band $n=0$,
with $m=0,1,2$ (black, blue, and red lines respectively)
for $\lambda=\eta=2.9$ and $\alpha=0.25$ and $0.5$ (solid and dashed lines respectively). 
For $m=2$, the decay length in NLS side for the larger value of $\alpha$ is close to diverge, signaling 
the merging of the state into the bulk band.}
\label{fig:sq2}
\end{figure}

Finally, in order to gain some insight into the spatial structure of the interface states, 
we plot in Fig.~\ref{fig:sq2} the squared modulus of the wave function as a function 
of the spatial coordinate $z$.
We observe that the probability density of the TDS has a single peak, 
while that of the $m^\text{th}$ DDSs has $m+1$ local maxima. Moreover, for increasing
magnetic field, the decay length in the NLS side increases 
until the interface states, one by one, merge with the bulk (reflected wave) solutions.
We close this section with an estimate for the material $\mathrm{Ca}_{3}\mathrm{P}_{2}$
\cite{Chan2016a}. Using a unit cell size $a \approx 5.3\,\text{Å}$ and $q_0=1$, 
for an interface varying over a distance of ten unit cells \mbox{($\ell\approx50\,\text{Å}$)}
one finds that a magnetic field $B \approx 1$\,T can be considered weak
and the number of interface states is of the order of $1/\alpha\approx5\times10^{2}$.

\section{Optical conductivity and absorption lineshapes in magnetic field}
\label{sec:Absorption}

As stated in Sec.~\ref{sec:Introduction}, two main experimental routes used 
to identify Volkov-Pankratov states in recent experiments are AC charge current measurements \cite{Inhofer2017} 
and absorption spectroscopy \cite{Bermejo2022}. We therefore now aim at quantifying the contribution 
of the interface states to the optical conductivity and to the absorption power. 
(See \cite{Barati2017,Pronin2020} for the analogous bulk quantities.)
In particular, we are interested in radiation with frequency $\omega \ll 10^{2}$\,THz
and propose infrared absorption spectroscopy in magnetic field
and the related optical conductivity
as ways to detect the presence of additional interface
states, which can be clearly identified via the presence of 
peaks in both absorption spectra and AC conductivity
and by their behavior in magnetic field.

In order to fix the notation, let us consider a monochromatic
uniform incident radiation of amplitude $A_0$, 
polarization $\hat{\mathbf{e}}$, frequency $\omega$, and wavevector $\boldsymbol{q}$, 
which can be introduced in the Hamiltonian \eqref{eq:dimensionlessHB} 
as a perturbation via minimal substitution.
In linear response, the oscillating electric field induces a current density 
$\mathbf{j}\left(\mathbf{q},\omega\right)=\sigma\left(\mathbf{q},
\omega\right)\mathbf{E}\left(\mathbf{q},\omega\right)$ proportional to the field 
via the complex dynamical conductivity $\sigma=\sigma_{1}+i\sigma_{2}$. 

Assuming the electrons to be in equilibrium with thermal distribution $f$ 
at given temperature $T$ and chemical potential $\mu$, the power dissipated 
in the system is expressed as 
$W(\mathbf{q},\omega)=2\pi\omega A_0^2\mathcal{I}(\mathbf{q},\omega)$ 
in terms of the function
\begin{equation}
    \mathcal{I}_c(\boldsymbol{q},\omega) = \frac{1}{\pi}\Im\sum_{a,b}
    \frac{\left|\left\langle \Psi_a\right|e^{i\mathbf{q}\cdot\mathbf{r}}\hat{\mathbf{e}}_c\cdot 
    \mathbf{j}\left|\Psi_b\right\rangle\right|^2}{\hbar\omega-E_a+E_b+i\Gamma}\left[
    f(E_a)-f(E_b)\right] ,
\label{eq:lineshape}
\end{equation}
in which $\mathbf{j}$ is the electric current  
and the polarization of the electric field $\hat{\mathbf{e}}_c$, $c=x,y,z$, appears. 
$\Psi_a$ and $\Psi_b$ are the states in Eq.~\eqref{eq:Factorizedstate}, 
where $a$ and $b$ stand for the set of quantum numbers labelling the states
(particle/hole index, Landau state indices $n$ and $s$, 
and interface index $m$).
The phenomenological broadening parameter $\Gamma$
encodes all the processes not explicitly modeled in our calculation, e.g.,
electron-electron interactions and scattering with localized impurities or phonons.
In the limit $\Gamma \to0$, one recovers the energy conservation delta function. 
We will refer to $\mathcal{I}$ as the lineshape function. 
The dissipated power per unit volume is 
also directly related to the electron loss function, 
recently investigated in type-I Weyl semimetals in \cite{Lu2021}.

For definiteness, let us consider that the semimetal 
sample occupies a volume $V=L\mathcal{S}$, where $L$ is a large length ($L \gg \ell,1/k_0$) 
in the $z$ direction and $\mathcal{S}$ is the area in the $xy$ plane. 
As the sum over $s$ produces the ratio $\sum_s=\Phi/\Phi_0$, 
where $\Phi=\mathcal{S}B$ is the magnetic flux through the sample and $\Phi_0=h/e$ 
the magnetic flux quantum \cite{Tong2016}, it is convenient to define the units 
$I_z^{(0)}= e^2 v_z V /2\pi \hbar \ell_B^2$
and $I_p^{(0)} = e^2 D_p^2 V  / 2\pi \hbar^3v_z\ell_B^4$
and use them to express the results of Eq.~\eqref{eq:lineshape} 
for the longitudinal and perpendicular parts, respectively. 
In order to calculate the lineshape function~\eqref{eq:lineshape}
we need the matrix elements of the current density operator 
between the initial and final states of the photoemission/absorption process. 
In the following section, we derive the selection rules for the possible transitions 
and their integral expressions.

Apart from the transitions involving interface states only, 
there can be processes which involve the bulk states as well. 
For these processes, 
the sum over the index $m$ 
implicit in the indices $a$, $b$ in Eq.~\eqref{eq:lineshape} 
is replaced by an integral over energies.
The pertinent matrix elements are constrained by the same selection rules as for the interface states, 
see below, and can be computed in an analogous manner.

The lineshape function \eqref{eq:lineshape} allows us to directly determine
the real part of the diagonal components of the dynamical conductivity.
The system has cylindrical symmetry and, in general, $\sigma_{x}=\sigma_{y}\ne\sigma_{z}$. 
In the dipole approximation, the real part of the conductivity reads
\begin{equation}
    \sigma_1(\omega) = \frac{\pi}{\omega V} \mathcal{I}(\omega)
    \label{eq:sigma}
\end{equation}
which is to be read as a tensor equality.
The magnitude of the conductivity along the longitudinal and transverse directions is 
therefore fixed by the factors $I_z^{(0)}/V$ and $I_p^{(0)}/V$, respectively. 
Its numerical value is the sum of two contributions
\begin{equation}
    \sigma_{c} = \sigma^{(\text{b})}_{c} + \frac{\ell}{L}\sigma^{(\text{i})}_{c}, \qquad c=x,y,z,
\end{equation}
originating from bulk states and interface states. 
With this definition, it is explicit that the contribution 
from the interface states scales as the inverse of the size $L$ along $z$ 
because of the factor  $L$ in the denominator of \eqref{eq:sigma}.

\subsection{Selection rules}    
\label{subsec:selection}

The current $\boldsymbol{j}= e \partial_{\boldsymbol{k}}H$ 
has the form ubiquitous in Dirac Hamiltonians. The component along
the main symmetry axis of the semimetal is $j_{z}=ev_{z}\tau_{y}$,
while the component in the nodal plane reads $\boldsymbol{j}_{p} = 2e D_p \boldsymbol{k}_{p}\tau_{z}$.
In the presence of a magnetic field $B\hat e_z$, $\boldsymbol{k}_p$ is
replaced by the gauge-invariant momentum  $\boldsymbol{\Pi}_{p} = \boldsymbol{k}_p + e\boldsymbol{A}$.
In the symmetric gauge, it is convenient to work with the two linear 
combinations of the current density components 
\begin{subequations}
\label{eq:jjbar}
    \begin{align}
\hat{\bar \jmath } & = \frac{1}{\sqrt{2}}\left( \hat{j}_{x}+i\hat{j}_{y}\right)  =  
2i\frac{eD_p}{\hbar\ell_{B}}\hat{a}^{\dagger}\tau_{z}, \\
\hat{\jmath} & = \frac{1}{\sqrt{2}}\left(\hat{j}_{x}-i\hat{j}_{y}\right)  =  
-2i\frac{eD_p}{\hbar\ell_{B}}\hat{a}\tau_{z} ,
\end{align}
\end{subequations}
where $\hat a^\dagger,\hat a$ are the standard ladder 
operators that change the Landau level index $n$ by $\pm 1$.

We are now in position to compute the matrix elements appearing in Eq.~\eqref{eq:lineshape}. 
We distinguish two situations, with the radiation 
linearly polarized either along the system's axis or in the nodal plane.
In the first case, the radiation couples to the $z$ component of the current
density operator, which does not change the Landau level.
As a consequence, we have the selection rule
\begin{equation}
\left\langle n_{1},s_{1}\right|\hat{j}_{z}
\left|n_{2},s_{2}\right\rangle \propto \delta_{n_{1},n_{2}}\delta_{s_{1},s_{2}},
\end{equation}
valid for both localized and bulk states. The full expressions of the matrix elements 
are combinations of integrals of hypergeometric functions, which we compute numerically. 

For the radiation field with linear polarization in the nodal plane, instead, 
the operators~\eqref{eq:jjbar} always couple neighboring Landau bands, 
analogously to what found in type-I Weyl semimetals \cite{Mukherjee2019}. 
Again, as the transverse part of the states is the same for localized states and reflected
waves, the selection rule $n_{2}=n_{1} \pm 1$ holds for both. 
The matrix elements then take the form 
\begin{equation}
\left\langle n_{1},s_{1} \right| \hat{\jmath}\left|n_{2},s_{2}\right\rangle 
\propto \sqrt{n_2} \, \delta_{s_{1},s_{2}}\delta_{n_{1},n_{2}-1}.
\end{equation}
From the above selection rules, it follows that the nonzero components
of the conductivity tensor are the diagonal ones $\sigma_{jj}$, with
$j=x,y,z$, plus the off-diagonal component $\sigma_{xy}$. 
We will focus on the former in this work and evaluate Eq.~\eqref{eq:lineshape}
numerically, using the exact eigenstates \eqref{eq:Psim} for the
bound states and \eqref{eq:PsiR} for the reflected waves.

\subsection{Absorption spectra and optical conductivity}
\label{absorptionspectra}

\begin{figure}
\centering{}
\includegraphics[height=0.18\textheight]{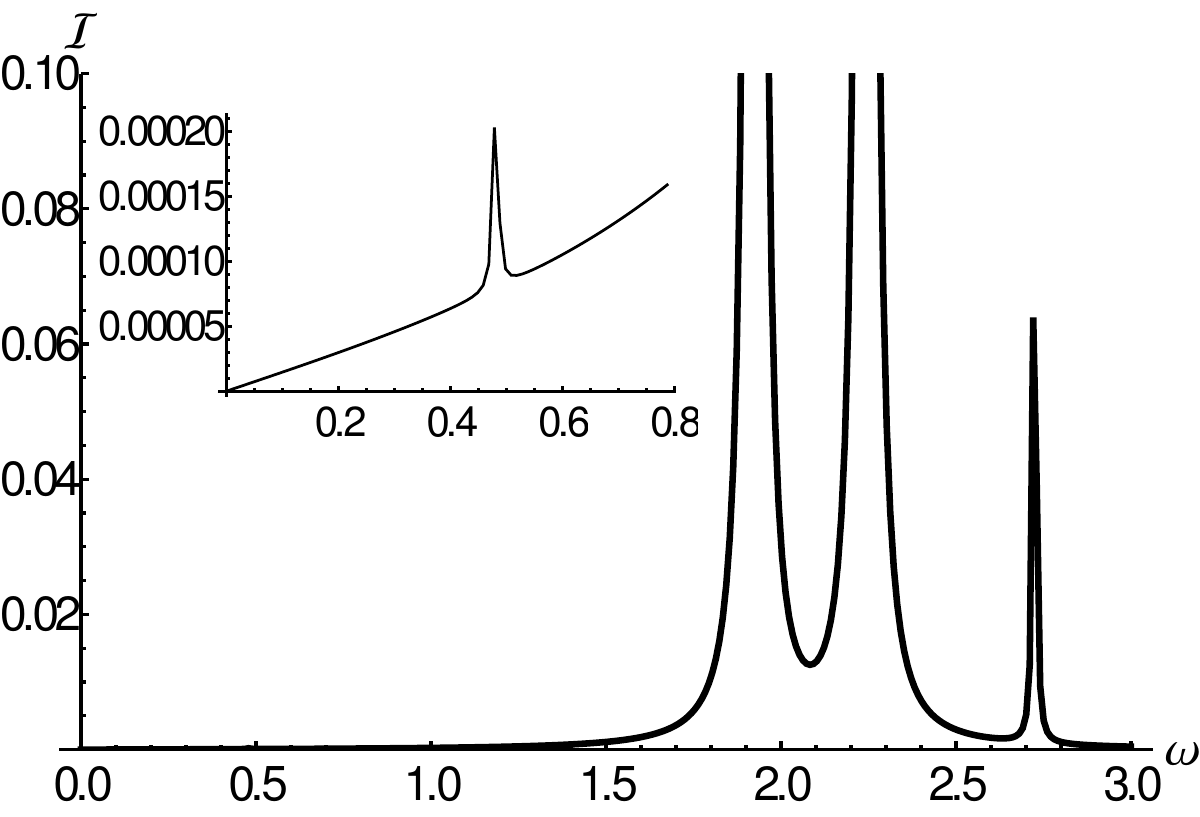}
\caption{Imaginary part of the axial component of the lineshape function $\mathcal{I}_{z}$  
(in units of $I_z^{(0)}\ell/L$) as a function of frequency for $\lambda=3.1$,
$\alpha=0.75$, $k_BT=0.25\hbar v_z/\ell $, $\mu=0$. 
Inset: detail highlighting the transition between two DDSs. The contribution of the 
interface-bulk transitions is not observable in this interval.}
\label{fig:Iz}
\end{figure}

\begin{figure}
\centering{}
\includegraphics[height=0.18\textheight]{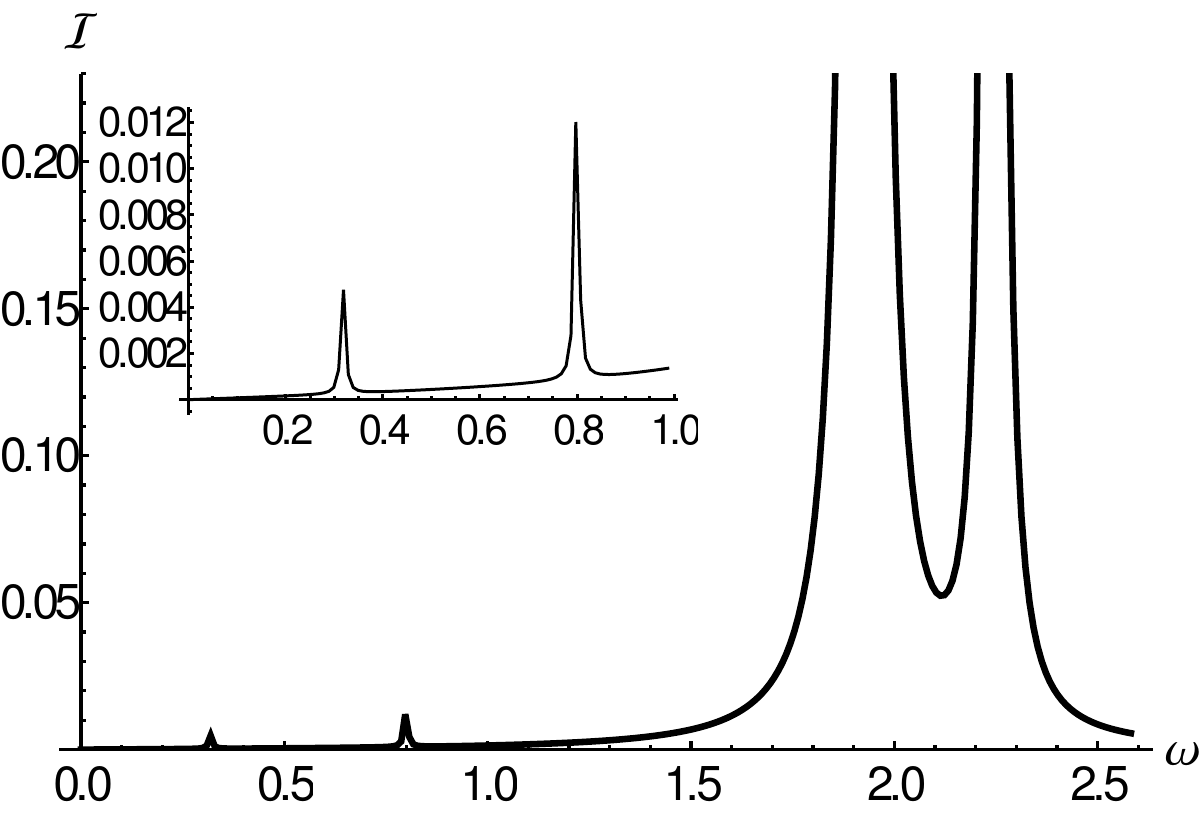}
\caption{Imaginary part of the component in the nodal-line plane 
of the lineshape function $\mathcal{I}_{x}=\mathcal{I}_{y}$ 
(in units of $I_p^{(0)}\ell/L$ as defined in the text)
as a function of the photon frequency (in units of $v_z/\ell$) for $\lambda=3.1$, 
$\alpha=0.75$, $k_BT=0.25\hbar v_z/\ell$ and chemical potential on the nodal line. 
The peaks originating from transitions between the TDS and the DDSs are dominant in this plot. 
The transitions between DDSs are highlighted in the inset. 
The interface-bulk transitions are of $\mathcal{O}\left(10^{-4}\right)$ in the shown interval.}
\label{fig:Ix}
\end{figure}

We focus on the scenario in which we have an insulator for $z<0$
with energy gap $2\varepsilon_\text{g}=2\eta$ (i.e., $\lambda=\eta$),
and on states with energies within the gap, i.e., $|\varepsilon|<\eta$.
Moreover, we choose representative values of $\lambda=0.31$ and $\alpha=0.75$, 
where there exist three interface states in the spectrum, as summarized in Table~\ref{tab:bs}.
\begin{table}
\centering{}%
\begin{tabular}{|c|c|c|}
\hline 
$m$ & $n$ & $\varepsilon_{m,n}$\tabularnewline
\hline 
\hline 
$1$ & $0$ & $2.2437$\tabularnewline
\hline 
$2$ & $0$ & $2.7247$\tabularnewline
\hline 
$1$ & $1$ & $1.9255$\tabularnewline
\hline 
\end{tabular}\hspace{1cm}%
\caption{Interface state energies for $\lambda=3.1$, $\alpha=0.75$, 
identified by the quantum number $m$ and Landau band $n$.
}
\label{tab:bs}
\end{table}
We exemplify the various contributions to the lineshape function
for the radiation polarized in the axial direction in Fig.~\ref{fig:Iz}
and for the radiation polarized in the nodal plane along $x$ in Fig.~\ref{fig:Ix}, with
the chemical potential set to the energy of the nodal line. 
 
An electric field in the $z$ direction  
will not induce transitions between different Landau levels,
and one observes the three peaks of Fig.~\ref{fig:Iz} at the transitions from the TDS
to the three states in Table~\ref{tab:bs}.  At finite temperature, a lower peak corresponding 
to the transition $\left(m=1,n=0\right)\leftrightarrow\left(m=2,n=0\right)$ 
also appears, as this is the only allowed transition between DDSs.

Let us consider now the case of electric field in the $x$ direction. 
The lowest-lying localized state is identified by $\left(m=1,n=1\right)$ 
and the transition from the TDS $\left(m=0,n=0\right)$ corresponds to the first peak 
in Fig.~\ref{fig:Ix}, as the current operators $\hat \jmath$ and $\hat{\bar \jmath}$ 
change the Landau level $n$ by one. The following peak corresponds to the transition 
$\left(m=0,n=1\right)\to\left(m=1,n=0\right)$, while the third peak, 
corresponding to the transition to the $\left(m=2,n=0\right)$ DDS and much smaller than the others, 
is not shown in the figure. Application of the current operators induce 
also the transitions $(m,n)=\left(1,1\right)\leftrightarrow\left(1,0\right)$ 
and $\left(1,1\right)\leftrightarrow\left(2,0\right)$, 
corresponding to the two peaks shown in the inset of Fig.~\ref{fig:Ix}.


In the numerical plots throughout this section, we express frequencies in units of $v_z/\ell$. 
For definiteness, with the velocity quoted in Sec.~\ref{sec:Model} and $\ell=25\,\text{Å}$, 
the peaks reported in the figures are at frequencies $\sim 10$~THz, 
in the infrared part of the spectrum. Importantly, the energy separation of 
the localized levels and therefore the position of the photo-absorption 
peaks decreases in first approximation as $1/\ell$.

As an example, we consider the compound $\mbox{Ca}_5\mbox{P}_3\mbox{H}$, 
a semiconductor with a band gap $\approx 0.7\,\mbox{eV}$ at the $\Gamma$ point \cite{Nie2021}, 
which is synthesized alongside the NLS $\mbox{Ca}_3\mbox{P}_2$ \cite{Xie2015}. 
A practical realization of our model would be an interface between these two materials, 
in which the two compounds are mixed over a region of a few unit cell width. 
In table \ref{tab:extable1}, we also provide the frequencies corresponding 
to the transition between the zero-energy states ($m=0$) and the interface states discussed above.
\begin{table}
\centering{}%
\begin{tabular}{|c|c|c|c|c|}
\hline 
$m$ & $n$ & $\varepsilon_{m,n}$ & $E_{m,n}$ (eV) & $\omega/2\pi$ ($10^{13}$ \mbox{Hz}) \tabularnewline
\hline 
\hline 
$1$ & $0$ & $1.8111$ & $0.19868$ & $4.8041$   \tabularnewline
\hline 
$1$ & $1$ & $1.7911$ & $0.19649$ & $4.7510$  \tabularnewline
\hline 
$1$ & $2$ & $1.7706$ & $0.19424$ & $4.6968$  \tabularnewline
\hline 
\end{tabular}\hspace{1cm}%
\caption{Interface states (besides the zero-energy states) in a 
$\mbox{Ca}_3\mbox{P}_2$/$\mbox{Ca}_5\mbox{P}_3\mbox{H}$ junction with $\ell=30\,\mbox{\AA}$  
in a magnetic field $B=0.5$\,T. The effective parameters for this configuration 
are  $\lambda \approx 4.1020$, $\eta \approx 1.8231$, $\alpha \approx 0.0209$. 
The first two columns list the quantum numbers specifying the states as described in the main text. 
The third and fourth columns contain 
the dimensionless energies (in units of $\hbar v_z/\ell\approx 0.110$\,eV) and 
the energies in \mbox{eV}, while the last column comprises the frequencies 
corresponding to the transitions between the zero-energy topological state and each interface state.
These transitions take place for any 
polarization of the radiation.
\label{tab:extable1}
}
\end{table}
Another possible platform to investigate the physics of the interface states is 
a junction between the half-Heusler semiconductor $\mbox{CaCdSi}$, 
with a band gap of $0.59\,\mbox{eV}$ \cite{Ould2021} and the NLS $\mbox{CaCdSn}$ \cite{Laha2020}, 
with $k_{0}\approx0.165\,\mbox{\AA}^{-1}$, $D_p k_0^2\approx 0.4\,\mbox{eV}$ and 
$\hbar v_F\approx 3.16\, \mbox{\AA}$. Engineering the interface in such a way that the relative 
concentration in the substitution $\mbox{Si}\to\mbox{Sn}$ varies over a few layers, 
as in \cite{Tchoumakov2017}, we find additional interface states, 
listed in Table \ref{tab:extable2} for sample values of the parameters.
\begin{table}
\centering{}%
\begin{tabular}{|c|c|c|c|c|}
\hline 
$m$ & $n$ & $\varepsilon_{m,n}$ & $E_{m,n}$ (\mbox{eV}) & $\omega/2\pi$ ($10^{13}$ \mbox{Hz}) \tabularnewline
\hline 
\hline 
$1$ & $0$ & $1.514$ & $0.3986$ & $9.638$   \tabularnewline
\hline 
$1$ & $1$ & $1.511$ & $0.3978$  & $9.619$  \tabularnewline
\hline 
$1$ & $2$ & $1.508$ & $0.3970$  & $9.600$  \tabularnewline
\hline 
$1$ & $3$ & $1.505$ & $0.3963$  & $9.582$  \tabularnewline
\hline 
$1$ & $4$ & $1.502$ & $0.3955$  & $9.562$  \tabularnewline
\hline 
$1$ & $5$ & $1.499$ & $0.3947$  & $9.543$  \tabularnewline
\hline 
$1$ & $6$ & $1.496$ & $0.3939$  & $9.523$  \tabularnewline
\hline 
$1$ & $7$ & $1.493$ & $0.3930$  & $9.504$  \tabularnewline
\hline 
$1$ & $8$ & $1.490$ & $0.3922$  & $9.484$  \tabularnewline
\hline 
$1$ & $9$ & $1.487$ & $0.3914$  & $9.463$  \tabularnewline
\hline 
$1$ & $10$ & $1.483$ & $0.3905$  & $9.443$  \tabularnewline
\hline 
$1$ & $11$ & $1.480$ & $0.3897$  & $9.422$  \tabularnewline
\hline 
$1$ & $12$ & $1.477$ & $0.3888$  & $9.401$  \tabularnewline
\hline 
$1$ & $13$ & $1.473$ & $0.3879$  & $9.380$  \tabularnewline
\hline 
$1$ & $14$ & $1.470$ & $0.3871$  & $9.359$  \tabularnewline
\hline 
\end{tabular}\hspace{1cm}%
\caption{Interface states (besides the zero-energy states) in a $\mbox{CaCdSn}$/$\mbox{CaCdSi}$ 
junction with $\ell=12$\,\AA ~in a magnetic field $B=2$\,T. The effective parameter 
for this configurations are $\lambda\approx1.8801$, $\eta\approx1.5193$, $\alpha\approx 0.0034$. 
The first two columns list the quantum numbers specifying the interface states, as described in the main text. 
The third and fourth column contain the dimensionless energies (expressed 
in units of $\hbar v_z/\ell\approx 0.2633$\,eV) and the energies in \mbox{eV}, 
while the last column comprises the frequencies corresponding 
to the transitions between the zero-energy topological state and each interface state. 
These transitions take place for any polarization or the radiation.
\label{tab:extable2}
}
\end{table}
Finally, we propose that a smooth NLS/NLS' interface could also be realized in compounds related to each other via chemical substitution, e.g., $\mbox{ZrSiS}$, $\mbox{ZrSiSe}$ and $\mbox{ZrSiTe}$ \cite{Hu2016,Hu2017,Hosen2017}.

\subsection{Bulk contributions}
\label{sec:bulk}

As discussed above, the current generated by the radiation in the semimetal has two contributions, 
respectively coming from the interface states and from the bulk reflected waves. 
It is then important to estimate the magnitude of the contribution from bulk-bulk transitions
and to classify the corresponding resonant peaks that can appear in absorption and conductivity measurements, 
possibly masking the contribution from the interface.

\begin{figure}
\centering{}
\includegraphics[height=0.18\textheight]{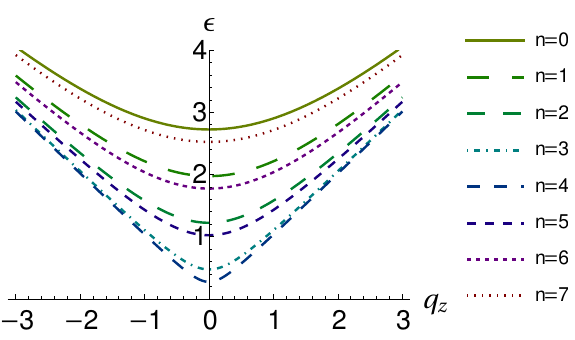}
\caption{Positive bulk Landau bands~\eqref{eq:ebulk} 
closest to the Fermi energy (set to zero)
as a function of the bulk momentum, for $\eta=3.1$, $\alpha=0.75$.}
\label{fig:Bands} 
\end{figure}

In order to estimate the contribution from bulk-bulk transitions, we consider the system far away from the interface, 
so that one can neglect the interface terms and focus on Eq.~\eqref{eq:H} directly. As detailed 
in App.~\ref{app:bulk}, since translation invariance is restored, 
the eigenstates are labeled by the momentum $q_z$, as well as the Landau band $n$ and the additional quantum number $s$. 
The spectrum is given by
\begin{equation}
    \varepsilon_{n,\pm}(q_z)=\pm\sqrt{q_z^2+M_n^2}, 
    \label{eq:ebulk}
\end{equation}
with $M_n=\omega_n-\eta$ in the units of Sec. \ref{sec:SpectrumB}. 
The first few bulk Landau bands are represented in Fig.~\ref{fig:Bands} for clarity. 
(Throughout this section, we continue to use the energy scale $\hbar v_z/\ell$, 
involving the characteristic length scale of the interface, even though the 
interface does not play any role in the bulk-bulk transitions.  
This choice allow us to compare the contribution of the bulk-bulk transitions 
with the contribution of the interface-bulk transitions.)
The same analysis as for the interface states can be applied to the bulk states, 
and one finds indeed the same selection rules on the quantum number $n$. 
Importantly, a series of thresholds corresponding to the separation 
between the Landau levels involved in the transitions appear, which we detail below, 
focusing on the absorption processes.

\begin{figure}
\centering{}
\includegraphics[height=0.18\textheight]{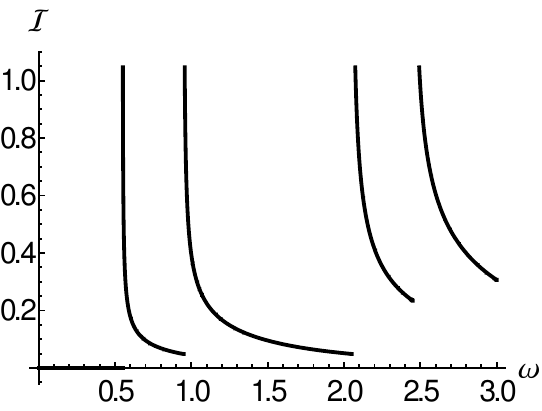}
\caption{Bulk contribution to the imaginary part of the longitudinal component of the 
lineshape function $\mathcal{I}_{z}$  (in units of $I_z^{(0)}$) 
as a function of the frequency for $\eta=3.1$ , $\alpha=0.75$, $k_BT=0.25\hbar v_z/\ell$, $\mu=0$. 
A series of divergences corresponding to the threshold for transitions between particle 
and hole bands with the same value of the Landau index $n$ is visible, see Figure \ref{fig:Bands}. 
The spacing between them is therefore proportional to the magnetic field.}
\label{fig:Ibz}
\end{figure}

For the electric field along $z$, 
the photon excites an electron from the hole band to the conduction band, 
without changing the Landau level $n$. 
This process has therefore a series of thresholds at $\omega=2 |M_n|$, 
of which the first is represented in Fig.~\ref{fig:Ibz}. 

\begin{figure}
\centering{}
\includegraphics[height=0.18\textheight]{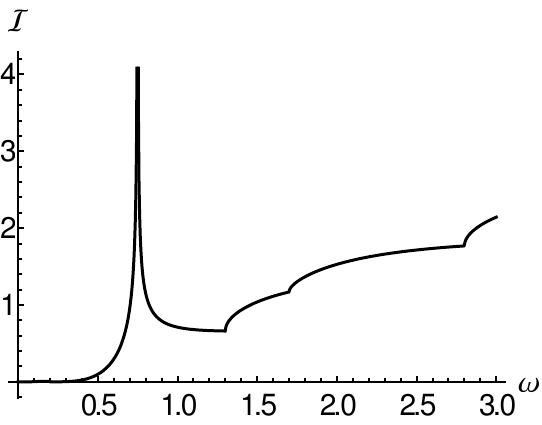}
\caption{Bulk contribution to the imaginary part of the lineshape function 
$\mathcal{I}_{x}$  (in units of $I_p^{(0)}$) 
as a function of the frequency. Here $\eta=3.1$,
$\alpha=0.75$, $k_BT=0.25\hbar v_z/\ell$, $\mu=0$. 
A divergence at $\omega=\alpha$ appears, corresponding to the separation between the minima
of neighboring Landau bands ($q_z=0$ in fig. \ref{fig:Bands}).}
\label{fig:Ibx} 
\end{figure}

Conversely, if the electric field is in the plane of the nodal line, 
the only allowed transitions change $n$ by one. There are here two types of processes, 
namely, the ones which excite electrons from below to above the Fermi energy 
and the ones that only change the Landau band, without changing the sign of the energy. 
The latter are strongly suppressed at low temperatures, because of the occupation 
factors of the bands, but nevertheless do not have a threshold frequency to be activated. 
This contribution is very small, except for a large peak at $\omega=\alpha$, 
corresponding to the separation between any pair of adjacent Landau levels at $q_z=0$. 
Finally, the transitions that flip the sign of the energy have an activation threshold $\omega=\alpha$, 
plus a series of features depending on the specific configuration of the Landau bands 
for the choice of parameters, see Fig.~\ref{fig:Bands}, which are as exemplified in Fig.~\ref{fig:Ibx}.
Our analysis suggests that there is ample room for observing the contributions to the optical conductivity 
arising from transitions to the DDS. 
For instance, comparing Figs.~\ref{fig:Ix} and~\ref{fig:Ibx}, one sees that the interface contribution 
is manifest in the presence of additional peaks, whose position has a distinctive functional dependence 
from those in the bulk, specified in Eqns. \eqref{eq:em} and \eqref{eq:omegan}. 
The height of the peaks of Fig.~\ref{fig:Ix} is proportional to the rate $\Gamma$ 
in the denominator of Eq. \eqref{eq:lineshape}. Therefore, we expect the contribution 
from the interface states in a NLS-insulator junction to remain clearly identifiable provided  
${\ell}/{L} \gtrsim {\hbar\Gamma}/{\left(E_\text{g}+E_0+\sqrt{E_\text{g}^2-E_0^2}\right)}$.
Analogous conclusions can be reached from the comparison of Figs.~\ref{fig:Iz} and~\ref{fig:Ibz}.

\begin{figure}
\centering{}
\includegraphics[height=0.18\textheight]{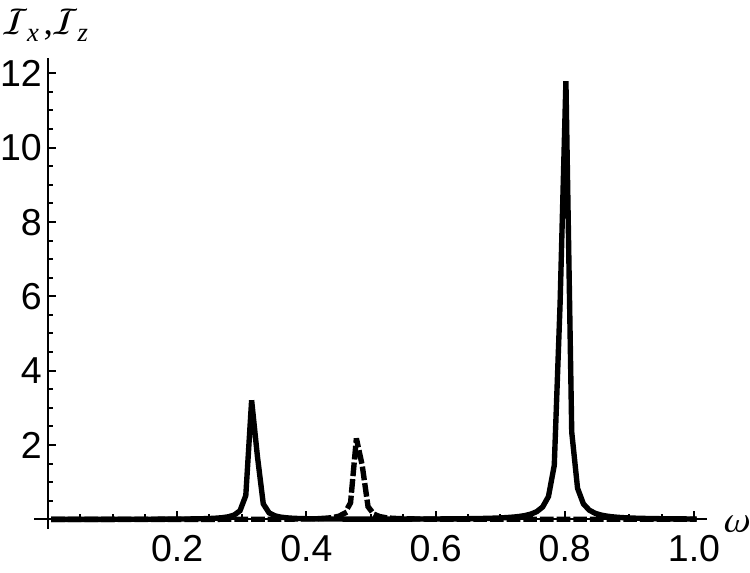}
\caption{Imaginary part of the functions $\mathcal{I}_{x}$ (solid line) and $\mathcal{I}_{z}$ (dashed line), 
in units of $I_p^{(0)}\ell/L$ and $I_z^{(0)}\ell/L$, respectively, as a function of the frequency. 
Here the parameters are $\lambda=\eta=3.1$, $\alpha=0.75$, $\mu=2.0$, $\Gamma=0.02$. 
The component $\mathcal{I}_{x}$ exhibits the transitions \mbox{$(1,1)\leftrightarrow(2,0)$} 
and \mbox{$(1,1)\leftrightarrow(1,0)$}, while the transitions to the bulk continuum (not shown) 
are $O(10^{-2})$. The component $\mathcal{I}_{z}$, multiplied for clarity by a factor $10$, 
shows the transition \mbox{$(1,1)\to(2,1)$} and, for this choice of parameters, 
the transitions to the bulk continuum produce a smooth background of the same order of magnitude of this peak.}
\label{fig:Im}
\end{figure}

We conclude this section by pointing out that, if the material allows somehow 
to adjust the chemical potential, one can also access a regime in which $\mu$ is 
close to the lowest-lying of the DDSs, 
thus, greatly enhancing their contribution. This scenario was explored
in the case of Weyl semimetals in \cite{Mukherjee2019}. In this case, the TDS lies deep 
within the valence band and can be safely neglected, while the most important transitions 
involve the DDSs alone, as exemplified in Fig.~\ref{fig:Im}, 
in which we observe the secondary peaks from Figs.~\ref{fig:Iz} and~\ref{fig:Ix} to be greatly enhanced. 
Nevertheless there are, also in this instance, contributions arising from transitions 
between the interface states and the continuum of bulk states. The bound state transitions 
are clearly distinguishable from transitions involving the continuum 
from the shape of the absorption peaks, even if this appears to be dependent 
on the direction of the electric field and the choice of parameters.

\section{Conclusions}
\label{sec:Conclusions}

In this work, we have shown that at the interfaces 
between NLSs and topologically trivial insulators,  
as well as between NLSs with different radii of the nodal ring,
in addition to topological drumhead states, 
a series of dispersive, non-topological states arise, 
if the interface is smooth enough. With the help of the exact solution of a model 
Hamiltonian, we have determined how
the parameters characterizing the interface smoothness and the nodal ring
determine the number of additional states, also in the presence of a magnetic field along
the system axis. The effective low-energy Hamiltonian used in this work is possibly the simplest model exhibiting 
a nodal line, but since our results mostly depend on the spectral features only, 
they hold qualitatively for a generic choice of material.

We have characterized the radiation-induced transitions 
between these dispersive drumhead states and the zero-energy topological states, 
as well as between them and the bulk states. 
For a generic NLS, and in particular for the specific choice of parameters 
corresponding to $\mbox{Ca}_2\mbox{P}_3$, we have found that the transitions 
can be clearly identified in optical conduction and absorption spectroscopy measurements, 
with radiation in the infrared part of the spectrum. 

Because of its flat energy dispersion,  
the  topological drumhead states are not directly observable in DC conductance experiments.
We have shown that AC measurements in the presence of additional interface states
are dominated by transitions from the drumhead state to the other localized states.
Our results, therefore, also provide an indirect way of detecting the presence of 
the topological state itself.

From a more general perspective, in topological semimetals, surface states 
can couple to massless bulk modes via various mechanisms, such as electron-impurity 
or electron-phonon scattering \cite{Roy2014,Breitkreiz2019,DeMartino2021,Buccheri2022a}. 
Importantly, the dispersive part of the interface spectrum does not enjoy the topological 
protection of the zero-energy mode and can strongly couple to the bulk in Dirac materials 
through strain \cite{Mahler2019}. While such hybridization can hinder their detection, 
it opens interesting possibilities. It is indeed possible, in principle, 
to have an attractive phonon-mediated 
electron-electron interaction  in topological semimetals \cite{Pereira2019}, 
which points toward the possibility that  the Volkov-Pankratov states could 
generate surface/interface superconductivity \cite{Gariglio2009}, 
while the bulk remains in its normal state.

With the sophisticated material synthesis techniques presently available, 
it appears likely that the prediction of this work may soon be tested experimentally, 
as already done for topological insulators.
As we have shown, the possibility of observing the transitions between interface states 
is mainly limited by the processes originating the linewidth broadening, 
which points toward the necessity of very clean samples.
Finally, the presence of both 
an interface and the magnetic field offers ample possibilities to tune the spectrum, 
that can be exploited to fully adjust the optical conductivity 
and the photo-absorption lineshapes for technological applications,
e.g, magnetic field detectors or magnetic-field-tunable laser sources exploiting the transition resonances.

\begin{acknowledgments}
We wish to thank D. Bercioux, N. Schröter and V. Dwivedi for interesting discussions. 
We acknowledge funding by the Deutsche Forschungsgemeinschaft (DFG, German Research Foundation) under
Projektnummer 277101999 - TRR 183 (project A02), as well as within Germany's 
Excellence Strategy-Cluster of Excellence ``Matter and Light for
Quantum Computing'' (ML4Q), EXC 2004/1-390534769.
\end{acknowledgments}

\appendix

\section{Absorbing the particle-hole symmetry breaking term}
\label{sec:gamma}

Let us consider the Dirac equation with the Hamiltonian~\eqref{eq:hgamma} 
including the particle-hole symmetry breaking term:
\begin{equation}
\left[-i\tau_{y} \partial_z + M(q_{p},z) \left( \tau_{z}+ \gamma \tau_0 \right) - \varepsilon \right]\psi=0,
\label{phham}
\end{equation}
where we use the notation $M(q_p,z)= d(q_p)  - \lambda \tanh z$.
Multiplying Eq.~\eqref{phham} by $\tau_y$, 
we rewrite it in the form
\begin{equation}
\left[ -i \partial_{z}+ M(q_{p},z)  \left( i\tau_{x}+\gamma \tau_{y} \right)-\tau_{y}\varepsilon\right]\psi=0.
\end{equation}
Taking into account that $D_p>-D_0>0$ \cite{Chan2016a}, we parameterize the ratio $\gamma=D_0/D_p$ 
as  
\begin{equation}
 \gamma = -\tanh \theta, 
\end{equation}
so that 
\begin{align}
     i\tau_{x}+\gamma\tau_{y}&=i \sqrt{1-\gamma^2}
    \begin{pmatrix}0 & e^{\theta}\\
e^{-\theta} & 0
\end{pmatrix}.
\end{align}
Then we introduce the transformation  $\tilde \psi=U\psi$, with
\begin{equation}
U=\begin{pmatrix}e^{-\theta/2} & 0\\
0 & e^{\theta/2}
\end{pmatrix},
\end{equation}
such that 
\begin{align*}
U\tau_{x}U^{-1} & =\tau_{x}\cosh \theta -i\tau_{y}\sinh \theta,\\
U\tau_{y}U^{-1} & =i\tau_{x}\sinh \theta+\tau_{y}\cosh \theta,\\
U\left(i\tau_{x} + \gamma \tau_{y}\right)U^{-1} & =i  \sqrt{1-\gamma^2 }\tau_{x}.
\end{align*}
Applying this transformation, Eq.~\eqref{phham} can be recast in the form 
\begin{align}
\left[-i \tau_{y}\partial_{z} + \tilde{M}(q_{p},z) \tau_{z} -\tilde \varepsilon \right] \tilde \psi=0,
\label{hamPHB}
\end{align}
where 
\begin{align}
\tilde{M}(q_p,z) & =\sqrt{1-\gamma^{2}}M(q_p,z)-\varepsilon \sinh \theta , \\
\tilde{\varepsilon}&= \varepsilon \cosh \theta. \nonumber
\end{align}
We see that Eq.~\eqref{hamPHB} has the same form as Eq.~\eqref{phham} with $\gamma=0$ and 
a redefinition of $M(q_p,z)$ and $\varepsilon$. Therefore, all results obtained in 
Sec.~\ref{sec:SpectrumNoField} for $\gamma=0$ can be easily adapted to the case $\gamma \neq 0$.

\section{Hypergeometric function}
\label{sec:Hypergeometric}

In this appendix, we collect few useful formulas involving the hypergeometric function
$F(a,b;c;u)$ (see, e.g., \cite{Lebedev}). $F(a,b;c;u)$ 
is a solution of the hypergeometric equation
\begin{equation}
 u\left(1-u\right)f''+\left[c-\left(a+b+1\right)u\right]f'-abf = 0.
\label{eq:hyper}
\end{equation}
The general solution of Eq.~\eqref{eq:hyper} can be expressed as 
\begin{align}
f(u) & =  A~F(a,b;c;u) + \nonumber \\ 
& + B\, u^{1-c} F(1+a-c,1+b-c;2-c;u), 
\label{eq:generalSolutio}
\end{align} 
or equivalently as
\begin{align}
f(u)  & =  C~F(a,b;a+b-c+1;1-u) + \label{eq:othergeneralSolutio} \\
& + D\left(1-u\right)^{c-a-b}F(c-a,c-b;c-a-b+1;1-u), \nonumber
\end{align}
where $A,B$ and $C,D$ are arbitrary complex coefficients. 
The two pairs of coefficients can be related by means of the following identity:
\begin{widetext}
\begin{align}
F(a,b;c;u) & = 
\frac{\Gamma (c) \, \Gamma(c-a-b)}{\Gamma (c-a ) \, \Gamma (c-b)}
F(a,b;a+b-c+1;1-u)~+ \nonumber \\
 &   +\frac{\Gamma (c) \, \Gamma (a+b-c )}{\Gamma (a )  \, \Gamma(b)}
 \left(1-u\right)^{c-a-b}
 F(c-a,c-b;c-a-b+1;1-u) .
 \label{eq:F(u)F(1-u)} 
\end{align}
\end{widetext}
The leading behavior of $F(a,b;c;u)$ for $u\rightarrow1^{-}$ is
\begin{equation}
F(a,b;c;u)\sim(1-u)^{c-a-b}\frac{\Gamma(c) \, \Gamma(a+b-c)}{\Gamma(a) \, \Gamma(b)},
\label{eq:F(u->1)}
\end{equation}
provided that $\Re(c-a-b)<0$.

\section{Solution of the Dirac-Schrödinger equation}
\label{sec:Equations}

In this appendix, we provide some technical details about the solution of the Dirac-Schrödinger 
equation with the Hamiltonian \eqref{eq:h}. In terms of $\phi_\pm$, defined in Eq.~\eqref{phi}, 
we have two coupled differential equations:
\begin{equation}
\left[ \mp \partial_z + d - \lambda \tanh z \right] \phi_{\pm}  =  
\varepsilon \phi_{\mp}.
\label{eq:SDeqpm}
\end{equation}
First, we observe that in the asymptotic region $z\to +\infty$, 
the general solution of Eq.~\eqref{eq:SDeqpm} takes the form
\begin{equation}
    \begin{pmatrix}
        \phi_+\\ \phi_-
    \end{pmatrix}\sim \sum_{r=\pm} c_r \begin{pmatrix}
\varepsilon \\ r \kappa_+ + d -\lambda
    \end{pmatrix} e^{-r\kappa_+z },
    \label{asymptotics+}
\end{equation}
while for $z\to-\infty$ it takes the form
\begin{equation}
    \begin{pmatrix}
        \phi_+\\ \phi_-
    \end{pmatrix}\sim \sum_{r=\pm} d_r \begin{pmatrix}
\varepsilon \\ - r \kappa_- + d + \lambda
    \end{pmatrix} e^{r\kappa_- z },
      \label{asymptotics-}
\end{equation}
with $c_\pm,d_\pm$ complex coefficients and $\kappa_{\pm}$ defined in \eqref{kappapm}.

\subsection{General solution}
\label{app:gensol}

By using the variable change~\eqref{eq:u} with the identities
$$
\partial_z= -2u(1-u)\partial_u, \quad \tanh z= 1-2u,
$$
and the factorization~\eqref{eq:xidef},
Eq.~\eqref{eq:SDeqpm} can be recast in the form of two coupled equations for $\chi_\pm$:
\begin{equation}
\left[ \pm 2u(1-u)\partial_u \pm \left( \kappa_+ \pm d \mp \lambda\right) 
 \mp 2(\kappa \mp \lambda)u\right] \chi_\pm =\varepsilon \chi_\mp, 
\label{eq:ReducedSchr}
\end{equation}
where  $\kappa=\frac{\kappa_{+}+\kappa_{-}}{2}$.
Notice that if we assume that $\chi_\pm$ stay finite for $u\to 0$ (resp. $u\to 1$),
Eqs.~\eqref{eq:ReducedSchr} reproduce the asymptotic behavior 
in~\eqref{asymptotics+} (resp.~\eqref{asymptotics-}), but including only the terms with $r=+1$. 
The equations~\eqref{eq:ReducedSchr}  can be decoupled and result in the second order equations
\begin{widetext}
\begin{align}
 \left[
 u\left(1-u\right) \partial_{u}^{2} + \left[ \kappa_{+}+1-\left(2\kappa+2\right)u\right] \partial_{u}
- \left(\kappa\mp\lambda\right)\left(\kappa\pm\lambda+1\right) \right] \chi_{\pm} =0 .
\label{eq:ueq3}
\end{align}
\end{widetext}
We recognize Eq.~\eqref{eq:ueq3} as the hypergeometric equation \eqref{eq:hyper} 
with $a=\kappa\mp\lambda$, $b=\kappa\pm\lambda+1$ and
$c=\kappa_{+}+1$. The general solution can then be written in
terms of hypergeometric functions in the form~\eqref{eq:generalSolutio} 
or~\eqref{eq:othergeneralSolutio}. 

\subsection{Bound states}
\label{app:bs}

In the regime in which $\kappa_\pm$ are both real, it is convenient to 
use the form~\eqref{eq:generalSolutio}. The condition of normalizability requires
to omit the second term in~\eqref{eq:generalSolutio}, and we obtain
\begin{equation}
\chi_\pm = A_\pm  F(\kappa \mp \lambda,\kappa \pm \lambda +1;\kappa_++1;u).
\label{chipmBS}
\end{equation}
The complex coefficients $A_{\pm}$ are determined by the first order equations~\eqref{eq:ReducedSchr} and by 
the overall normalization of the wave function. In fact, 
from the limit $u\to 0$ of Eq.~\eqref{eq:ReducedSchr}, 
one finds the relation
\begin{equation}
\left( \kappa_+ + d -\lambda \right)A_+ = \varepsilon A_-,
\end{equation}
hence we can write
\begin{align}
& \begin{pmatrix}
    \phi_+ \\ \phi_- 
    \end{pmatrix}
     =  u^{\kappa_{+}/2}\left(1-u\right)^{\kappa_{-}/2} \times \label{chiBS}\\
& \times  \begin{pmatrix}
      F(\kappa - \lambda,\kappa + \lambda +1;\kappa_++1;u) \\ 
   \frac{ \kappa_+ + d -\lambda}{\varepsilon}
   F(\kappa + \lambda,\kappa - \lambda +1;\kappa_++1;u)
\end{pmatrix}. \nonumber 
\end{align}
The wave function in Eq.~\eqref{chiBS} are normalizable only if 
either of the first two arguments of the hypergeometric functions is a non-positive integer, 
which gives the quantization condition discussed in Sec.~\ref{sec:SpectrumNoField}.
The full wave function of the interface states reads
\begin{align}
&\psi_m  =  
\mathcal{N}_{m}  u^{\kappa_{+,m}/2}\left(1-u\right)^{\kappa_{-,m}/2}\times  \label{eq:Psim}\\
& \times \Big[
 F(-m,-m+2\lambda+1;\kappa_{+,m} +1 ; u )  \, | + \rangle \, + \nonumber  \\
& +\frac{ \kappa_{+,m} + d -\lambda }{\varepsilon_m} 
F(-m +2\lambda , - m + 1;\kappa_{+,m)}+1 ; u )
\, | - \rangle \Big] , \nonumber 
\end{align}
where $\varepsilon_m$ and $\kappa_{\pm,m}$ are given in Eq.~\eqref{eq:em} and~\eqref{eq:kappapm} respectively,
and $\mathcal{N}_{m}$  is the overall normalization.
The normalization constant has a particularly simple form  for the zero-energy solution~\eqref{eq:Psi0}, 
where we find  
\begin{align}
& \mathcal{N}_{0}   = 
\frac{1}{2^{\lambda-1/2} \sqrt{\ell}}
\Big[  
(-1)^{d-\lambda} B(-1,\lambda - d,1-2\lambda) \, +  \nonumber \\
& + \Gamma\left(\lambda + d \right)
F_\text{reg}(2\lambda,\lambda+d,1+\lambda + d, -1) \Big]^{-1/2},
\label{eq:N0}
\end{align}
where $F_\text{reg}(a,b,c,z)$ is the regularized hypergeometric function and 
$B(z,a,b)$ the incomplete Beta function  \cite{NIST:DLMF}.

\subsection{Scattering states}
\label{app:ss}

In the regime in which $\kappa_-$ is real and $\kappa_+=ik_+$ is imaginary,
it is convenient to use the general solution in the form~\eqref{eq:othergeneralSolutio}.
The condition of normalizability requires again to omit the second term and we obtain 
\begin{equation}
\chi_\pm = C_\pm  F(\kappa \mp \lambda,\kappa \pm \lambda +1;\kappa_-+1;1-u),
\label{chipmRW}
\end{equation}
with $\kappa= \frac{\kappa_-+ik_+}{2}$.
From the asymptotic behavior of Eq.~\eqref{eq:ReducedSchr} for $u\rightarrow 1$, 
we find the relation
\begin{equation}
     \left(- \kappa_- +   d + \lambda  \right) C_+ = \varepsilon C_- ,
\end{equation}
and we arrive at  
\begin{align}  
& \begin{pmatrix}
    \phi_+ \\ \phi_- 
    \end{pmatrix}
    = u^{ik_+/2}\left(1-u\right)^{\kappa_-/2} \times  \label{propatatingphi}  \\
& \times  \begin{pmatrix}
 F(\kappa - \lambda,\kappa + \lambda +1;\kappa_-+1;1-u) \\ 
  \frac{ -\kappa_- + d +\lambda}{  \varepsilon }
   F(\kappa + \lambda,\kappa - \lambda +1;\kappa_-+1;1-u)
\end{pmatrix}. \nonumber
\end{align}
Far away from the interface, in the limit $z \to +\infty$ ($u\to0$),
the wave function assumes the form of a superposition of a left-moving and a right-moving wave, 
thus describing a propagating state in the NLS incoming from the right, incident on the interface 
and reflected back. To see this, we use $u \sim e^{-2z}$ and the identity~\eqref{eq:F(u->1)} to
arrive at the asymptotic form for $z\to +\infty$ 
\begin{align}
    \begin{pmatrix}
    \phi_+ \\ \phi_-
    \end{pmatrix}
    &\sim \frac{\Gamma(\kappa_-+1) \, \Gamma(-ik_+) }{ \Gamma(\kappa' -\lambda) \, \Gamma(\kappa' + \lambda+1)  
    } \varphi_+ (k_+) e^{-ik_+z} + \nonumber \\
    &+    \frac{\Gamma(\kappa_-+1) \, \Gamma(ik_+) }{ \Gamma(\kappa-\lambda) \, \Gamma(\kappa +\lambda+1) }
        \varphi_+(-k_+) e^{ik_+z} , 
\label{asym}
\end{align}
where $\kappa'=\frac{\kappa_--ik_+}{2}$ and the spinors $\varphi_\pm(k)$ are given by
\begin{equation}
    \varphi_\pm (k) = \begin{pmatrix}
\varepsilon \\ i k  + d \mp \lambda
    \end{pmatrix},
    \label{asymptotic+spinors}
\end{equation}
in agreement with Eq.~\eqref{asymptotics+}. 
From Eq.~\eqref{asym} we can read the reflection amplitude $\mathcal{R} $ given in Eq.~\eqref{reflectionampl}.
Since in this case $\kappa'$ coincides with the complex conjugate of $\kappa$, 
we find $|\mathcal{R}|^2=1$, and the transmission coefficient is zero. 
The normalization constant is obtained by imposing the condition 
$$
\int_0^\infty dz \, \psi^\dagger_{k_+}\psi_{k'_+} =  \delta(k_+-k_+').
$$
Then, the normalized scattering state reads
\begin{align}
 &   \psi  =
    \frac{\Gamma(1+\kappa'+\lambda) \, \Gamma(\kappa'-\lambda)}{ 2\sqrt{\pi} \Gamma(\kappa_-+1) \, 
    \Gamma[-ik_+) }   u^{ik_+/2}  (1-u)^{\kappa_-/2} \times     \label{eq:PsiR} \\
 & \times  
 \Big[ F(\kappa - \lambda,\kappa + \lambda +1 ; \kappa_- +1 ; 1-u) \, |+\rangle + \nonumber \\
  &  + 
 \frac{-\kappa_- + d + \lambda}{\varepsilon} 
   F( \kappa + \lambda, \kappa - \lambda + 1 ;  \kappa_- +1 ; 1-u) \, |-\rangle \Big]. \nonumber
\end{align} 

Finally, in the regime in which both $\kappa_+$ and $\kappa_-$ are imaginary, the
scattering state~\eqref{propatatingphi} describes an incident wave incoming from the right, 
which at the interface is partially reflected back and partially transmitted to the left, 
with the asymptotic form for $z\to -\infty$ given by
\begin{align}
    \begin{pmatrix}
    \phi_+ \\ \phi_-
    \end{pmatrix} \sim 
   \varphi_-(k_-) e^{-ik_-z}  .
\label{asym2}
\end{align}
The full wave function can be obtained from Eq.~\eqref{eq:PsiR} 
by the analytic continuation $\kappa_-=-ik_-$, 
but the overall normalization must be recalculated to include the 
contribution of the propagating wave in the left region $z\to -\infty$.
The transmission coefficient is now finite and given in Eq.~\eqref{transmission}.

\section{Interface with vacuum}
\label{sec:Vacuum}

An interesting case of the interface problem is realized when the NLS occupies the half space $z>0$
and has an open surface at $z=0$. This can be regarded as a special case of the system studied 
in this work, where the gap in the insulating side $z<0$ diverges. In this section, 
we discuss this limit using the exact solution of Sec.~\ref{sec:SpectrumNoField}. 
We have also studied the exact solution of a different version of this problem, where the
infinite mass boundary condition at $z=0$ is imposed in the standard way. As expected, the 
qualitative features of the surface states spectrum are the same, although the detailed 
shape of the energy dispersions shows small quantitative differences.

We consider our model with $\gamma=0$ for simplicity and restrict our discussion to the surface states. 
It is straightforward to calculate the limit $\lambda \rightarrow +\infty$ of the interface states energies.
Using the formulas in Sec.~\ref{subsec:drumheadstates}, we obtain:
\begin{align}
\kappa_{+,m} &= \eta - \eta q_p^2 -2m, 
\label{limitK}\\  
\varepsilon_m & = 2  \sqrt{m\left( \eta - \eta q_p^2 -m \right)},  
\label{limitE}
\end{align}
with $m=0,1,2,\dots, m_\text{max}$.
The surface state dispersions merge in the NLS bulk band at
$$
q_{0,m}=\sqrt{1 - 2m/\eta}, \quad  \varepsilon_m^{(0)}= 2m.
$$
Interestingly, the number of surface states only depends on the nodal radius 
through the parameter $\eta$, and we find 
\begin{equation}
    m_\text{max} = \lfloor\eta/2 \rfloor.
    \label{infinitegapmmax}
\end{equation}

In order to calculate the limit of the wave functions, we can proceed as follows.
The mass function is
\begin{equation}
    M(q_p^2,z) = \eta q_p^2- \eta +\lambda  \left( 1-\tanh z \right).
\end{equation}
We want to take the limit $\lambda \rightarrow +\infty$ keeping $\eta$ fixed and with a prescribed 
value of the mass at $z=0$. To achieve this, rather than taking directly the limit $\lambda \rightarrow +\infty$,
we set $\lambda=\frac{1}{2} \tilde \lambda e^{2z_0}$, shift the coordinate $z$ by $z_0$,
and perform the limit $z_0\rightarrow +\infty$, keeping $\tilde \lambda$ fixed. 
This procedure  gives a finite limit for the mass function:
\begin{equation}
   \lim_{z_0\rightarrow +\infty} M(q_p^2,z+z_0) = \eta q_p^2 - \eta + \tilde \lambda e^{-2z}.
  \label{limitmass}
\end{equation}
From this expression, we see that $\sqrt{1 -\tilde \lambda/\eta}$ represents the radius of the nodal line at $z=0$
(in units of $k_0$), 
as long as $\tilde \lambda<\eta$; if $\tilde \lambda>\eta$, this limit describes a system with a gap at the surface $z=0$.
Applying this procedure to the zero-energy state in Eq.~\eqref{eq:Psi0}, we find
\begin{align}
\psi_0 (z)
= \mathcal{N}_0 e^{(\eta q_p^2-\eta)z - (\tilde \lambda/2)e^{-2z}} | +\rangle, \quad q_p<1,
\label{zeroenergyVacuum}
\end{align}
with normalization 
\begin{equation}
    \mathcal{N}_0=\left( 2^{(\eta q_p^2-\eta)-1} \ell (\tilde \lambda/2)^{(\eta q_p^2-\eta)} 
    \Gamma(\eta - \eta q_p^2)\right)^{-\frac{1}{2}}.
\end{equation}
This result can also be obtained directly by solving the Dirac equation with the 
mass function~\eqref{limitmass} at zero energy.
To find the other surface states, we first observe that in the limit $z_0 \rightarrow + \infty$, 
we have $u \rightarrow e^{-2(z+z_0)}$ and
$$
u^{\kappa_+/2} \rightarrow  e^{-\kappa_+(z+z_0)}, \quad 
(1-u)^{\kappa_-/2}  \rightarrow e^{-(\tilde \lambda/2)e^{-2z}}.
$$
Moreover, we use the identity
$$
\lim_{z_0 \rightarrow \infty} F(a,\tilde \lambda e^{2z_0};c;e^{-2(z+z_0)}) = M(a,c, \tilde \lambda e^{-2z}),
$$
where $M(a,c,z)$ is the confluent hypergeometric function \cite{NIST:DLMF} 
defined by 
$$
M(a,c,u) =1 + \frac{a}{c}u + \frac{a(a+1)}{c(c+1)} \frac{u^2}{2!} + \dots 
$$
Therefore, the eigenstates~\eqref{eigenstates} become
\begin{align}
\begin{pmatrix}
\phi_+ \\ \phi_- 
\end{pmatrix}
&=  e^{-\kappa_{+,m}z- \frac{\tilde \lambda}{2}e^{-2z}} \times  \nonumber \\
& \times \begin{pmatrix}
  M( -m , 1 + \kappa_{+,m} +1 , \tilde \lambda e^{-2z})
 \\ \frac{-2 m}{\varepsilon_m}  M ( -m + 1 ,  \kappa_{+,m} +1 , \tilde \lambda e^{-2z})
 \end{pmatrix} ,
\end{align}
with $\kappa_{+,m}$ and  $\varepsilon_m$ given in Eqs.~\eqref{limitK} and~\eqref{limitE}.

\section{Estimation of the bulk contributions}
\label{app:bulk}

The bulk transitions are most efficiently estimated using a purely bulk theory, 
in which the transverse part of the eigenstates is a Landau state labelled by the indices $n,s$ as in
Eq.~\eqref{eq:Factorizedstate}, while the longitudinal part  is 
simply a plane wave labeled by the component of the momentum along $z$, $q_z$,
and by the particle/hole index $r=\pm$. 
The longitudinal part takes the form $\psi_{n,q_z,r}(z)=u_{n,r}(q_z) e^{iq_zz}/\sqrt{L}$ 
with
\begin{equation}
    u_{n,r}(q_z)	=	\mathcal{\mathcal{N}}_{n,q_{z},r}
    \begin{pmatrix}
-iq_{z}\\
r\varepsilon_{n}(q_{z}) - M_{n}
\end{pmatrix},
\end{equation}
where $\varepsilon_{n}(q_z)$
has the expression given in~\eqref{eq:ebulk}  
with the positive sign and $M_n$ is also provided after Eq. \eqref{eq:ebulk}. 
The normalization factor is 
\begin{equation}
\mathcal{\mathcal{N}}_{n,q_{z},r}=
\frac{1}{\sqrt{2r\varepsilon_{n}(q_{z})\left[r\varepsilon_{n} (q_{z})-M_{n} \right]}}.
\end{equation}

In the limit in which the momentum of the radiation vanishes, 
one has the so-called vertical transitions only, where $q_{z}$ is conserved. 
Hence, for notational simplicity, throughout this section we will often omit the argument $q_z$.
Moreover, we consider here the limit $\Gamma\to 0$ in Eq.~\eqref{eq:lineshape}, 
where the energy conservation is strictly enforced. 

In the case  with the electric field in the $z$ direction, 
the relevant matrix elements are 
\begin{equation}
    \left\langle \Psi_{n',s',\mp} \right| \hat j_{z} \left| \Psi_{n,s,\pm} \right\rangle 
    = -\frac{ev_z q_{z}M_{n}}{\varepsilon_{n}\left|q_{z}\right|} \delta_{s,s'}\delta_{n,n'}.
\end{equation}
As the allowed transitions are those between states in the same Landau orbital 
but with opposite energy, the energy conservation  implies $|\omega|=2\varepsilon_n$.
We then find the expression
\begin{equation}
   \frac{\mathcal{I}_z (\omega)}{I_z^{(0)}}=\frac{1}{2\pi} \sum_{n}
    \frac{\Theta\left(\frac{\omega^{2}}{4}-M_{n}^{2}\right) M_n^2 }{\frac{|\omega|}{2}\sqrt{\frac{\omega^{2}}{4}-M_{n}^{2}}}
    \frac{\sinh(\frac{\beta\omega}{2})}{\cosh(\frac{\beta\omega}{2}) + \cosh(\beta\mu)},
\end{equation}
where, following our conventions, 
the chemical potential is measured in units of $\hbar v_z /\ell$ and $\beta=\hbar v_z/\ell k_BT$, accordingly.

When the electric field is in the plane of the nodal line, 
the relevant matrix elements of the current density operator take the form
\begin{widetext}
\begin{align}
 \left\langle \Psi_{n',s',r'}  \right| \hat{\bar \jmath} \left|\Psi_{n,s,r} \right\rangle &=
 \frac{2ieD_p}{\hbar \ell_B}  \left\langle u_{n',r'} \right| \tau_z \left|u_{n,r} \right\rangle
\delta_{s,s'}\delta_{n',n+1} \sqrt{n+1}, \\
  \left\langle \Psi_{n',s',r'}  \right| \hat \jmath \left|\Psi_{n,s,r} \right\rangle &=
 \frac{2eD_p}{i\hbar \ell_B}  \left\langle u_{n',r'} \right| \tau_z \left|u_{n,r} \right\rangle
\delta_{s,s'}\delta_{n',n-1} \sqrt{n},
\end{align}
with
\begin{align}
    \left\langle u_{n',r'}\right|\tau_{z}\left|u_{n,r}\right\rangle &
    =\frac{\left(r'\varepsilon_{n'}-r\varepsilon_{n}\right)^{2}-\left(M_{n}+M_{n'}\right)^{2}
    +2\left(r'\varepsilon_{n'}M_{n}+r\varepsilon_{n}M_{n'}\right)}{4\sqrt{rr'\varepsilon_{n}\varepsilon_{n'}
    \left(r\varepsilon_{n}-M_{n}\right)\left(r'\varepsilon_{n'}-M_{n'}\right)}}.
\end{align}
In these transitions, the Landau level index changes by one. By using the energy conservation 
$\omega=r\varepsilon_n-r'\varepsilon_{n'}$, one can rewrite the square modulus of the matrix element as

\begin{align}\label{eq:sqmetau}
    \left|\left\langle u_{n-1,r'}\right|\tau_{z}\left|u_{n,r}\right\rangle \right|^{2} &=	
    \frac{\omega^{2} \left[\omega^2-\left(M_{n}+M_{n-1}\right)^2\right] }
    {\omega^{4}-\left(M_{n}^{2}-M_{n-1}^{2}\right)^2}.
\end{align}
In this case, two types of processes contribute to $\mathcal{I}_x$: 
those in which the energy changes sign ($r'=-r$) and those in which 
the energy doesn't change sign ($r'=r$).
In the first case, the transition can only take place if the photon carries more energy 
than the minimal separation between the involved bands, $|\omega|>{|M_{n}|+|M_{n-1}|}$. 
We arrive at the expression
\begin{align}\label{eq:Ixinter}
   \frac{\mathcal{I}^{(1)}_x(\omega)}{I_p^{(0)}} &= 4\sum_{n\ge0} \Theta\left(|\omega|-\left|M_{n}\right|-\left|M_{n-1}\right|\right) 
   \frac{n}{2\pi}
   \sqrt{\frac{\omega^2-\left(M_{n}+M_{n-1}\right)^2}{\omega^2-\left(M_{n}-M_{n-1}\right)^2}}\frac{\sinh\frac{\beta\omega}{2}}{\cosh\frac{\beta\omega}{2}+\cosh\frac{\beta\left(M_{n}^{2}-M_{n-1}^{2}\right)}{2\omega}},
\end{align}
written for $\mu=0$ for simplicity.
In the second case, because of the form of the bands (see Fig.~\ref{fig:Bands}), 
there is no activation threshold, but instead an upper bound 
for the frequency, $|\omega|<\big|\left|M_{n}\right|-\left|M_{n-1}\right|\big|$, 
and one arrives at the expression
\begin{align}\label{eq:Ixintra}
\frac{\mathcal{I}^{(2)}_{x}\left(\omega\right)}{I_p^{(0)}}&= 4
\sum_{n\ge0}\Theta\left(\big|\left|M_{n}\right|-\left|M_{n-1}\right|\big|-|\omega|\right)
\frac{n}{2\pi}\sqrt{\frac{\left(M_{n}+M_{n-1}\right)^2-\omega^2}{\left(M_{n}-M_{n-1}\right)^2-\omega^2}}\frac{\sinh\frac{\beta\omega}{2}}{\cosh\frac{\beta\omega}{2}+\cosh\frac{\beta\left(M_{n}^{2}-M_{n-1}^{2}\right)}{2\omega}}
\end{align}
which is the same as \eqref{eq:Ixinter}, except for the argument of the Heaviside theta function. 

\end{widetext}

\bibliography{NLI}

\begin{thebibliography}{72}%
\makeatletter
\providecommand \@ifxundefined [1]{%
 \@ifx{#1\undefined}
}%
\providecommand \@ifnum [1]{%
 \ifnum #1\expandafter \@firstoftwo
 \else \expandafter \@secondoftwo
 \fi
}%
\providecommand \@ifx [1]{%
 \ifx #1\expandafter \@firstoftwo
 \else \expandafter \@secondoftwo
 \fi
}%
\providecommand \natexlab [1]{#1}%
\providecommand \enquote  [1]{``#1''}%
\providecommand \bibnamefont  [1]{#1}%
\providecommand \bibfnamefont [1]{#1}%
\providecommand \citenamefont [1]{#1}%
\providecommand \href@noop [0]{\@secondoftwo}%
\providecommand \href [0]{\begingroup \@sanitize@url \@href}%
\providecommand \@href[1]{\@@startlink{#1}\@@href}%
\providecommand \@@href[1]{\endgroup#1\@@endlink}%
\providecommand \@sanitize@url [0]{\catcode `\\12\catcode `\$12\catcode
  `\&12\catcode `\#12\catcode `\^12\catcode `\_12\catcode `\%12\relax}%
\providecommand \@@startlink[1]{}%
\providecommand \@@endlink[0]{}%
\providecommand \url  [0]{\begingroup\@sanitize@url \@url }%
\providecommand \@url [1]{\endgroup\@href {#1}{\urlprefix }}%
\providecommand \urlprefix  [0]{URL }%
\providecommand \Eprint [0]{\href }%
\providecommand \doibase [0]{https://doi.org/}%
\providecommand \selectlanguage [0]{\@gobble}%
\providecommand \bibinfo  [0]{\@secondoftwo}%
\providecommand \bibfield  [0]{\@secondoftwo}%
\providecommand \translation [1]{[#1]}%
\providecommand \BibitemOpen [0]{}%
\providecommand \bibitemStop [0]{}%
\providecommand \bibitemNoStop [0]{.\EOS\space}%
\providecommand \EOS [0]{\spacefactor3000\relax}%
\providecommand \BibitemShut  [1]{\csname bibitem#1\endcsname}%
\let\auto@bib@innerbib\@empty
\bibitem [{\citenamefont {Syzranov}\ and\ \citenamefont
  {Skinner}(2017)}]{Syzranov2017}%
  \BibitemOpen
  \bibfield  {author} {\bibinfo {author} {\bibfnamefont {S.~V.}\ \bibnamefont
  {Syzranov}}\ and\ \bibinfo {author} {\bibfnamefont {B.}~\bibnamefont
  {Skinner}},\ }\bibfield  {title} {\bibinfo {title} {Electron transport in
  nodal-line semimetals},\ }\href {https://doi.org/10.1103/PhysRevB.96.161105}
  {\bibfield  {journal} {\bibinfo  {journal} {Phys. Rev. B}\ }\textbf {\bibinfo
  {volume} {96}},\ \bibinfo {pages} {161105} (\bibinfo {year}
  {2017})}\BibitemShut {NoStop}%
\bibitem [{\citenamefont {Laha}\ \emph {et~al.}(2020)\citenamefont {Laha},
  \citenamefont {Mardanya}, \citenamefont {Singh}, \citenamefont {Lin},
  \citenamefont {Bansil}, \citenamefont {Agarwal},\ and\ \citenamefont
  {Hossain}}]{Laha2020}%
  \BibitemOpen
  \bibfield  {author} {\bibinfo {author} {\bibfnamefont {A.}~\bibnamefont
  {Laha}}, \bibinfo {author} {\bibfnamefont {S.}~\bibnamefont {Mardanya}},
  \bibinfo {author} {\bibfnamefont {B.}~\bibnamefont {Singh}}, \bibinfo
  {author} {\bibfnamefont {H.}~\bibnamefont {Lin}}, \bibinfo {author}
  {\bibfnamefont {A.}~\bibnamefont {Bansil}}, \bibinfo {author} {\bibfnamefont
  {A.}~\bibnamefont {Agarwal}},\ and\ \bibinfo {author} {\bibfnamefont
  {Z.}~\bibnamefont {Hossain}},\ }\bibfield  {title} {\bibinfo {title}
  {{Magnetotransport properties of the topological nodal-line semimetal
  CaCdSn}},\ }\href {https://doi.org/10.1103/PhysRevB.102.035164} {\bibfield
  {journal} {\bibinfo  {journal} {Phys. Rev. B}\ }\textbf {\bibinfo {volume}
  {102}},\ \bibinfo {pages} {035164} (\bibinfo {year} {2020})}\BibitemShut
  {NoStop}%
\bibitem [{\citenamefont {Barati}\ and\ \citenamefont
  {Abedinpour}(2020)}]{Barati2020}%
  \BibitemOpen
  \bibfield  {author} {\bibinfo {author} {\bibfnamefont {S.}~\bibnamefont
  {Barati}}\ and\ \bibinfo {author} {\bibfnamefont {S.~H.}\ \bibnamefont
  {Abedinpour}},\ }\bibfield  {title} {\bibinfo {title} {{Thermoelectric
  response of nodal-line semimetals: Probing the Fermi surface topology}},\
  }\href {https://doi.org/10.1103/PhysRevB.102.125139} {\bibfield  {journal}
  {\bibinfo  {journal} {Phys. Rev. B}\ }\textbf {\bibinfo {volume} {102}},\
  \bibinfo {pages} {125139} (\bibinfo {year} {2020})}\BibitemShut {NoStop}%
\bibitem [{\citenamefont {Yang}\ \emph {et~al.}(2022)\citenamefont {Yang},
  \citenamefont {Luo},\ and\ \citenamefont {Chen}}]{Yang2022}%
  \BibitemOpen
  \bibfield  {author} {\bibinfo {author} {\bibfnamefont {M.-X.}\ \bibnamefont
  {Yang}}, \bibinfo {author} {\bibfnamefont {W.}~\bibnamefont {Luo}},\ and\
  \bibinfo {author} {\bibfnamefont {W.}~\bibnamefont {Chen}},\ }\bibfield
  {title} {\bibinfo {title} {Quantum transport in topological nodal-line
  semimetals},\ }\href {https://doi.org/10.1080/23746149.2022.2065216}
  {\bibfield  {journal} {\bibinfo  {journal} {Advances in Physics: X}\ }\textbf
  {\bibinfo {volume} {7}},\ \bibinfo {pages} {2065216} (\bibinfo {year}
  {2022})}\BibitemShut {NoStop}%
\bibitem [{\citenamefont {Lau}\ \emph {et~al.}(2021)\citenamefont {Lau},
  \citenamefont {Hyart}, \citenamefont {Autieri}, \citenamefont {Chen},\ and\
  \citenamefont {Pikulin}}]{Lau2021}%
  \BibitemOpen
  \bibfield  {author} {\bibinfo {author} {\bibfnamefont {A.}~\bibnamefont
  {Lau}}, \bibinfo {author} {\bibfnamefont {T.}~\bibnamefont {Hyart}}, \bibinfo
  {author} {\bibfnamefont {C.}~\bibnamefont {Autieri}}, \bibinfo {author}
  {\bibfnamefont {A.}~\bibnamefont {Chen}},\ and\ \bibinfo {author}
  {\bibfnamefont {D.~I.}\ \bibnamefont {Pikulin}},\ }\bibfield  {title}
  {\bibinfo {title} {{Designing Three-Dimensional Flat Bands in Nodal-Line
  Semimetals}},\ }\href {https://doi.org/10.1103/PhysRevX.11.031017} {\bibfield
   {journal} {\bibinfo  {journal} {Phys. Rev. X}\ }\textbf {\bibinfo {volume}
  {11}},\ \bibinfo {pages} {031017} (\bibinfo {year} {2021})}\BibitemShut
  {NoStop}%
\bibitem [{\citenamefont {Wang}\ \emph
  {et~al.}(2022{\natexlab{a}})\citenamefont {Wang}, \citenamefont {B\"omerich},
  \citenamefont {Park}, \citenamefont {Legg}, \citenamefont {Taskin},
  \citenamefont {Rosch},\ and\ \citenamefont {Ando}}]{Wang2022b}%
  \BibitemOpen
  \bibfield  {author} {\bibinfo {author} {\bibfnamefont {Y.}~\bibnamefont
  {Wang}}, \bibinfo {author} {\bibfnamefont {T.}~\bibnamefont {B\"omerich}},
  \bibinfo {author} {\bibfnamefont {J.}~\bibnamefont {Park}}, \bibinfo {author}
  {\bibfnamefont {H.~F.}\ \bibnamefont {Legg}}, \bibinfo {author}
  {\bibfnamefont {A.~A.}\ \bibnamefont {Taskin}}, \bibinfo {author}
  {\bibfnamefont {A.}~\bibnamefont {Rosch}},\ and\ \bibinfo {author}
  {\bibfnamefont {Y.}~\bibnamefont {Ando}},\ }\href@noop {} {\bibinfo {title}
  {{Nonlinear transport due to magnetic-field-induced flat bands in the
  nodal-line semimetal ZrTe$_5$}}} (\bibinfo {year} {2022}{\natexlab{a}}),\
  \Eprint {https://arxiv.org/abs/2208.10314} {arXiv:2208.10314
  [cond-mat.mtrl-sci]} \BibitemShut {NoStop}%
\bibitem [{\citenamefont {Xie}\ \emph {et~al.}(2015)\citenamefont {Xie},
  \citenamefont {Schoop}, \citenamefont {Seibel}, \citenamefont {Gibson},
  \citenamefont {Xie},\ and\ \citenamefont {Cava}}]{Xie2015}%
  \BibitemOpen
  \bibfield  {author} {\bibinfo {author} {\bibfnamefont {L.~S.}\ \bibnamefont
  {Xie}}, \bibinfo {author} {\bibfnamefont {L.~M.}\ \bibnamefont {Schoop}},
  \bibinfo {author} {\bibfnamefont {E.~M.}\ \bibnamefont {Seibel}}, \bibinfo
  {author} {\bibfnamefont {Q.~D.}\ \bibnamefont {Gibson}}, \bibinfo {author}
  {\bibfnamefont {W.}~\bibnamefont {Xie}},\ and\ \bibinfo {author}
  {\bibfnamefont {R.~J.}\ \bibnamefont {Cava}},\ }\bibfield  {title} {\bibinfo
  {title} {{A new form of Ca$_3$P$_2$ with a ring of Dirac nodes}},\ }\href
  {https://doi.org/10.1063/1.4926545} {\bibfield  {journal} {\bibinfo
  {journal} {APL Materials}\ }\textbf {\bibinfo {volume} {3}},\ \bibinfo
  {pages} {083602} (\bibinfo {year} {2015})}\BibitemShut {NoStop}%
\bibitem [{\citenamefont {Abedi}\ \emph {et~al.}(2022)\citenamefont {Abedi},
  \citenamefont {Taghizadeh~Sisakht}, \citenamefont {Hashemifar}, \citenamefont
  {Ghafari~Cherati}, \citenamefont {Abdolhosseini~Sarsari},\ and\ \citenamefont
  {Peeters}}]{Abedi2022}%
  \BibitemOpen
  \bibfield  {author} {\bibinfo {author} {\bibfnamefont {S.}~\bibnamefont
  {Abedi}}, \bibinfo {author} {\bibfnamefont {E.}~\bibnamefont
  {Taghizadeh~Sisakht}}, \bibinfo {author} {\bibfnamefont {S.~J.}\ \bibnamefont
  {Hashemifar}}, \bibinfo {author} {\bibfnamefont {N.}~\bibnamefont
  {Ghafari~Cherati}}, \bibinfo {author} {\bibfnamefont {I.}~\bibnamefont
  {Abdolhosseini~Sarsari}},\ and\ \bibinfo {author} {\bibfnamefont {F.~M.}\
  \bibnamefont {Peeters}},\ }\bibfield  {title} {\bibinfo {title} {Prediction
  of novel two-dimensional {Dirac} nodal line semimetals in {Al}$_2${B}$_2$ and
  {Al}{B}$_4$ monolayers},\ }\href {https://doi.org/10.1039/D2NR00888B}
  {\bibfield  {journal} {\bibinfo  {journal} {Nanoscale}\ }\textbf {\bibinfo
  {volume} {14}},\ \bibinfo {pages} {11270} (\bibinfo {year}
  {2022})}\BibitemShut {NoStop}%
\bibitem [{\citenamefont {{Yu}}\ \emph {et~al.}(2017)\citenamefont {{Yu}},
  \citenamefont {{Fang}}, \citenamefont {{Dai}},\ and\ \citenamefont
  {{Weng}}}]{Yu2017}%
  \BibitemOpen
  \bibfield  {author} {\bibinfo {author} {\bibfnamefont {R.}~\bibnamefont
  {{Yu}}}, \bibinfo {author} {\bibfnamefont {Z.}~\bibnamefont {{Fang}}},
  \bibinfo {author} {\bibfnamefont {X.}~\bibnamefont {{Dai}}},\ and\ \bibinfo
  {author} {\bibfnamefont {H.}~\bibnamefont {{Weng}}},\ }\bibfield  {title}
  {\bibinfo {title} {Topological nodal line semimetals predicted from
  first-principles calculations},\ }\href
  {https://doi.org/10.1007/s11467-016-0630-1} {\bibfield  {journal} {\bibinfo
  {journal} {Frontiers of Physics}\ }\textbf {\bibinfo {volume} {12}},\
  \bibinfo {eid} {127202} (\bibinfo {year} {2017})}\BibitemShut {NoStop}%
\bibitem [{\citenamefont {Weng}\ \emph {et~al.}(2016)\citenamefont {Weng},
  \citenamefont {Dai},\ and\ \citenamefont {Fang}}]{Weng2016}%
  \BibitemOpen
  \bibfield  {author} {\bibinfo {author} {\bibfnamefont {H.}~\bibnamefont
  {Weng}}, \bibinfo {author} {\bibfnamefont {X.}~\bibnamefont {Dai}},\ and\
  \bibinfo {author} {\bibfnamefont {Z.}~\bibnamefont {Fang}},\ }\bibfield
  {title} {\bibinfo {title} {Topological semimetals predicted from
  first-principles calculations},\ }\href
  {https://doi.org/10.1088/0953-8984/28/30/303001} {\bibfield  {journal}
  {\bibinfo  {journal} {Journal of Physics: Condensed Matter}\ }\textbf
  {\bibinfo {volume} {28}},\ \bibinfo {pages} {303001} (\bibinfo {year}
  {2016})}\BibitemShut {NoStop}%
\bibitem [{\citenamefont {Mele}(2019)}]{Mele2019}%
  \BibitemOpen
  \bibfield  {author} {\bibinfo {author} {\bibfnamefont {E.~J.}\ \bibnamefont
  {Mele}},\ }\bibfield  {title} {\bibinfo {title} {Dowsing for nodal lines in a
  topological semimetal},\ }\href {https://doi.org/10.1073/pnas.1820331116}
  {\bibfield  {journal} {\bibinfo  {journal} {Proceedings of the National
  Academy of Sciences}\ }\textbf {\bibinfo {volume} {116}},\ \bibinfo {pages}
  {1084} (\bibinfo {year} {2019})}\BibitemShut {NoStop}%
\bibitem [{\citenamefont {Wu}\ \emph {et~al.}(2022)\citenamefont {Wu},
  \citenamefont {Wang}, \citenamefont {Kuang}, \citenamefont {Xu},\ and\
  \citenamefont {Kuang}}]{Wu2022}%
  \BibitemOpen
  \bibfield  {author} {\bibinfo {author} {\bibfnamefont {M.-X.}\ \bibnamefont
  {Wu}}, \bibinfo {author} {\bibfnamefont {P.}~\bibnamefont {Wang}}, \bibinfo
  {author} {\bibfnamefont {A.-L.}\ \bibnamefont {Kuang}}, \bibinfo {author}
  {\bibfnamefont {X.-H.}\ \bibnamefont {Xu}},\ and\ \bibinfo {author}
  {\bibfnamefont {M.-Q.}\ \bibnamefont {Kuang}},\ }\bibfield  {title} {\bibinfo
  {title} {{The topological nodal lines and drum-head-like surface states in
  semimetals CrSi$_2$, MoSi$_2$ and WSi$_2$}},\ }\href
  {https://doi.org/10.1016/j.physb.2022.413928} {\bibfield  {journal} {\bibinfo
   {journal} {Physica B: Condensed Matter}\ }\textbf {\bibinfo {volume}
  {639}},\ \bibinfo {pages} {413928} (\bibinfo {year} {2022})}\BibitemShut
  {NoStop}%
\bibitem [{\citenamefont {Gao}\ \emph {et~al.}(2023)\citenamefont {Gao},
  \citenamefont {Wu},\ and\ \citenamefont {An}}]{Gao2023}%
  \BibitemOpen
  \bibfield  {author} {\bibinfo {author} {\bibfnamefont {M.-J.}\ \bibnamefont
  {Gao}}, \bibinfo {author} {\bibfnamefont {H.}~\bibnamefont {Wu}},\ and\
  \bibinfo {author} {\bibfnamefont {J.-H.}\ \bibnamefont {An}},\ }\bibfield
  {title} {\bibinfo {title} {Engineering second-order nodal-line semimetals by
  breaking $\mathcal{PT}$ symmetry and periodic driving},\ }\href
  {https://doi.org/10.1103/PhysRevB.107.035128} {\bibfield  {journal} {\bibinfo
   {journal} {Phys. Rev. B}\ }\textbf {\bibinfo {volume} {107}},\ \bibinfo
  {pages} {035128} (\bibinfo {year} {2023})}\BibitemShut {NoStop}%
\bibitem [{\citenamefont {Chang}\ \emph {et~al.}(2019)\citenamefont {Chang},
  \citenamefont {Pletikosic}, \citenamefont {Kong}, \citenamefont {Bian},
  \citenamefont {Huang}, \citenamefont {Denlinger}, \citenamefont {Kushwaha},
  \citenamefont {Sinkovic}, \citenamefont {Jeng}, \citenamefont {Valla},
  \citenamefont {Xie},\ and\ \citenamefont {Cava}}]{Chang2019}%
  \BibitemOpen
  \bibfield  {author} {\bibinfo {author} {\bibfnamefont {T.-R.}\ \bibnamefont
  {Chang}}, \bibinfo {author} {\bibfnamefont {I.}~\bibnamefont {Pletikosic}},
  \bibinfo {author} {\bibfnamefont {T.}~\bibnamefont {Kong}}, \bibinfo {author}
  {\bibfnamefont {G.}~\bibnamefont {Bian}}, \bibinfo {author} {\bibfnamefont
  {A.}~\bibnamefont {Huang}}, \bibinfo {author} {\bibfnamefont
  {J.}~\bibnamefont {Denlinger}}, \bibinfo {author} {\bibfnamefont {S.~K.}\
  \bibnamefont {Kushwaha}}, \bibinfo {author} {\bibfnamefont {B.}~\bibnamefont
  {Sinkovic}}, \bibinfo {author} {\bibfnamefont {H.-T.}\ \bibnamefont {Jeng}},
  \bibinfo {author} {\bibfnamefont {T.}~\bibnamefont {Valla}}, \bibinfo
  {author} {\bibfnamefont {W.}~\bibnamefont {Xie}},\ and\ \bibinfo {author}
  {\bibfnamefont {R.~J.}\ \bibnamefont {Cava}},\ }\bibfield  {title} {\bibinfo
  {title} {{Realization of a Type-II Nodal-Line Semimetal in Mg$_3$Bi$_2$}},\
  }\href {https://doi.org/10.1002/advs.201800897} {\bibfield  {journal}
  {\bibinfo  {journal} {Advanced Science}\ }\textbf {\bibinfo {volume} {6}},\
  \bibinfo {pages} {1800897} (\bibinfo {year} {2019})}\BibitemShut {NoStop}%
\bibitem [{\citenamefont {Hu}\ \emph {et~al.}(2017)\citenamefont {Hu},
  \citenamefont {Tang}, \citenamefont {Liu}, \citenamefont {Zhu}, \citenamefont
  {Wei},\ and\ \citenamefont {Mao}}]{Hu2017}%
  \BibitemOpen
  \bibfield  {author} {\bibinfo {author} {\bibfnamefont {J.}~\bibnamefont
  {Hu}}, \bibinfo {author} {\bibfnamefont {Z.}~\bibnamefont {Tang}}, \bibinfo
  {author} {\bibfnamefont {J.}~\bibnamefont {Liu}}, \bibinfo {author}
  {\bibfnamefont {Y.}~\bibnamefont {Zhu}}, \bibinfo {author} {\bibfnamefont
  {J.}~\bibnamefont {Wei}},\ and\ \bibinfo {author} {\bibfnamefont
  {Z.}~\bibnamefont {Mao}},\ }\bibfield  {title} {\bibinfo {title} {{Nearly
  massless Dirac fermions and strong Zeeman splitting in the nodal-line
  semimetal ZrSiS probed by de Haas--van Alphen quantum oscillations}},\ }\href
  {https://doi.org/10.1103/PhysRevB.96.045127} {\bibfield  {journal} {\bibinfo
  {journal} {Phys. Rev. B}\ }\textbf {\bibinfo {volume} {96}},\ \bibinfo
  {pages} {045127} (\bibinfo {year} {2017})}\BibitemShut {NoStop}%
\bibitem [{\citenamefont {Laha}\ \emph {et~al.}(2019)\citenamefont {Laha},
  \citenamefont {Malick}, \citenamefont {Singha}, \citenamefont {Mandal},
  \citenamefont {Rambabu}, \citenamefont {Kanchana},\ and\ \citenamefont
  {Hossain}}]{Laha2019}%
  \BibitemOpen
  \bibfield  {author} {\bibinfo {author} {\bibfnamefont {A.}~\bibnamefont
  {Laha}}, \bibinfo {author} {\bibfnamefont {S.}~\bibnamefont {Malick}},
  \bibinfo {author} {\bibfnamefont {R.}~\bibnamefont {Singha}}, \bibinfo
  {author} {\bibfnamefont {P.}~\bibnamefont {Mandal}}, \bibinfo {author}
  {\bibfnamefont {P.}~\bibnamefont {Rambabu}}, \bibinfo {author} {\bibfnamefont
  {V.}~\bibnamefont {Kanchana}},\ and\ \bibinfo {author} {\bibfnamefont
  {Z.}~\bibnamefont {Hossain}},\ }\bibfield  {title} {\bibinfo {title}
  {Magnetotransport properties of the correlated topological nodal-line
  semimetal ybcdge},\ }\href {https://doi.org/10.1103/PhysRevB.99.241102}
  {\bibfield  {journal} {\bibinfo  {journal} {Phys. Rev. B}\ }\textbf {\bibinfo
  {volume} {99}},\ \bibinfo {pages} {241102} (\bibinfo {year}
  {2019})}\BibitemShut {NoStop}%
\bibitem [{\citenamefont {Emmanouilidou}\ \emph {et~al.}(2017)\citenamefont
  {Emmanouilidou}, \citenamefont {Shen}, \citenamefont {Deng}, \citenamefont
  {Chang}, \citenamefont {Shi}, \citenamefont {Kotliar}, \citenamefont {Xu},\
  and\ \citenamefont {Ni}}]{Emmanouilidou2017}%
  \BibitemOpen
  \bibfield  {author} {\bibinfo {author} {\bibfnamefont {E.}~\bibnamefont
  {Emmanouilidou}}, \bibinfo {author} {\bibfnamefont {B.}~\bibnamefont {Shen}},
  \bibinfo {author} {\bibfnamefont {X.}~\bibnamefont {Deng}}, \bibinfo {author}
  {\bibfnamefont {T.-R.}\ \bibnamefont {Chang}}, \bibinfo {author}
  {\bibfnamefont {A.}~\bibnamefont {Shi}}, \bibinfo {author} {\bibfnamefont
  {G.}~\bibnamefont {Kotliar}}, \bibinfo {author} {\bibfnamefont {S.-Y.}\
  \bibnamefont {Xu}},\ and\ \bibinfo {author} {\bibfnamefont {N.}~\bibnamefont
  {Ni}},\ }\bibfield  {title} {\bibinfo {title} {{Magnetotransport properties
  of the single-crystalline nodal-line semimetal candidates
  $\mathrm{Ca}TX(T=\text{Ag},\text{Cd};X=\text{As},\text{Ge})$}},\ }\href
  {https://doi.org/10.1103/PhysRevB.95.245113} {\bibfield  {journal} {\bibinfo
  {journal} {Phys. Rev. B}\ }\textbf {\bibinfo {volume} {95}},\ \bibinfo
  {pages} {245113} (\bibinfo {year} {2017})}\BibitemShut {NoStop}%
\bibitem [{\citenamefont {Pezzini}\ \emph {et~al.}(2017)\citenamefont
  {Pezzini}, \citenamefont {van Delft}, \citenamefont {Schoop}, \citenamefont
  {Lotsch}, \citenamefont {Carrington}, \citenamefont {Katsnelson},
  \citenamefont {Hussey},\ and\ \citenamefont {Wiedmann}}]{Pezzini2017}%
  \BibitemOpen
  \bibfield  {author} {\bibinfo {author} {\bibfnamefont {S.}~\bibnamefont
  {Pezzini}}, \bibinfo {author} {\bibfnamefont {M.~R.}\ \bibnamefont {van
  Delft}}, \bibinfo {author} {\bibfnamefont {L.~M.}\ \bibnamefont {Schoop}},
  \bibinfo {author} {\bibfnamefont {B.~V.}\ \bibnamefont {Lotsch}}, \bibinfo
  {author} {\bibfnamefont {A.}~\bibnamefont {Carrington}}, \bibinfo {author}
  {\bibfnamefont {M.~I.}\ \bibnamefont {Katsnelson}}, \bibinfo {author}
  {\bibfnamefont {N.~E.}\ \bibnamefont {Hussey}},\ and\ \bibinfo {author}
  {\bibfnamefont {S.}~\bibnamefont {Wiedmann}},\ }\bibfield  {title} {\bibinfo
  {title} {Unconventional mass enhancement around the dirac nodal loop in
  {ZrSiS}},\ }\href {https://doi.org/10.1038/nphys4306} {\bibfield  {journal}
  {\bibinfo  {journal} {Nature Physics}\ }\textbf {\bibinfo {volume} {14}},\
  \bibinfo {pages} {178} (\bibinfo {year} {2017})}\BibitemShut {NoStop}%
\bibitem [{\citenamefont {Wang}\ \emph
  {et~al.}(2022{\natexlab{b}})\citenamefont {Wang}, \citenamefont {Legg},
  \citenamefont {B\"omerich}, \citenamefont {Park}, \citenamefont {Biesenkamp},
  \citenamefont {Taskin}, \citenamefont {Braden}, \citenamefont {Rosch},\ and\
  \citenamefont {Ando}}]{Wang2022a}%
  \BibitemOpen
  \bibfield  {author} {\bibinfo {author} {\bibfnamefont {Y.}~\bibnamefont
  {Wang}}, \bibinfo {author} {\bibfnamefont {H.~F.}\ \bibnamefont {Legg}},
  \bibinfo {author} {\bibfnamefont {T.}~\bibnamefont {B\"omerich}}, \bibinfo
  {author} {\bibfnamefont {J.}~\bibnamefont {Park}}, \bibinfo {author}
  {\bibfnamefont {S.}~\bibnamefont {Biesenkamp}}, \bibinfo {author}
  {\bibfnamefont {A.~A.}\ \bibnamefont {Taskin}}, \bibinfo {author}
  {\bibfnamefont {M.}~\bibnamefont {Braden}}, \bibinfo {author} {\bibfnamefont
  {A.}~\bibnamefont {Rosch}},\ and\ \bibinfo {author} {\bibfnamefont
  {Y.}~\bibnamefont {Ando}},\ }\bibfield  {title} {\bibinfo {title} {{Gigantic
  Magnetochiral Anisotropy in the Topological Semimetal ZrTe$_5$}},\ }\href
  {https://doi.org/10.1103/PhysRevLett.128.176602} {\bibfield  {journal}
  {\bibinfo  {journal} {Phys. Rev. Lett.}\ }\textbf {\bibinfo {volume} {128}},\
  \bibinfo {pages} {176602} (\bibinfo {year} {2022}{\natexlab{b}})}\BibitemShut
  {NoStop}%
\bibitem [{\citenamefont {Guo}\ \emph {et~al.}(2019)\citenamefont {Guo},
  \citenamefont {Chen}, \citenamefont {Chen}, \citenamefont {Chen},
  \citenamefont {Zhang}, \citenamefont {Gao}, \citenamefont {Yang},
  \citenamefont {Li}, \citenamefont {Zhao}, \citenamefont {Dong},\ and\
  \citenamefont {Zheng}}]{Guo2019}%
  \BibitemOpen
  \bibfield  {author} {\bibinfo {author} {\bibfnamefont {L.}~\bibnamefont
  {Guo}}, \bibinfo {author} {\bibfnamefont {T.-W.}\ \bibnamefont {Chen}},
  \bibinfo {author} {\bibfnamefont {C.}~\bibnamefont {Chen}}, \bibinfo {author}
  {\bibfnamefont {L.}~\bibnamefont {Chen}}, \bibinfo {author} {\bibfnamefont
  {Y.}~\bibnamefont {Zhang}}, \bibinfo {author} {\bibfnamefont {G.-Y.}\
  \bibnamefont {Gao}}, \bibinfo {author} {\bibfnamefont {J.}~\bibnamefont
  {Yang}}, \bibinfo {author} {\bibfnamefont {X.-G.}\ \bibnamefont {Li}},
  \bibinfo {author} {\bibfnamefont {W.-Y.}\ \bibnamefont {Zhao}}, \bibinfo
  {author} {\bibfnamefont {S.}~\bibnamefont {Dong}},\ and\ \bibinfo {author}
  {\bibfnamefont {R.-K.}\ \bibnamefont {Zheng}},\ }\bibfield  {title} {\bibinfo
  {title} {Electronic transport evidence for topological nodal-line semimetals
  of {ZrGeSe} single crystals},\ }\href
  {https://doi.org/10.1021/acsaelm.9b00061} {\bibfield  {journal} {\bibinfo
  {journal} {{ACS} Applied Electronic Materials}\ }\textbf {\bibinfo {volume}
  {1}},\ \bibinfo {pages} {869} (\bibinfo {year} {2019})}\BibitemShut {NoStop}%
\bibitem [{\citenamefont {Hu}\ \emph {et~al.}(2016)\citenamefont {Hu},
  \citenamefont {Tang}, \citenamefont {Liu}, \citenamefont {Liu}, \citenamefont
  {Zhu}, \citenamefont {Graf}, \citenamefont {Myhro}, \citenamefont {Tran},
  \citenamefont {Lau}, \citenamefont {Wei},\ and\ \citenamefont
  {Mao}}]{Hu2016}%
  \BibitemOpen
  \bibfield  {author} {\bibinfo {author} {\bibfnamefont {J.}~\bibnamefont
  {Hu}}, \bibinfo {author} {\bibfnamefont {Z.}~\bibnamefont {Tang}}, \bibinfo
  {author} {\bibfnamefont {J.}~\bibnamefont {Liu}}, \bibinfo {author}
  {\bibfnamefont {X.}~\bibnamefont {Liu}}, \bibinfo {author} {\bibfnamefont
  {Y.}~\bibnamefont {Zhu}}, \bibinfo {author} {\bibfnamefont {D.}~\bibnamefont
  {Graf}}, \bibinfo {author} {\bibfnamefont {K.}~\bibnamefont {Myhro}},
  \bibinfo {author} {\bibfnamefont {S.}~\bibnamefont {Tran}}, \bibinfo {author}
  {\bibfnamefont {C.~N.}\ \bibnamefont {Lau}}, \bibinfo {author} {\bibfnamefont
  {J.}~\bibnamefont {Wei}},\ and\ \bibinfo {author} {\bibfnamefont
  {Z.}~\bibnamefont {Mao}},\ }\bibfield  {title} {\bibinfo {title} {{Evidence
  of Topological Nodal-Line Fermions in ZrSiSe and ZrSiTe}},\ }\href
  {https://doi.org/10.1103/PhysRevLett.117.016602} {\bibfield  {journal}
  {\bibinfo  {journal} {Phys. Rev. Lett.}\ }\textbf {\bibinfo {volume} {117}},\
  \bibinfo {pages} {016602} (\bibinfo {year} {2016})}\BibitemShut {NoStop}%
\bibitem [{\citenamefont {Stuart}\ \emph {et~al.}(2022)\citenamefont {Stuart},
  \citenamefont {Choi}, \citenamefont {Kim}, \citenamefont {Muechler},
  \citenamefont {Queiroz}, \citenamefont {Oudah}, \citenamefont {Schoop},
  \citenamefont {Bonn},\ and\ \citenamefont {Burke}}]{Stuart2022}%
  \BibitemOpen
  \bibfield  {author} {\bibinfo {author} {\bibfnamefont {B.~A.}\ \bibnamefont
  {Stuart}}, \bibinfo {author} {\bibfnamefont {S.}~\bibnamefont {Choi}},
  \bibinfo {author} {\bibfnamefont {J.}~\bibnamefont {Kim}}, \bibinfo {author}
  {\bibfnamefont {L.}~\bibnamefont {Muechler}}, \bibinfo {author}
  {\bibfnamefont {R.}~\bibnamefont {Queiroz}}, \bibinfo {author} {\bibfnamefont
  {M.}~\bibnamefont {Oudah}}, \bibinfo {author} {\bibfnamefont {L.~M.}\
  \bibnamefont {Schoop}}, \bibinfo {author} {\bibfnamefont {D.~A.}\
  \bibnamefont {Bonn}},\ and\ \bibinfo {author} {\bibfnamefont {S.~A.}\
  \bibnamefont {Burke}},\ }\bibfield  {title} {\bibinfo {title} {{Quasiparticle
  interference observation of the topologically nontrivial drumhead surface
  state in ZrSiTe}},\ }\href {https://doi.org/10.1103/PhysRevB.105.L121111}
  {\bibfield  {journal} {\bibinfo  {journal} {Phys. Rev. B}\ }\textbf {\bibinfo
  {volume} {105}},\ \bibinfo {pages} {L121111} (\bibinfo {year}
  {2022})}\BibitemShut {NoStop}%
\bibitem [{\citenamefont {Deng}\ \emph {et~al.}(2019)\citenamefont {Deng},
  \citenamefont {Lu}, \citenamefont {Li}, \citenamefont {Huang}, \citenamefont
  {Yan}, \citenamefont {Ma},\ and\ \citenamefont {Liu}}]{Deng2019}%
  \BibitemOpen
  \bibfield  {author} {\bibinfo {author} {\bibfnamefont {W.}~\bibnamefont
  {Deng}}, \bibinfo {author} {\bibfnamefont {J.}~\bibnamefont {Lu}}, \bibinfo
  {author} {\bibfnamefont {F.}~\bibnamefont {Li}}, \bibinfo {author}
  {\bibfnamefont {X.}~\bibnamefont {Huang}}, \bibinfo {author} {\bibfnamefont
  {M.}~\bibnamefont {Yan}}, \bibinfo {author} {\bibfnamefont {J.}~\bibnamefont
  {Ma}},\ and\ \bibinfo {author} {\bibfnamefont {Z.}~\bibnamefont {Liu}},\
  }\bibfield  {title} {\bibinfo {title} {Nodal rings and drumhead surface
  states in phononic crystals},\ }\href
  {https://doi.org/10.1038/s41467-019-09820-8} {\bibfield  {journal} {\bibinfo
  {journal} {Nature Communications}\ }\textbf {\bibinfo {volume} {10}},\
  \bibinfo {pages} {1769} (\bibinfo {year} {2019})}\BibitemShut {NoStop}%
\bibitem [{\citenamefont {Park}\ \emph {et~al.}(2022)\citenamefont {Park},
  \citenamefont {Gao}, \citenamefont {Zhang},\ and\ \citenamefont
  {Oh}}]{Park2022}%
  \BibitemOpen
  \bibfield  {author} {\bibinfo {author} {\bibfnamefont {H.}~\bibnamefont
  {Park}}, \bibinfo {author} {\bibfnamefont {W.}~\bibnamefont {Gao}}, \bibinfo
  {author} {\bibfnamefont {X.}~\bibnamefont {Zhang}},\ and\ \bibinfo {author}
  {\bibfnamefont {S.~S.}\ \bibnamefont {Oh}},\ }\bibfield  {title} {\bibinfo
  {title} {Nodal lines in momentum space: topological invariants and recent
  realizations in photonic and other systems},\ }\href
  {https://doi.org/10.1515/nanoph-2021-0692} {\bibfield  {journal} {\bibinfo
  {journal} {Nanophotonics}\ }\textbf {\bibinfo {volume} {11}},\ \bibinfo
  {pages} {2779} (\bibinfo {year} {2022})}\BibitemShut {NoStop}%
\bibitem [{\citenamefont {Nguyen}\ \emph {et~al.}(2023)\citenamefont {Nguyen},
  \citenamefont {Devescovi}, \citenamefont {Nguyen}, \citenamefont {Nguyen},\
  and\ \citenamefont {Bercioux}}]{Bercioux2023}%
  \BibitemOpen
  \bibfield  {author} {\bibinfo {author} {\bibfnamefont {D.-H.-M.}\
  \bibnamefont {Nguyen}}, \bibinfo {author} {\bibfnamefont {C.}~\bibnamefont
  {Devescovi}}, \bibinfo {author} {\bibfnamefont {D.~X.}\ \bibnamefont
  {Nguyen}}, \bibinfo {author} {\bibfnamefont {H.~S.}\ \bibnamefont {Nguyen}},\
  and\ \bibinfo {author} {\bibfnamefont {D.}~\bibnamefont {Bercioux}},\
  }\bibfield  {title} {\bibinfo {title} {Fermi arc reconstruction in synthetic
  photonic lattice},\ }\href {https://doi.org/10.1103/PhysRevLett.131.053602}
  {\bibfield  {journal} {\bibinfo  {journal} {Phys. Rev. Lett.}\ }\textbf
  {\bibinfo {volume} {131}},\ \bibinfo {pages} {053602} (\bibinfo {year}
  {2023})}\BibitemShut {NoStop}%
\bibitem [{\citenamefont {Burkov}\ \emph {et~al.}(2011)\citenamefont {Burkov},
  \citenamefont {Hook},\ and\ \citenamefont {Balents}}]{Burkov2011}%
  \BibitemOpen
  \bibfield  {author} {\bibinfo {author} {\bibfnamefont {A.~A.}\ \bibnamefont
  {Burkov}}, \bibinfo {author} {\bibfnamefont {M.~D.}\ \bibnamefont {Hook}},\
  and\ \bibinfo {author} {\bibfnamefont {L.}~\bibnamefont {Balents}},\
  }\bibfield  {title} {\bibinfo {title} {Topological nodal semimetals},\ }\href
  {https://doi.org/10.1103/PhysRevB.84.235126} {\bibfield  {journal} {\bibinfo
  {journal} {Phys. Rev. B}\ }\textbf {\bibinfo {volume} {84}},\ \bibinfo
  {pages} {235126} (\bibinfo {year} {2011})}\BibitemShut {NoStop}%
\bibitem [{\citenamefont {Bianchi}\ \emph {et~al.}(2010)\citenamefont
  {Bianchi}, \citenamefont {Guan}, \citenamefont {Bao}, \citenamefont {Mi},
  \citenamefont {Iversen}, \citenamefont {King},\ and\ \citenamefont
  {Hofmann}}]{Bianchi2010}%
  \BibitemOpen
  \bibfield  {author} {\bibinfo {author} {\bibfnamefont {M.}~\bibnamefont
  {Bianchi}}, \bibinfo {author} {\bibfnamefont {D.}~\bibnamefont {Guan}},
  \bibinfo {author} {\bibfnamefont {S.}~\bibnamefont {Bao}}, \bibinfo {author}
  {\bibfnamefont {J.}~\bibnamefont {Mi}}, \bibinfo {author} {\bibfnamefont
  {B.~B.}\ \bibnamefont {Iversen}}, \bibinfo {author} {\bibfnamefont {P.~D.}\
  \bibnamefont {King}},\ and\ \bibinfo {author} {\bibfnamefont
  {P.}~\bibnamefont {Hofmann}},\ }\bibfield  {title} {\bibinfo {title}
  {{Coexistence of the topological state and a two-dimensional electron gas on
  the surface of Bi$_2$Se$_3$}},\ }\href {https://doi.org/10.1038/ncomms1131}
  {\bibfield  {journal} {\bibinfo  {journal} {Nature Communications}\ }\textbf
  {\bibinfo {volume} {1}},\ \bibinfo {pages} {128} (\bibinfo {year}
  {2010})}\BibitemShut {NoStop}%
\bibitem [{\citenamefont {Chen}\ \emph {et~al.}(2012)\citenamefont {Chen},
  \citenamefont {He}, \citenamefont {Weng}, \citenamefont {Zhang},
  \citenamefont {Zhao}, \citenamefont {Liu}, \citenamefont {Jia}, \citenamefont
  {Mou}, \citenamefont {Liu}, \citenamefont {He}, \citenamefont {Peng},
  \citenamefont {Feng}, \citenamefont {Xie}, \citenamefont {Liu}, \citenamefont
  {Dong}, \citenamefont {Zhang}, \citenamefont {Wang}, \citenamefont {Peng},
  \citenamefont {Wang}, \citenamefont {Zhang}, \citenamefont {Yang},
  \citenamefont {Chen}, \citenamefont {Xu}, \citenamefont {Dai}, \citenamefont
  {Fang},\ and\ \citenamefont {Zhou}}]{Chen2012}%
  \BibitemOpen
  \bibfield  {author} {\bibinfo {author} {\bibfnamefont {C.}~\bibnamefont
  {Chen}}, \bibinfo {author} {\bibfnamefont {S.}~\bibnamefont {He}}, \bibinfo
  {author} {\bibfnamefont {H.}~\bibnamefont {Weng}}, \bibinfo {author}
  {\bibfnamefont {W.}~\bibnamefont {Zhang}}, \bibinfo {author} {\bibfnamefont
  {L.}~\bibnamefont {Zhao}}, \bibinfo {author} {\bibfnamefont {H.}~\bibnamefont
  {Liu}}, \bibinfo {author} {\bibfnamefont {X.}~\bibnamefont {Jia}}, \bibinfo
  {author} {\bibfnamefont {D.}~\bibnamefont {Mou}}, \bibinfo {author}
  {\bibfnamefont {S.}~\bibnamefont {Liu}}, \bibinfo {author} {\bibfnamefont
  {J.}~\bibnamefont {He}}, \bibinfo {author} {\bibfnamefont {Y.}~\bibnamefont
  {Peng}}, \bibinfo {author} {\bibfnamefont {Y.}~\bibnamefont {Feng}}, \bibinfo
  {author} {\bibfnamefont {Z.}~\bibnamefont {Xie}}, \bibinfo {author}
  {\bibfnamefont {G.}~\bibnamefont {Liu}}, \bibinfo {author} {\bibfnamefont
  {X.}~\bibnamefont {Dong}}, \bibinfo {author} {\bibfnamefont {J.}~\bibnamefont
  {Zhang}}, \bibinfo {author} {\bibfnamefont {X.}~\bibnamefont {Wang}},
  \bibinfo {author} {\bibfnamefont {Q.}~\bibnamefont {Peng}}, \bibinfo {author}
  {\bibfnamefont {Z.}~\bibnamefont {Wang}}, \bibinfo {author} {\bibfnamefont
  {S.}~\bibnamefont {Zhang}}, \bibinfo {author} {\bibfnamefont
  {F.}~\bibnamefont {Yang}}, \bibinfo {author} {\bibfnamefont {C.}~\bibnamefont
  {Chen}}, \bibinfo {author} {\bibfnamefont {Z.}~\bibnamefont {Xu}}, \bibinfo
  {author} {\bibfnamefont {X.}~\bibnamefont {Dai}}, \bibinfo {author}
  {\bibfnamefont {Z.}~\bibnamefont {Fang}},\ and\ \bibinfo {author}
  {\bibfnamefont {X.~J.}\ \bibnamefont {Zhou}},\ }\bibfield  {title} {\bibinfo
  {title} {Robustness of topological order and formation of quantum well states
  in topological insulators exposed to ambient environment},\ }\href
  {https://doi.org/10.1073/pnas.1115555109} {\bibfield  {journal} {\bibinfo
  {journal} {Proceedings of the National Academy of Sciences}\ }\textbf
  {\bibinfo {volume} {109}},\ \bibinfo {pages} {3694} (\bibinfo {year}
  {2012})}\BibitemShut {NoStop}%
\bibitem [{\citenamefont {Alspaugh}\ \emph {et~al.}(2020)\citenamefont
  {Alspaugh}, \citenamefont {Sheehy}, \citenamefont {Goerbig},\ and\
  \citenamefont {Simon}}]{Alspaugh2020}%
  \BibitemOpen
  \bibfield  {author} {\bibinfo {author} {\bibfnamefont {D.~J.}\ \bibnamefont
  {Alspaugh}}, \bibinfo {author} {\bibfnamefont {D.~E.}\ \bibnamefont
  {Sheehy}}, \bibinfo {author} {\bibfnamefont {M.~O.}\ \bibnamefont
  {Goerbig}},\ and\ \bibinfo {author} {\bibfnamefont {P.}~\bibnamefont
  {Simon}},\ }\bibfield  {title} {\bibinfo {title} {{Volkov-Pankratov states in
  topological superconductors}},\ }\href
  {https://doi.org/10.1103/PhysRevResearch.2.023146} {\bibfield  {journal}
  {\bibinfo  {journal} {Phys. Rev. Research}\ }\textbf {\bibinfo {volume}
  {2}},\ \bibinfo {pages} {023146} (\bibinfo {year} {2020})}\BibitemShut
  {NoStop}%
\bibitem [{\citenamefont {Volkov}\ and\ \citenamefont
  {Pankratov}(1985)}]{Volkov1985}%
  \BibitemOpen
  \bibfield  {author} {\bibinfo {author} {\bibfnamefont {B.~A.}\ \bibnamefont
  {Volkov}}\ and\ \bibinfo {author} {\bibfnamefont {O.~A.}\ \bibnamefont
  {Pankratov}},\ }\bibfield  {title} {\bibinfo {title} {Two-dimensional
  massless electrons in an inverted contact},\ }\href@noop {} {\bibfield
  {journal} {\bibinfo  {journal} {JETP Letters}\ }\textbf {\bibinfo {volume}
  {42}},\ \bibinfo {pages} {145} (\bibinfo {year} {1985})}\BibitemShut
  {NoStop}%
\bibitem [{\citenamefont {Volkov}\ and\ \citenamefont
  {Pankratov}(1986)}]{Volkov1986}%
  \BibitemOpen
  \bibfield  {author} {\bibinfo {author} {\bibfnamefont {B.~A.}\ \bibnamefont
  {Volkov}}\ and\ \bibinfo {author} {\bibfnamefont {O.~A.}\ \bibnamefont
  {Pankratov}},\ }\bibfield  {title} {\bibinfo {title} {Inverted contact in
  semiconductors--a new inhomogeneous structure with a two-dimensional gas of
  zero-mass electrons},\ }\href
  {https://doi.org/10.1070/PU1986v029n06ABEH003428} {\bibfield  {journal}
  {\bibinfo  {journal} {Soviet Physics Uspekhi}\ }\textbf {\bibinfo {volume}
  {29}},\ \bibinfo {pages} {579} (\bibinfo {year} {1986})}\BibitemShut
  {NoStop}%
\bibitem [{\citenamefont {Lu}\ and\ \citenamefont {Goerbig}(2019)}]{Lu2019}%
  \BibitemOpen
  \bibfield  {author} {\bibinfo {author} {\bibfnamefont {X.}~\bibnamefont
  {Lu}}\ and\ \bibinfo {author} {\bibfnamefont {M.~O.}\ \bibnamefont
  {Goerbig}},\ }\bibfield  {title} {\bibinfo {title} {{Magneto-optical
  signatures of Volkov-Pankratov states in topological insulators}},\ }\href
  {https://doi.org/10.1209/0295-5075/126/67004} {\bibfield  {journal} {\bibinfo
   {journal} {Europhysics Letters}\ }\textbf {\bibinfo {volume} {126}},\
  \bibinfo {pages} {67004} (\bibinfo {year} {2019})}\BibitemShut {NoStop}%
\bibitem [{\citenamefont {Inhofer}\ \emph {et~al.}(2017)\citenamefont
  {Inhofer}, \citenamefont {Tchoumakov}, \citenamefont {Assaf}, \citenamefont
  {F\`eve}, \citenamefont {Berroir}, \citenamefont {Jouffrey}, \citenamefont
  {Carpentier}, \citenamefont {Goerbig}, \citenamefont
  {Pla\ifmmode~\mbox{\c{c}}\else \c{c}\fi{}ais}, \citenamefont {Bendias},
  \citenamefont {Mahler}, \citenamefont {Bocquillon}, \citenamefont
  {Schlereth}, \citenamefont {Br\"une}, \citenamefont {Buhmann},\ and\
  \citenamefont {Molenkamp}}]{Inhofer2017}%
  \BibitemOpen
  \bibfield  {author} {\bibinfo {author} {\bibfnamefont {A.}~\bibnamefont
  {Inhofer}}, \bibinfo {author} {\bibfnamefont {S.}~\bibnamefont {Tchoumakov}},
  \bibinfo {author} {\bibfnamefont {B.~A.}\ \bibnamefont {Assaf}}, \bibinfo
  {author} {\bibfnamefont {G.}~\bibnamefont {F\`eve}}, \bibinfo {author}
  {\bibfnamefont {J.~M.}\ \bibnamefont {Berroir}}, \bibinfo {author}
  {\bibfnamefont {V.}~\bibnamefont {Jouffrey}}, \bibinfo {author}
  {\bibfnamefont {D.}~\bibnamefont {Carpentier}}, \bibinfo {author}
  {\bibfnamefont {M.~O.}\ \bibnamefont {Goerbig}}, \bibinfo {author}
  {\bibfnamefont {B.}~\bibnamefont {Pla\ifmmode~\mbox{\c{c}}\else
  \c{c}\fi{}ais}}, \bibinfo {author} {\bibfnamefont {K.}~\bibnamefont
  {Bendias}}, \bibinfo {author} {\bibfnamefont {D.~M.}\ \bibnamefont {Mahler}},
  \bibinfo {author} {\bibfnamefont {E.}~\bibnamefont {Bocquillon}}, \bibinfo
  {author} {\bibfnamefont {R.}~\bibnamefont {Schlereth}}, \bibinfo {author}
  {\bibfnamefont {C.}~\bibnamefont {Br\"une}}, \bibinfo {author} {\bibfnamefont
  {H.}~\bibnamefont {Buhmann}},\ and\ \bibinfo {author} {\bibfnamefont {L.~W.}\
  \bibnamefont {Molenkamp}},\ }\bibfield  {title} {\bibinfo {title}
  {{Observation of Volkov-Pankratov states in topological HgTe heterojunctions
  using high-frequency compressibility}},\ }\href
  {https://doi.org/10.1103/PhysRevB.96.195104} {\bibfield  {journal} {\bibinfo
  {journal} {Phys. Rev. B}\ }\textbf {\bibinfo {volume} {96}},\ \bibinfo
  {pages} {195104} (\bibinfo {year} {2017})}\BibitemShut {NoStop}%
\bibitem [{\citenamefont {Bermejo-Ortiz}\ \emph {et~al.}(2023)\citenamefont
  {Bermejo-Ortiz}, \citenamefont {Krizman}, \citenamefont {Jakiela},
  \citenamefont {Khosravizadeh}, \citenamefont {Hajlaoui}, \citenamefont
  {Bauer}, \citenamefont {Springholz}, \citenamefont {de~Vaulchier},\ and\
  \citenamefont {Guldner}}]{Bermejo2022}%
  \BibitemOpen
  \bibfield  {author} {\bibinfo {author} {\bibfnamefont {J.}~\bibnamefont
  {Bermejo-Ortiz}}, \bibinfo {author} {\bibfnamefont {G.}~\bibnamefont
  {Krizman}}, \bibinfo {author} {\bibfnamefont {R.}~\bibnamefont {Jakiela}},
  \bibinfo {author} {\bibfnamefont {Z.}~\bibnamefont {Khosravizadeh}}, \bibinfo
  {author} {\bibfnamefont {M.}~\bibnamefont {Hajlaoui}}, \bibinfo {author}
  {\bibfnamefont {G.}~\bibnamefont {Bauer}}, \bibinfo {author} {\bibfnamefont
  {G.}~\bibnamefont {Springholz}}, \bibinfo {author} {\bibfnamefont {L.-A.}\
  \bibnamefont {de~Vaulchier}},\ and\ \bibinfo {author} {\bibfnamefont
  {Y.}~\bibnamefont {Guldner}},\ }\bibfield  {title} {\bibinfo {title}
  {{Observation of Weyl and Dirac fermions at smooth topological
  Volkov-Pankratov heterojunctions}},\ }\href
  {https://doi.org/10.1103/PhysRevB.107.075129} {\bibfield  {journal} {\bibinfo
   {journal} {Phys. Rev. B}\ }\textbf {\bibinfo {volume} {107}},\ \bibinfo
  {pages} {075129} (\bibinfo {year} {2023})}\BibitemShut {NoStop}%
\bibitem [{\citenamefont {Nilforoushan}\ \emph {et~al.}(2021)\citenamefont
  {Nilforoushan}, \citenamefont {Casula}, \citenamefont {Amaricci},
  \citenamefont {Caputo}, \citenamefont {Caillaux}, \citenamefont {Khalil},
  \citenamefont {Papalazarou}, \citenamefont {Simon}, \citenamefont {Perfetti},
  \citenamefont {Vobornik}, \citenamefont {Das}, \citenamefont {Fujii},
  \citenamefont {Barinov}, \citenamefont {Santos-Cottin}, \citenamefont
  {Klein}, \citenamefont {Fabrizio}, \citenamefont {Gauzzi},\ and\
  \citenamefont {Marsi}}]{Nilforoushan2021}%
  \BibitemOpen
  \bibfield  {author} {\bibinfo {author} {\bibfnamefont {N.}~\bibnamefont
  {Nilforoushan}}, \bibinfo {author} {\bibfnamefont {M.}~\bibnamefont
  {Casula}}, \bibinfo {author} {\bibfnamefont {A.}~\bibnamefont {Amaricci}},
  \bibinfo {author} {\bibfnamefont {M.}~\bibnamefont {Caputo}}, \bibinfo
  {author} {\bibfnamefont {J.}~\bibnamefont {Caillaux}}, \bibinfo {author}
  {\bibfnamefont {L.}~\bibnamefont {Khalil}}, \bibinfo {author} {\bibfnamefont
  {E.}~\bibnamefont {Papalazarou}}, \bibinfo {author} {\bibfnamefont
  {P.}~\bibnamefont {Simon}}, \bibinfo {author} {\bibfnamefont
  {L.}~\bibnamefont {Perfetti}}, \bibinfo {author} {\bibfnamefont
  {I.}~\bibnamefont {Vobornik}}, \bibinfo {author} {\bibfnamefont {P.~K.}\
  \bibnamefont {Das}}, \bibinfo {author} {\bibfnamefont {J.}~\bibnamefont
  {Fujii}}, \bibinfo {author} {\bibfnamefont {A.}~\bibnamefont {Barinov}},
  \bibinfo {author} {\bibfnamefont {D.}~\bibnamefont {Santos-Cottin}}, \bibinfo
  {author} {\bibfnamefont {Y.}~\bibnamefont {Klein}}, \bibinfo {author}
  {\bibfnamefont {M.}~\bibnamefont {Fabrizio}}, \bibinfo {author}
  {\bibfnamefont {A.}~\bibnamefont {Gauzzi}},\ and\ \bibinfo {author}
  {\bibfnamefont {M.}~\bibnamefont {Marsi}},\ }\bibfield  {title} {\bibinfo
  {title} {{Moving Dirac nodes by chemical substitution}},\ }\href
  {https://doi.org/10.1073/pnas.2108617118} {\bibfield  {journal} {\bibinfo
  {journal} {Proceedings of the National Academy of Sciences}\ }\textbf
  {\bibinfo {volume} {118}},\ \bibinfo {pages} {e2108617118} (\bibinfo {year}
  {2021})}\BibitemShut {NoStop}%
\bibitem [{\citenamefont {Hao}\ \emph {et~al.}(2013)\citenamefont {Hao},
  \citenamefont {Richter}, \citenamefont {Erhai}, \citenamefont {Bonevich},
  \citenamefont {Kimes}, \citenamefont {Hyuk-Jae}, \citenamefont {Hui},
  \citenamefont {Haitao}, \citenamefont {Abbas}, \citenamefont {Oleg},
  \citenamefont {Maslar}, \citenamefont {Ioannou},\ and\ \citenamefont
  {Qiliang}}]{zhu2013}%
  \BibitemOpen
  \bibfield  {author} {\bibinfo {author} {\bibfnamefont {Z.}~\bibnamefont
  {Hao}}, \bibinfo {author} {\bibfnamefont {C.~A.}\ \bibnamefont {Richter}},
  \bibinfo {author} {\bibfnamefont {Z.}~\bibnamefont {Erhai}}, \bibinfo
  {author} {\bibfnamefont {J.~E.}\ \bibnamefont {Bonevich}}, \bibinfo {author}
  {\bibfnamefont {W.~A.}\ \bibnamefont {Kimes}}, \bibinfo {author}
  {\bibfnamefont {J.}~\bibnamefont {Hyuk-Jae}}, \bibinfo {author}
  {\bibfnamefont {Y.}~\bibnamefont {Hui}}, \bibinfo {author} {\bibfnamefont
  {L.}~\bibnamefont {Haitao}}, \bibinfo {author} {\bibfnamefont
  {A.}~\bibnamefont {Abbas}}, \bibinfo {author} {\bibfnamefont
  {K.}~\bibnamefont {Oleg}}, \bibinfo {author} {\bibfnamefont {J.~E.}\
  \bibnamefont {Maslar}}, \bibinfo {author} {\bibfnamefont {D.~E.}\
  \bibnamefont {Ioannou}},\ and\ \bibinfo {author} {\bibfnamefont
  {L.}~\bibnamefont {Qiliang}},\ }\bibfield  {title} {\bibinfo {title}
  {{Topological Insulator Bi$_2$Se$_3$ Nanowire High Performance Field-Effect
  Transistors}},\ }\href {https://doi.org/10.1038/srep01757} {\bibfield
  {journal} {\bibinfo  {journal} {Scientific Reports}\ }\textbf {\bibinfo
  {volume} {3}},\ \bibinfo {pages} {1757} (\bibinfo {year} {2013})}\BibitemShut
  {NoStop}%
\bibitem [{\citenamefont {Gorbar}\ \emph {et~al.}(2016)\citenamefont {Gorbar},
  \citenamefont {Miransky}, \citenamefont {Shovkovy},\ and\ \citenamefont
  {Sukhachov}}]{Gorbar2016}%
  \BibitemOpen
  \bibfield  {author} {\bibinfo {author} {\bibfnamefont {E.~V.}\ \bibnamefont
  {Gorbar}}, \bibinfo {author} {\bibfnamefont {V.~A.}\ \bibnamefont
  {Miransky}}, \bibinfo {author} {\bibfnamefont {I.~A.}\ \bibnamefont
  {Shovkovy}},\ and\ \bibinfo {author} {\bibfnamefont {P.~O.}\ \bibnamefont
  {Sukhachov}},\ }\bibfield  {title} {\bibinfo {title} {{Origin of dissipative
  Fermi arc transport in Weyl semimetals}},\ }\href
  {https://doi.org/10.1103/PhysRevB.93.235127} {\bibfield  {journal} {\bibinfo
  {journal} {Phys. Rev. B}\ }\textbf {\bibinfo {volume} {93}},\ \bibinfo
  {pages} {235127} (\bibinfo {year} {2016})}\BibitemShut {NoStop}%
\bibitem [{\citenamefont {Breitkreiz}\ and\ \citenamefont
  {Brouwer}(2019)}]{Breitkreiz2019}%
  \BibitemOpen
  \bibfield  {author} {\bibinfo {author} {\bibfnamefont {M.}~\bibnamefont
  {Breitkreiz}}\ and\ \bibinfo {author} {\bibfnamefont {P.~W.}\ \bibnamefont
  {Brouwer}},\ }\bibfield  {title} {\bibinfo {title} {{Large Contribution of
  Fermi Arcs to the Conductivity of Topological Metals}},\ }\href
  {https://doi.org/10.1103/PhysRevLett.123.066804} {\bibfield  {journal}
  {\bibinfo  {journal} {Phys. Rev. Lett.}\ }\textbf {\bibinfo {volume} {123}},\
  \bibinfo {pages} {066804} (\bibinfo {year} {2019})}\BibitemShut {NoStop}%
\bibitem [{\citenamefont {De~Martino}\ \emph {et~al.}(2021)\citenamefont
  {De~Martino}, \citenamefont {Dorn}, \citenamefont {Buccheri},\ and\
  \citenamefont {Egger}}]{DeMartino2021}%
  \BibitemOpen
  \bibfield  {author} {\bibinfo {author} {\bibfnamefont {A.}~\bibnamefont
  {De~Martino}}, \bibinfo {author} {\bibfnamefont {K.}~\bibnamefont {Dorn}},
  \bibinfo {author} {\bibfnamefont {F.}~\bibnamefont {Buccheri}},\ and\
  \bibinfo {author} {\bibfnamefont {R.}~\bibnamefont {Egger}},\ }\bibfield
  {title} {\bibinfo {title} {{Phonon-induced magnetoresistivity of Weyl
  semimetal nanowires}},\ }\href {https://doi.org/10.1103/PhysRevB.104.155425}
  {\bibfield  {journal} {\bibinfo  {journal} {Phys. Rev. B}\ }\textbf {\bibinfo
  {volume} {104}},\ \bibinfo {pages} {155425} (\bibinfo {year}
  {2021})}\BibitemShut {NoStop}%
\bibitem [{\citenamefont {Buccheri}\ \emph
  {et~al.}(2022{\natexlab{a}})\citenamefont {Buccheri}, \citenamefont
  {De~Martino}, \citenamefont {Pereira}, \citenamefont {Brouwer},\ and\
  \citenamefont {Egger}}]{Buccheri2022a}%
  \BibitemOpen
  \bibfield  {author} {\bibinfo {author} {\bibfnamefont {F.}~\bibnamefont
  {Buccheri}}, \bibinfo {author} {\bibfnamefont {A.}~\bibnamefont
  {De~Martino}}, \bibinfo {author} {\bibfnamefont {R.~G.}\ \bibnamefont
  {Pereira}}, \bibinfo {author} {\bibfnamefont {P.~W.}\ \bibnamefont
  {Brouwer}},\ and\ \bibinfo {author} {\bibfnamefont {R.}~\bibnamefont
  {Egger}},\ }\bibfield  {title} {\bibinfo {title} {{Phonon-limited transport
  and Fermi arc lifetime in Weyl semimetals}},\ }\href
  {https://doi.org/10.1103/PhysRevB.105.085410} {\bibfield  {journal} {\bibinfo
   {journal} {Phys. Rev. B}\ }\textbf {\bibinfo {volume} {105}},\ \bibinfo
  {pages} {085410} (\bibinfo {year} {2022}{\natexlab{a}})}\BibitemShut
  {NoStop}%
\bibitem [{\citenamefont {Chen}\ \emph {et~al.}(2018)\citenamefont {Chen},
  \citenamefont {Luo}, \citenamefont {Li},\ and\ \citenamefont
  {Zilberberg}}]{Chen2018}%
  \BibitemOpen
  \bibfield  {author} {\bibinfo {author} {\bibfnamefont {W.}~\bibnamefont
  {Chen}}, \bibinfo {author} {\bibfnamefont {K.}~\bibnamefont {Luo}}, \bibinfo
  {author} {\bibfnamefont {L.}~\bibnamefont {Li}},\ and\ \bibinfo {author}
  {\bibfnamefont {O.}~\bibnamefont {Zilberberg}},\ }\bibfield  {title}
  {\bibinfo {title} {{Proposal for Detecting Nodal-Line Semimetal Surface
  States with Resonant Spin-Flipped Reflection}},\ }\href
  {https://doi.org/10.1103/PhysRevLett.121.166802} {\bibfield  {journal}
  {\bibinfo  {journal} {Phys. Rev. Lett.}\ }\textbf {\bibinfo {volume} {121}},\
  \bibinfo {pages} {166802} (\bibinfo {year} {2018})}\BibitemShut {NoStop}%
\bibitem [{\citenamefont {Goerbig}(2023)}]{Goerbig2023}%
  \BibitemOpen
  \bibfield  {author} {\bibinfo {author} {\bibfnamefont {M.~O.}\ \bibnamefont
  {Goerbig}},\ }\href@noop {} {\bibinfo {title} {{Topological interface states
  -- a possible path towards a Landau-level laser in the THz regime}}}
  (\bibinfo {year} {2023}),\ \Eprint {https://arxiv.org/abs/2307.05116}
  {arXiv:2307.05116 [cond-mat.mes-hall]} \BibitemShut {NoStop}%
\bibitem [{\citenamefont {van~den Berg}\ \emph
  {et~al.}(2020{\natexlab{a}})\citenamefont {van~den Berg}, \citenamefont
  {De~Martino}, \citenamefont {Calvo},\ and\ \citenamefont
  {Bercioux}}]{vandenBerg2020}%
  \BibitemOpen
  \bibfield  {author} {\bibinfo {author} {\bibfnamefont {T.~L.}\ \bibnamefont
  {van~den Berg}}, \bibinfo {author} {\bibfnamefont {A.}~\bibnamefont
  {De~Martino}}, \bibinfo {author} {\bibfnamefont {M.~R.}\ \bibnamefont
  {Calvo}},\ and\ \bibinfo {author} {\bibfnamefont {D.}~\bibnamefont
  {Bercioux}},\ }\bibfield  {title} {\bibinfo {title} {{Volkov-Pankratov states
  in topological graphene nanoribbons}},\ }\href
  {https://doi.org/10.1103/PhysRevResearch.2.023373} {\bibfield  {journal}
  {\bibinfo  {journal} {Phys. Rev. Research}\ }\textbf {\bibinfo {volume}
  {2}},\ \bibinfo {pages} {023373} (\bibinfo {year}
  {2020}{\natexlab{a}})}\BibitemShut {NoStop}%
\bibitem [{\citenamefont {Kawakami}\ \emph {et~al.}(2023)\citenamefont
  {Kawakami}, \citenamefont {Tamaki},\ and\ \citenamefont
  {Koshino}}]{Kawakami2023}%
  \BibitemOpen
  \bibfield  {author} {\bibinfo {author} {\bibfnamefont {T.}~\bibnamefont
  {Kawakami}}, \bibinfo {author} {\bibfnamefont {G.}~\bibnamefont {Tamaki}},\
  and\ \bibinfo {author} {\bibfnamefont {M.}~\bibnamefont {Koshino}},\
  }\bibfield  {title} {\bibinfo {title} {Topological domain walls in graphene
  nanoribbons with carrier doping},\ }\href
  {https://doi.org/10.1103/PhysRevB.108.045401} {\bibfield  {journal} {\bibinfo
   {journal} {Phys. Rev. B}\ }\textbf {\bibinfo {volume} {108}},\ \bibinfo
  {pages} {045401} (\bibinfo {year} {2023})}\BibitemShut {NoStop}%
\bibitem [{\citenamefont {Tchoumakov}\ \emph {et~al.}(2017)\citenamefont
  {Tchoumakov}, \citenamefont {Jouffrey}, \citenamefont {Inhofer},
  \citenamefont {Bocquillon}, \citenamefont {Pla\ifmmode~\mbox{\c{c}}\else
  \c{c}\fi{}ais}, \citenamefont {Carpentier},\ and\ \citenamefont
  {Goerbig}}]{Tchoumakov2017}%
  \BibitemOpen
  \bibfield  {author} {\bibinfo {author} {\bibfnamefont {S.}~\bibnamefont
  {Tchoumakov}}, \bibinfo {author} {\bibfnamefont {V.}~\bibnamefont
  {Jouffrey}}, \bibinfo {author} {\bibfnamefont {A.}~\bibnamefont {Inhofer}},
  \bibinfo {author} {\bibfnamefont {E.}~\bibnamefont {Bocquillon}}, \bibinfo
  {author} {\bibfnamefont {B.}~\bibnamefont {Pla\ifmmode~\mbox{\c{c}}\else
  \c{c}\fi{}ais}}, \bibinfo {author} {\bibfnamefont {D.}~\bibnamefont
  {Carpentier}},\ and\ \bibinfo {author} {\bibfnamefont {M.~O.}\ \bibnamefont
  {Goerbig}},\ }\bibfield  {title} {\bibinfo {title} {{Volkov-Pankratov} states
  in topological heterojunctions},\ }\href
  {https://doi.org/10.1103/PhysRevB.96.201302} {\bibfield  {journal} {\bibinfo
  {journal} {Phys. Rev. B}\ }\textbf {\bibinfo {volume} {96}},\ \bibinfo
  {pages} {201302} (\bibinfo {year} {2017})}\BibitemShut {NoStop}%
\bibitem [{\citenamefont {van~den Berg}\ \emph
  {et~al.}(2020{\natexlab{b}})\citenamefont {van~den Berg}, \citenamefont
  {Calvo},\ and\ \citenamefont {Bercioux}}]{Bercioux2020}%
  \BibitemOpen
  \bibfield  {author} {\bibinfo {author} {\bibfnamefont {T.~L.}\ \bibnamefont
  {van~den Berg}}, \bibinfo {author} {\bibfnamefont {M.~R.}\ \bibnamefont
  {Calvo}},\ and\ \bibinfo {author} {\bibfnamefont {D.}~\bibnamefont
  {Bercioux}},\ }\bibfield  {title} {\bibinfo {title} {Living on the edge:
  Topology, electrostatics, and disorder},\ }\href
  {https://doi.org/10.1103/PhysRevResearch.2.013171} {\bibfield  {journal}
  {\bibinfo  {journal} {Phys. Rev. Res.}\ }\textbf {\bibinfo {volume} {2}},\
  \bibinfo {pages} {013171} (\bibinfo {year} {2020}{\natexlab{b}})}\BibitemShut
  {NoStop}%
\bibitem [{\citenamefont {Mukherjee}\ \emph {et~al.}(2019)\citenamefont
  {Mukherjee}, \citenamefont {Carpentier},\ and\ \citenamefont
  {Goerbig}}]{Mukherjee2019}%
  \BibitemOpen
  \bibfield  {author} {\bibinfo {author} {\bibfnamefont {D.~K.}\ \bibnamefont
  {Mukherjee}}, \bibinfo {author} {\bibfnamefont {D.}~\bibnamefont
  {Carpentier}},\ and\ \bibinfo {author} {\bibfnamefont {M.~O.}\ \bibnamefont
  {Goerbig}},\ }\bibfield  {title} {\bibinfo {title} {{Dynamical conductivity
  of the Fermi arc and the Volkov-Pankratov states on the surface of Weyl
  semimetals}},\ }\href {https://doi.org/10.1103/PhysRevB.100.195412}
  {\bibfield  {journal} {\bibinfo  {journal} {Phys. Rev. B}\ }\textbf {\bibinfo
  {volume} {100}},\ \bibinfo {pages} {195412} (\bibinfo {year}
  {2019})}\BibitemShut {NoStop}%
\bibitem [{\citenamefont {Chan}\ \emph {et~al.}(2016)\citenamefont {Chan},
  \citenamefont {Chiu}, \citenamefont {Chou},\ and\ \citenamefont
  {Schnyder}}]{Chan2016a}%
  \BibitemOpen
  \bibfield  {author} {\bibinfo {author} {\bibfnamefont {Y.-H.}\ \bibnamefont
  {Chan}}, \bibinfo {author} {\bibfnamefont {C.-K.}\ \bibnamefont {Chiu}},
  \bibinfo {author} {\bibfnamefont {M.~Y.}\ \bibnamefont {Chou}},\ and\
  \bibinfo {author} {\bibfnamefont {A.~P.}\ \bibnamefont {Schnyder}},\
  }\bibfield  {title} {\bibinfo {title} {{Ca$_3$P$_2$ and other topological
  semimetals with line nodes and drumhead surface states}},\ }\href
  {https://doi.org/10.1103/PhysRevB.93.205132} {\bibfield  {journal} {\bibinfo
  {journal} {Phys. Rev. B}\ }\textbf {\bibinfo {volume} {93}},\ \bibinfo
  {pages} {205132} (\bibinfo {year} {2016})}\BibitemShut {NoStop}%
\bibitem [{\citenamefont {Fang}\ \emph {et~al.}(2015)\citenamefont {Fang},
  \citenamefont {Chen}, \citenamefont {Kee},\ and\ \citenamefont
  {Fu}}]{Fang2015}%
  \BibitemOpen
  \bibfield  {author} {\bibinfo {author} {\bibfnamefont {C.}~\bibnamefont
  {Fang}}, \bibinfo {author} {\bibfnamefont {Y.}~\bibnamefont {Chen}}, \bibinfo
  {author} {\bibfnamefont {H.-Y.}\ \bibnamefont {Kee}},\ and\ \bibinfo {author}
  {\bibfnamefont {L.}~\bibnamefont {Fu}},\ }\bibfield  {title} {\bibinfo
  {title} {Topological nodal line semimetals with and without spin-orbital
  coupling},\ }\href {https://doi.org/10.1103/PhysRevB.92.081201} {\bibfield
  {journal} {\bibinfo  {journal} {Phys. Rev. B}\ }\textbf {\bibinfo {volume}
  {92}},\ \bibinfo {pages} {081201} (\bibinfo {year} {2015})}\BibitemShut
  {NoStop}%
\bibitem [{\citenamefont {Bernevig}\ and\ \citenamefont
  {Hughes}(2013)}]{Bernevig2013}%
  \BibitemOpen
  \bibfield  {author} {\bibinfo {author} {\bibfnamefont {B.}~\bibnamefont
  {Bernevig}}\ and\ \bibinfo {author} {\bibfnamefont {T.}~\bibnamefont
  {Hughes}},\ }\href@noop {} {\emph {\bibinfo {title} {{Topological Insulators
  and Topological Superconductors}}}}\ (\bibinfo  {publisher} {Princeton
  University Press},\ \bibinfo {year} {2013})\BibitemShut {NoStop}%
\bibitem [{\citenamefont {Chiu}\ and\ \citenamefont
  {Schnyder}(2014)}]{Chiu2014}%
  \BibitemOpen
  \bibfield  {author} {\bibinfo {author} {\bibfnamefont {C.-K.}\ \bibnamefont
  {Chiu}}\ and\ \bibinfo {author} {\bibfnamefont {A.~P.}\ \bibnamefont
  {Schnyder}},\ }\bibfield  {title} {\bibinfo {title} {Classification of
  reflection-symmetry-protected topological semimetals and nodal
  superconductors},\ }\href {https://doi.org/10.1103/PhysRevB.90.205136}
  {\bibfield  {journal} {\bibinfo  {journal} {Phys. Rev. B}\ }\textbf {\bibinfo
  {volume} {90}},\ \bibinfo {pages} {205136} (\bibinfo {year}
  {2014})}\BibitemShut {NoStop}%
\bibitem [{\citenamefont {Fang}\ \emph {et~al.}(2016)\citenamefont {Fang},
  \citenamefont {Weng}, \citenamefont {Dai},\ and\ \citenamefont
  {Fang}}]{Fang2016}%
  \BibitemOpen
  \bibfield  {author} {\bibinfo {author} {\bibfnamefont {C.}~\bibnamefont
  {Fang}}, \bibinfo {author} {\bibfnamefont {H.}~\bibnamefont {Weng}}, \bibinfo
  {author} {\bibfnamefont {X.}~\bibnamefont {Dai}},\ and\ \bibinfo {author}
  {\bibfnamefont {Z.}~\bibnamefont {Fang}},\ }\bibfield  {title} {\bibinfo
  {title} {Topological nodal line semimetals},\ }\href
  {https://doi.org/10.1088/1674-1056/25/11/117106} {\bibfield  {journal}
  {\bibinfo  {journal} {Chinese Physics B}\ }\textbf {\bibinfo {volume} {25}},\
  \bibinfo {pages} {117106} (\bibinfo {year} {2016})}\BibitemShut {NoStop}%
\bibitem [{\citenamefont {Buccheri}\ \emph
  {et~al.}(2022{\natexlab{b}})\citenamefont {Buccheri}, \citenamefont {Egger},\
  and\ \citenamefont {De~Martino}}]{Buccheri2022b}%
  \BibitemOpen
  \bibfield  {author} {\bibinfo {author} {\bibfnamefont {F.}~\bibnamefont
  {Buccheri}}, \bibinfo {author} {\bibfnamefont {R.}~\bibnamefont {Egger}},\
  and\ \bibinfo {author} {\bibfnamefont {A.}~\bibnamefont {De~Martino}},\
  }\bibfield  {title} {\bibinfo {title} {{Transport, refraction, and interface
  arcs in junctions of Weyl semimetals}},\ }\href
  {https://doi.org/10.1103/PhysRevB.106.045413} {\bibfield  {journal} {\bibinfo
   {journal} {Phys. Rev. B}\ }\textbf {\bibinfo {volume} {106}},\ \bibinfo
  {pages} {045413} (\bibinfo {year} {2022}{\natexlab{b}})}\BibitemShut
  {NoStop}%
\bibitem [{\citenamefont {Hasan}\ and\ \citenamefont {Kane}(2010)}]{Hasan2010}%
  \BibitemOpen
  \bibfield  {author} {\bibinfo {author} {\bibfnamefont {M.~Z.}\ \bibnamefont
  {Hasan}}\ and\ \bibinfo {author} {\bibfnamefont {C.~L.}\ \bibnamefont
  {Kane}},\ }\bibfield  {title} {\bibinfo {title} {Colloquium: Topological
  insulators},\ }\href {https://doi.org/10.1103/RevModPhys.82.3045} {\bibfield
  {journal} {\bibinfo  {journal} {Rev. Mod. Phys.}\ }\textbf {\bibinfo {volume}
  {82}},\ \bibinfo {pages} {3045} (\bibinfo {year} {2010})}\BibitemShut
  {NoStop}%
\bibitem [{{\relax DLMF}()}]{NIST:DLMF}%
  \BibitemOpen
  {\relax DLMF},\ \href@noop {} {\bibinfo {title} {{\it NIST Digital Library of
  Mathematical Functions}}},\ \bibinfo {howpublished}
  {\url{https://dlmf.nist.gov/}, Release 1.1.10 of 2023-06-15},\ \bibinfo
  {note} {{F}.~W.~J. Olver, A.~B. {Olde Daalhuis}, D.~W. Lozier, B.~I.
  Schneider, R.~F. Boisvert, C.~W. Clark, B.~R. Miller, B.~V. Saunders, H.~S.
  Cohl, and M.~A. McClain, eds.}\BibitemShut {Stop}%
\bibitem [{Ca3()}]{Ca3p2materials}%
  \BibitemOpen
  \href {https://materialsproject.org/materials/mp-1013547\#crystal\_structure}
  {\bibinfo {title} {Materials project, mp-1013547: Ca3p2 (cubic, pm-3m,
  221)}}\BibitemShut {NoStop}%
\bibitem [{\citenamefont {Li}\ \emph {et~al.}(2009)\citenamefont {Li},
  \citenamefont {Luican}, \citenamefont {dos Santos}, \citenamefont {Neto},
  \citenamefont {Reina}, \citenamefont {Kong},\ and\ \citenamefont
  {Andrei}}]{Li2009}%
  \BibitemOpen
  \bibfield  {author} {\bibinfo {author} {\bibfnamefont {G.}~\bibnamefont
  {Li}}, \bibinfo {author} {\bibfnamefont {A.}~\bibnamefont {Luican}}, \bibinfo
  {author} {\bibfnamefont {J.~M. B.~L.}\ \bibnamefont {dos Santos}}, \bibinfo
  {author} {\bibfnamefont {A.~H.~C.}\ \bibnamefont {Neto}}, \bibinfo {author}
  {\bibfnamefont {A.}~\bibnamefont {Reina}}, \bibinfo {author} {\bibfnamefont
  {J.}~\bibnamefont {Kong}},\ and\ \bibinfo {author} {\bibfnamefont {E.~Y.}\
  \bibnamefont {Andrei}},\ }\bibfield  {title} {\bibinfo {title} {{Observation
  of Van Hove singularities in twisted graphene layers}},\ }\href
  {https://doi.org/10.1038/nphys1463} {\bibfield  {journal} {\bibinfo
  {journal} {Nature Physics}\ }\textbf {\bibinfo {volume} {6}},\ \bibinfo
  {pages} {109} (\bibinfo {year} {2009})}\BibitemShut {NoStop}%
\bibitem [{\citenamefont {Brihuega}\ \emph {et~al.}(2012)\citenamefont
  {Brihuega}, \citenamefont {Mallet}, \citenamefont {Gonz\'alez-Herrero},
  \citenamefont {Trambly~de Laissardi\`ere}, \citenamefont {Ugeda},
  \citenamefont {Magaud}, \citenamefont {G\'omez-Rodr\'{\i}guez}, \citenamefont
  {Yndur\'ain},\ and\ \citenamefont {Veuillen}}]{Brihuega2009}%
  \BibitemOpen
  \bibfield  {author} {\bibinfo {author} {\bibfnamefont {I.}~\bibnamefont
  {Brihuega}}, \bibinfo {author} {\bibfnamefont {P.}~\bibnamefont {Mallet}},
  \bibinfo {author} {\bibfnamefont {H.}~\bibnamefont {Gonz\'alez-Herrero}},
  \bibinfo {author} {\bibfnamefont {G.}~\bibnamefont {Trambly~de
  Laissardi\`ere}}, \bibinfo {author} {\bibfnamefont {M.~M.}\ \bibnamefont
  {Ugeda}}, \bibinfo {author} {\bibfnamefont {L.}~\bibnamefont {Magaud}},
  \bibinfo {author} {\bibfnamefont {J.~M.}\ \bibnamefont
  {G\'omez-Rodr\'{\i}guez}}, \bibinfo {author} {\bibfnamefont {F.}~\bibnamefont
  {Yndur\'ain}},\ and\ \bibinfo {author} {\bibfnamefont {J.-Y.}\ \bibnamefont
  {Veuillen}},\ }\bibfield  {title} {\bibinfo {title} {{Unraveling the
  Intrinsic and Robust Nature of van Hove Singularities in Twisted Bilayer
  Graphene by Scanning Tunneling Microscopy and Theoretical Analysis}},\ }\href
  {https://doi.org/10.1103/PhysRevLett.109.196802} {\bibfield  {journal}
  {\bibinfo  {journal} {Phys. Rev. Lett.}\ }\textbf {\bibinfo {volume} {109}},\
  \bibinfo {pages} {196802} (\bibinfo {year} {2012})}\BibitemShut {NoStop}%
\bibitem [{\citenamefont {Liu}\ \emph {et~al.}(2019)\citenamefont {Liu},
  \citenamefont {Xin}, \citenamefont {Fu}, \citenamefont {Liu}, \citenamefont
  {Song}, \citenamefont {Cui}, \citenamefont {Zhao},\ and\ \citenamefont
  {Zhao}}]{Liu2018}%
  \BibitemOpen
  \bibfield  {author} {\bibinfo {author} {\bibfnamefont {Z.}~\bibnamefont
  {Liu}}, \bibinfo {author} {\bibfnamefont {H.}~\bibnamefont {Xin}}, \bibinfo
  {author} {\bibfnamefont {L.}~\bibnamefont {Fu}}, \bibinfo {author}
  {\bibfnamefont {Y.}~\bibnamefont {Liu}}, \bibinfo {author} {\bibfnamefont
  {T.}~\bibnamefont {Song}}, \bibinfo {author} {\bibfnamefont {X.}~\bibnamefont
  {Cui}}, \bibinfo {author} {\bibfnamefont {G.}~\bibnamefont {Zhao}},\ and\
  \bibinfo {author} {\bibfnamefont {J.}~\bibnamefont {Zhao}},\ }\bibfield
  {title} {\bibinfo {title} {{All-Silicon Topological Semimetals with Closed
  Nodal Line}},\ }\href {https://doi.org/10.1021/acs.jpclett.8b03345}
  {\bibfield  {journal} {\bibinfo  {journal} {J. Phys. Chem. Lett.}\ }\textbf
  {\bibinfo {volume} {10}},\ \bibinfo {pages} {244} (\bibinfo {year}
  {2019})}\BibitemShut {NoStop}%
\bibitem [{\citenamefont {Ramshaw}\ \emph {et~al.}(2018)\citenamefont
  {Ramshaw}, \citenamefont {Modic}, \citenamefont {Shekhter}, \citenamefont
  {Zhang}, \citenamefont {Kim}, \citenamefont {Moll}, \citenamefont {Bachmann},
  \citenamefont {Chan}, \citenamefont {Betts}, \citenamefont {Balakirev},
  \citenamefont {Migliori}, \citenamefont {Ghimire}, \citenamefont {Bauer},
  \citenamefont {Ronning},\ and\ \citenamefont {McDonald}}]{Ramshaw2018}%
  \BibitemOpen
  \bibfield  {author} {\bibinfo {author} {\bibfnamefont {B.~J.}\ \bibnamefont
  {Ramshaw}}, \bibinfo {author} {\bibfnamefont {K.~A.}\ \bibnamefont {Modic}},
  \bibinfo {author} {\bibfnamefont {A.}~\bibnamefont {Shekhter}}, \bibinfo
  {author} {\bibfnamefont {Y.}~\bibnamefont {Zhang}}, \bibinfo {author}
  {\bibfnamefont {E.-A.}\ \bibnamefont {Kim}}, \bibinfo {author} {\bibfnamefont
  {P.~J.~W.}\ \bibnamefont {Moll}}, \bibinfo {author} {\bibfnamefont {M.~D.}\
  \bibnamefont {Bachmann}}, \bibinfo {author} {\bibfnamefont {M.~K.}\
  \bibnamefont {Chan}}, \bibinfo {author} {\bibfnamefont {J.~B.}\ \bibnamefont
  {Betts}}, \bibinfo {author} {\bibfnamefont {F.}~\bibnamefont {Balakirev}},
  \bibinfo {author} {\bibfnamefont {A.}~\bibnamefont {Migliori}}, \bibinfo
  {author} {\bibfnamefont {N.~J.}\ \bibnamefont {Ghimire}}, \bibinfo {author}
  {\bibfnamefont {E.~D.}\ \bibnamefont {Bauer}}, \bibinfo {author}
  {\bibfnamefont {F.}~\bibnamefont {Ronning}},\ and\ \bibinfo {author}
  {\bibfnamefont {R.~D.}\ \bibnamefont {McDonald}},\ }\bibfield  {title}
  {\bibinfo {title} {{Quantum limit transport and destruction of the Weyl nodes
  in TaAs}},\ }\href {https://doi.org/10.1038/s41467-018-04542-9} {\bibfield
  {journal} {\bibinfo  {journal} {Nature Communications}\ }\textbf {\bibinfo
  {volume} {9}},\ \bibinfo {pages} {2217} (\bibinfo {year} {2018})}\BibitemShut
  {NoStop}%
\bibitem [{\citenamefont {Barati}\ and\ \citenamefont
  {Abedinpour}(2017)}]{Barati2017}%
  \BibitemOpen
  \bibfield  {author} {\bibinfo {author} {\bibfnamefont {S.}~\bibnamefont
  {Barati}}\ and\ \bibinfo {author} {\bibfnamefont {S.~H.}\ \bibnamefont
  {Abedinpour}},\ }\bibfield  {title} {\bibinfo {title} {Optical conductivity
  of three and two dimensional topological nodal-line semimetals},\ }\href
  {https://doi.org/10.1103/PhysRevB.96.155150} {\bibfield  {journal} {\bibinfo
  {journal} {Phys. Rev. B}\ }\textbf {\bibinfo {volume} {96}},\ \bibinfo
  {pages} {155150} (\bibinfo {year} {2017})}\BibitemShut {NoStop}%
\bibitem [{\citenamefont {Pronin}\ and\ \citenamefont
  {Dressel}(2021)}]{Pronin2020}%
  \BibitemOpen
  \bibfield  {author} {\bibinfo {author} {\bibfnamefont {A.~V.}\ \bibnamefont
  {Pronin}}\ and\ \bibinfo {author} {\bibfnamefont {M.}~\bibnamefont
  {Dressel}},\ }\bibfield  {title} {\bibinfo {title} {{Nodal Semimetals: A
  Survey on Optical Conductivity}},\ }\href
  {https://doi.org/10.1002/pssb.202000027} {\bibfield  {journal} {\bibinfo
  {journal} {physica status solidi (b)}\ }\textbf {\bibinfo {volume} {258}},\
  \bibinfo {pages} {2000027} (\bibinfo {year} {2021})}\BibitemShut {NoStop}%
\bibitem [{\citenamefont {Lu}\ \emph {et~al.}(2021)\citenamefont {Lu},
  \citenamefont {Mukherjee},\ and\ \citenamefont {Goerbig}}]{Lu2021}%
  \BibitemOpen
  \bibfield  {author} {\bibinfo {author} {\bibfnamefont {X.}~\bibnamefont
  {Lu}}, \bibinfo {author} {\bibfnamefont {D.~K.}\ \bibnamefont {Mukherjee}},\
  and\ \bibinfo {author} {\bibfnamefont {M.~O.}\ \bibnamefont {Goerbig}},\
  }\bibfield  {title} {\bibinfo {title} {{Surface plasmonics of Weyl
  semimetals}},\ }\href {https://doi.org/10.1103/PhysRevB.104.155103}
  {\bibfield  {journal} {\bibinfo  {journal} {Phys. Rev. B}\ }\textbf {\bibinfo
  {volume} {104}},\ \bibinfo {pages} {155103} (\bibinfo {year}
  {2021})}\BibitemShut {NoStop}%
\bibitem [{\citenamefont {Tong}(2016)}]{Tong2016}%
  \BibitemOpen
  \bibfield  {author} {\bibinfo {author} {\bibfnamefont {D.}~\bibnamefont
  {Tong}},\ }\href@noop {} {\bibinfo {title} {{Lectures on the Quantum Hall
  Effect}}} (\bibinfo {year} {2016}),\ \Eprint
  {https://arxiv.org/abs/1606.06687} {arXiv:1606.06687 [hep-th]} \BibitemShut
  {NoStop}%
\bibitem [{\citenamefont {Nie}\ \emph {et~al.}(2021)\citenamefont {Nie},
  \citenamefont {Qian}, \citenamefont {Gao}, \citenamefont {Fang},
  \citenamefont {Weng},\ and\ \citenamefont {Wang}}]{Nie2021}%
  \BibitemOpen
  \bibfield  {author} {\bibinfo {author} {\bibfnamefont {S.}~\bibnamefont
  {Nie}}, \bibinfo {author} {\bibfnamefont {Y.}~\bibnamefont {Qian}}, \bibinfo
  {author} {\bibfnamefont {J.}~\bibnamefont {Gao}}, \bibinfo {author}
  {\bibfnamefont {Z.}~\bibnamefont {Fang}}, \bibinfo {author} {\bibfnamefont
  {H.}~\bibnamefont {Weng}},\ and\ \bibinfo {author} {\bibfnamefont
  {Z.}~\bibnamefont {Wang}},\ }\bibfield  {title} {\bibinfo {title}
  {Application of topological quantum chemistry in electrides},\ }\href
  {https://doi.org/10.1103/PhysRevB.103.205133} {\bibfield  {journal} {\bibinfo
   {journal} {Phys. Rev. B}\ }\textbf {\bibinfo {volume} {103}},\ \bibinfo
  {pages} {205133} (\bibinfo {year} {2021})}\BibitemShut {NoStop}%
\bibitem [{\citenamefont {Ould-Mohamed}\ \emph {et~al.}(2022)\citenamefont
  {Ould-Mohamed}, \citenamefont {Boukri},\ and\ \citenamefont
  {Ouahrani}}]{Ould2021}%
  \BibitemOpen
  \bibfield  {author} {\bibinfo {author} {\bibfnamefont {M.}~\bibnamefont
  {Ould-Mohamed}}, \bibinfo {author} {\bibfnamefont {K.}~\bibnamefont
  {Boukri}},\ and\ \bibinfo {author} {\bibfnamefont {T.}~\bibnamefont
  {Ouahrani}},\ }\bibfield  {title} {\bibinfo {title} {Structural, vibrational,
  electronic, thermodynamic and thermoelectric properties of {CaCdSi} half
  {Heusler} compound: A first-principles study},\ }\href
  {https://doi.org/https://doi.org/10.1016/j.mtcomm.2022.104668} {\bibfield
  {journal} {\bibinfo  {journal} {Materials Today Communications}\ }\textbf
  {\bibinfo {volume} {33}},\ \bibinfo {pages} {104668} (\bibinfo {year}
  {2022})}\BibitemShut {NoStop}%
\bibitem [{\citenamefont {Hosen}\ \emph {et~al.}(2017)\citenamefont {Hosen},
  \citenamefont {Dimitri}, \citenamefont {Belopolski}, \citenamefont
  {Maldonado}, \citenamefont {Sankar}, \citenamefont {Dhakal}, \citenamefont
  {Dhakal}, \citenamefont {Cole}, \citenamefont {Oppeneer}, \citenamefont
  {Kaczorowski}, \citenamefont {Chou}, \citenamefont {Hasan}, \citenamefont
  {Durakiewicz},\ and\ \citenamefont {Neupane}}]{Hosen2017}%
  \BibitemOpen
  \bibfield  {author} {\bibinfo {author} {\bibfnamefont {M.~M.}\ \bibnamefont
  {Hosen}}, \bibinfo {author} {\bibfnamefont {K.}~\bibnamefont {Dimitri}},
  \bibinfo {author} {\bibfnamefont {I.}~\bibnamefont {Belopolski}}, \bibinfo
  {author} {\bibfnamefont {P.}~\bibnamefont {Maldonado}}, \bibinfo {author}
  {\bibfnamefont {R.}~\bibnamefont {Sankar}}, \bibinfo {author} {\bibfnamefont
  {N.}~\bibnamefont {Dhakal}}, \bibinfo {author} {\bibfnamefont
  {G.}~\bibnamefont {Dhakal}}, \bibinfo {author} {\bibfnamefont
  {T.}~\bibnamefont {Cole}}, \bibinfo {author} {\bibfnamefont {P.~M.}\
  \bibnamefont {Oppeneer}}, \bibinfo {author} {\bibfnamefont {D.}~\bibnamefont
  {Kaczorowski}}, \bibinfo {author} {\bibfnamefont {F.}~\bibnamefont {Chou}},
  \bibinfo {author} {\bibfnamefont {M.~Z.}\ \bibnamefont {Hasan}}, \bibinfo
  {author} {\bibfnamefont {T.}~\bibnamefont {Durakiewicz}},\ and\ \bibinfo
  {author} {\bibfnamefont {M.}~\bibnamefont {Neupane}},\ }\bibfield  {title}
  {\bibinfo {title} {Tunability of the topological nodal-line semimetal phase
  in $\mathrm{ZrSi}x$-type materials ($x=\mathrm{S}, \mathrm{Se},
  \mathrm{Te}$)},\ }\href {https://doi.org/10.1103/PhysRevB.95.161101}
  {\bibfield  {journal} {\bibinfo  {journal} {Phys. Rev. B}\ }\textbf {\bibinfo
  {volume} {95}},\ \bibinfo {pages} {161101} (\bibinfo {year}
  {2017})}\BibitemShut {NoStop}%
\bibitem [{\citenamefont {Roy}\ \emph {et~al.}(2014)\citenamefont {Roy},
  \citenamefont {Sau},\ and\ \citenamefont {Das~Sarma}}]{Roy2014}%
  \BibitemOpen
  \bibfield  {author} {\bibinfo {author} {\bibfnamefont {B.}~\bibnamefont
  {Roy}}, \bibinfo {author} {\bibfnamefont {J.~D.}\ \bibnamefont {Sau}},\ and\
  \bibinfo {author} {\bibfnamefont {S.}~\bibnamefont {Das~Sarma}},\ }\bibfield
  {title} {\bibinfo {title} {{Migdal's theorem and electron-phonon vertex
  corrections in Dirac materials}},\ }\href
  {https://doi.org/10.1103/PhysRevB.89.165119} {\bibfield  {journal} {\bibinfo
  {journal} {Phys. Rev. B}\ }\textbf {\bibinfo {volume} {89}},\ \bibinfo
  {pages} {165119} (\bibinfo {year} {2014})}\BibitemShut {NoStop}%
\bibitem [{\citenamefont {Mahler}\ \emph {et~al.}(2019)\citenamefont {Mahler},
  \citenamefont {Mayer}, \citenamefont {Leubner}, \citenamefont {Lunczer},
  \citenamefont {Di~Sante}, \citenamefont {Sangiovanni}, \citenamefont
  {Thomale}, \citenamefont {Hankiewicz}, \citenamefont {Buhmann}, \citenamefont
  {Gould},\ and\ \citenamefont {Molenkamp}}]{Mahler2019}%
  \BibitemOpen
  \bibfield  {author} {\bibinfo {author} {\bibfnamefont {D.~M.}\ \bibnamefont
  {Mahler}}, \bibinfo {author} {\bibfnamefont {J.-B.}\ \bibnamefont {Mayer}},
  \bibinfo {author} {\bibfnamefont {P.}~\bibnamefont {Leubner}}, \bibinfo
  {author} {\bibfnamefont {L.}~\bibnamefont {Lunczer}}, \bibinfo {author}
  {\bibfnamefont {D.}~\bibnamefont {Di~Sante}}, \bibinfo {author}
  {\bibfnamefont {G.}~\bibnamefont {Sangiovanni}}, \bibinfo {author}
  {\bibfnamefont {R.}~\bibnamefont {Thomale}}, \bibinfo {author} {\bibfnamefont
  {E.~M.}\ \bibnamefont {Hankiewicz}}, \bibinfo {author} {\bibfnamefont
  {H.}~\bibnamefont {Buhmann}}, \bibinfo {author} {\bibfnamefont
  {C.}~\bibnamefont {Gould}},\ and\ \bibinfo {author} {\bibfnamefont {L.~W.}\
  \bibnamefont {Molenkamp}},\ }\bibfield  {title} {\bibinfo {title} {{Interplay
  of Dirac Nodes and Volkov-Pankratov Surface States in Compressively Strained
  HgTe}},\ }\href {https://doi.org/10.1103/PhysRevX.9.031034} {\bibfield
  {journal} {\bibinfo  {journal} {Phys. Rev. X}\ }\textbf {\bibinfo {volume}
  {9}},\ \bibinfo {pages} {031034} (\bibinfo {year} {2019})}\BibitemShut
  {NoStop}%
\bibitem [{\citenamefont {Pereira}\ \emph {et~al.}(2019)\citenamefont
  {Pereira}, \citenamefont {Buccheri}, \citenamefont {De~Martino},\ and\
  \citenamefont {Egger}}]{Pereira2019}%
  \BibitemOpen
  \bibfield  {author} {\bibinfo {author} {\bibfnamefont {R.~G.}\ \bibnamefont
  {Pereira}}, \bibinfo {author} {\bibfnamefont {F.}~\bibnamefont {Buccheri}},
  \bibinfo {author} {\bibfnamefont {A.}~\bibnamefont {De~Martino}},\ and\
  \bibinfo {author} {\bibfnamefont {R.}~\bibnamefont {Egger}},\ }\bibfield
  {title} {\bibinfo {title} {{Superconductivity from piezoelectric interactions
  in Weyl semimetals}},\ }\href {https://doi.org/10.1103/PhysRevB.100.035106}
  {\bibfield  {journal} {\bibinfo  {journal} {Phys. Rev. B}\ }\textbf {\bibinfo
  {volume} {100}},\ \bibinfo {pages} {035106} (\bibinfo {year}
  {2019})}\BibitemShut {NoStop}%
\bibitem [{\citenamefont {Gariglio}\ \emph {et~al.}(2009)\citenamefont
  {Gariglio}, \citenamefont {Reyren}, \citenamefont {Caviglia},\ and\
  \citenamefont {Triscone}}]{Gariglio2009}%
  \BibitemOpen
  \bibfield  {author} {\bibinfo {author} {\bibfnamefont {S.}~\bibnamefont
  {Gariglio}}, \bibinfo {author} {\bibfnamefont {N.}~\bibnamefont {Reyren}},
  \bibinfo {author} {\bibfnamefont {A.~D.}\ \bibnamefont {Caviglia}},\ and\
  \bibinfo {author} {\bibfnamefont {J.-M.}\ \bibnamefont {Triscone}},\
  }\bibfield  {title} {\bibinfo {title} {{Superconductivity at the
  LaAlO$_3$/SrTiO$_3$ interface}},\ }\href
  {https://doi.org/10.1088/0953-8984/21/16/164213} {\bibfield  {journal}
  {\bibinfo  {journal} {Journal of Physics: Condensed Matter}\ }\textbf
  {\bibinfo {volume} {21}},\ \bibinfo {pages} {164213} (\bibinfo {year}
  {2009})}\BibitemShut {NoStop}%
\bibitem [{\citenamefont {Lebedev}\ and\ \citenamefont
  {Silverman}(1972)}]{Lebedev}%
  \BibitemOpen
  \bibfield  {author} {\bibinfo {author} {\bibfnamefont {N.}~\bibnamefont
  {Lebedev}}\ and\ \bibinfo {author} {\bibfnamefont {R.}~\bibnamefont
  {Silverman}},\ }\href@noop {} {\emph {\bibinfo {title} {Special Functions and
  Their Applications}}}\ (\bibinfo  {publisher} {Dover Publications},\ \bibinfo
  {year} {1972})\BibitemShut {NoStop}%
\end{thebibliography}%

\end{document}